\documentclass[12pt]{article}
\pdfoutput=1
\usepackage{jheppub}
\usepackage{amsmath}
\usepackage{amsfonts}
\usepackage{amssymb}
\usepackage{latexsym}
\usepackage{tkz-graph}
\usepackage{tkz-euclide}
\RequirePackage{amssymb,amsmath,amsthm,amsfonts}
\RequirePackage{bbm}
\numberwithin{equation}{section}

\usepackage{mathtools}

\usepackage{epsf}
\usepackage{color}
\usepackage{graphicx}
\usepackage{subfig}
\usepackage{dsfont}

\usepackage[export]{adjustbox}
\usepackage{hyperref}
\definecolor{darkred}{rgb}{0.8,0.1,0.1}
\hypersetup{colorlinks=true, linkcolor=darkred, citecolor=blue, linktoc=page}

\usepackage[utf8]{inputenc} 

\AtBeginDocument{}

\def\tr{{\rm tr}}

\newcommand{\Ham}{\mathcal{H}}
\newcommand{\RR}{\ensuremath{\mathbb R}}
\newcommand{\NN}{\ensuremath{\mathbb N}}
\newcommand{\ZZ}{\ensuremath{\mathbb Z}}

\newcommand{\ra}{\ensuremath{\rightarrow}}
\newcommand{\bpm}{\ensuremath{\begin{pmatrix}}}
\newcommand{\epm}{\ensuremath{\end{pmatrix}}}
\newcommand{\ket}[1]{\left| #1 \right> }
\newcommand{\bra}[1]{\left< #1 \right| }
\newcommand{\expt}[1]{\left< #1 \right> }
\newcommand{\innn}[3]{\left< #1 \left| #2 \right| #3 \right>}
\newcommand{\inn}[2]{\left< #1 \left| #2 \right. \right>}

\DeclareMathOperator{\intinf}{\int_{-\infty}^{\infty}}

\usepackage{comment}

\title{The double scaled limit of Super--Symmetric SYK models}
\author[a]{Micha Berkooz,}
\author[a]{Nadav Brukner,}
\author[a,b]{Vladimir Narovlansky,}
\author[a]{and Amir Raz}
\emailAdd{micha.berkooz@weizmann.ac.il}
\emailAdd{nadav.brukner@weizmann.ac.il}
\emailAdd{n.vladi@gmail.com}
\emailAdd{a.raz@weizmann.ac.il}
\affiliation[a]{Department of Particle Physics and Astrophysics, \\ Weizmann Institute of Science, Rehovot 7610001, Israel}
\affiliation[b]{Princeton Center for Theoretical Science, \\  Princeton University, Princeton, NJ 08544, USA}

\abstract{ We compute the exact density of states and 2-point function of the $\mathcal{N} =2$ super-symmetric SYK model in the large $N$ double-scaled limit, by using combinatorial tools that relate the moments of the distribution
to sums over oriented chord diagrams. In particular we show how SUSY is realized on the (highly degenerate) Hilbert space of chords. We further calculate analytically the number of ground states of the model in each charge sector at finite $N$, and compare it to the results from the double-scaled limit.  Our results reduce to the super-Schwarzian action in the low energy short interaction length limit. They imply that the conformal ansatz of the 2-point function is inconsistent due to the large number of ground states, and we show how to add this contribution. We also discuss the relation of the model to $SL_q(2|1)$. For completeness we present an overview of the $\mathcal{N}=1$ super-symmetric SYK model in the large $N$ double-scaled limit.
}

\begin{document}

\setcounter{footnote}{0}

\maketitle

\setcounter{equation}{0}
\setcounter{footnote}{0}

\section{Introduction}

The Sachdev-Ye-Kitaev (SYK) model consists of $N$ Majorana fermions with random all-to-all interactions \cite{Sachdev_1993,Kitaev_talk}. It has recently gained substantial attention as a simple toy model that is both solvable and maximally chaotic \cite{MaldecenaStanford,Polchinski_2016,Rosenhaus_2019,Bagrets_2016,Bagrets_2017}. Moreover, the SYK model has a nearly conformal symmetry in the IR, and fluctuations around it are described by a Schwarzian effective action, which is the same dynamics that describes JT gravity on $AdS_2$ \cite{NADSbreaking,Almheiri2015,Stanford_Witten2017,Sachdev_2019}. Thus the SYK model has emerged as a tractable example of $AdS_2/CFT_1$ holography \cite{Kitaev_talk,SYK_bulk_dual,Kitaev_2018}, creating a basic setup to study the problems of quantum gravity, including black hole thermodynamics and the information paradox \cite{Jensen_2016,Kitaev_2019,Cotler:2016fpe,Davison_2017,Garc_a_Garc_a_2016,Lam_2018}.

The SYK model is typically studied in the large $N$ limit, with the length of the interaction is taken to be fixed, where to leading order in $N$ only melonic diagrams contribute to the 2-point function, and ladder diagrams to the 4-point function. Higher point correlation functions in the conformal limit were also computed \cite{Gross_2017}. Additionally, the model has seen several generalizations, including complex fermions \cite{Sachdev_2015,gu2019notes}, fermions in higher dimensions \cite{Liu_2019,Murugan_2017,HigherSYK,Gu_2017,Chaturvedi_2018}, similar tensor models without disorder \cite{WittenSYK,Klebanov_2017}, and others \cite{GenSYK,cSYK}. The fine grain level spacing of the model (after unfolding the spectrum) has also been studied, with a complete classification of the adjacent level spacing statistics through random matrix theory universality classes  \cite{sun2019periodic,Kanazawa_2017,stanford2019jt}, and its applications to the long time behavior of the spectral form factor \cite{Cotler:2016fpe,Altland_2018,saad2018}.

The SYK model has also been studied in {\it the double scaled limit}, where the length of the interaction is taken to scale as $\sqrt{N}$. In this limit the model has a well defined asymptotic density of states, which can be calculated through the tools of random matrix theory \cite{Cotler:2016fpe,feng2018spectrum,Erdos14,Micha2018}.  Correlation functions have been calculated in this limit using the technique of chord diagrams \cite{Micha2018,Berkooz_2019}. We note that recently this limit has been connected to $q$-Brownian motion processes \cite{Speicher2019}.


The SYK model has natural $\mathcal{N} =1$ and $\mathcal{N} =2$ supersymmetric extensions \cite{SusySYK}, which have applications to the study of supersymmetric black holes. The $\mathcal{N} =1$ model is very similar to the regular SYK model, as the supersymmetric charge is simply the regular SYK Hamiltonian with an odd interaction length. As such, its correlation functions, asymptotic density of states, and classifications of level spacing statistics are similar to the regular SYK model, and have been studied extensively \cite{SusySYK,Garc_a_Garc_a_2018,Narayan_2018,Li_2017,feng2018spectrum,Murugan_2017,Kanazawa_2017,sun2019periodic}. The $\mathcal{N} = 2$ supersymmetric SYK model, on the other hand, has not been studied in the double scaling limit, and is much less understood. This model has many interesting features that are absent from the $\mathcal{N} =1$ model, including a $U(1)$ R-symmetry and a large amount of exact ground states which leave the supersymmetry unbroken at finite $N$.

In this paper we primarily focus on the $\mathcal{N} = 2$ supersymmetric SYK model in the double scaled limit, extending the chord diagram and transfer matrix methods in \cite{Micha2018,Berkooz_2019} to treat this model as well. This requires the introduction of new tools from q-brownian motion, namely the Hilbert space metric associated with such processes, which turns out to be highly degenerate in our case (which is a key fact in solving the model). Our main result is an analytic expression for the asymptotic density of states in the double scaled limit.  We also present an analysis of the number of ground states at finite $N$.

Additionally, we use this formalism to calculate the 2-point functions in the double scaled limit. We note that the simple conformal ansatz assumed for such correlators - $1/x^{2\Delta}$ and its finite temperature counterpart - does not hold in this model due to the large number of degenerate ground states, and we show how to correct it.
Finally, we connect our results to the quantum group $sl_q(2|1)$.

The paper is organized as followed: In \textbf{Section 2} we review the chord diagrammatic treatment of the (Majorana) SYK model in the double scaled limit, which directly generalizes to the $\mathcal{N}=1$ supersymmetric SYK model.  The main novelty in the latter case, is that the effective Hamiltonian is built from $q$-deformed fermionic creation and annihilation operators.

 We proceed to define the  $\mathcal{N}=2$ supersymmetric SYK model in \textbf{Section 3}, followed by a brief summary of known results. In \textbf{Section 4} we build the chord diagram and transfer matrix formalism for the $\mathcal{N}=2$ model. A priori the chord Hilbert space is exponentially more complex than the $\mathcal{N}=0,1$ cases. It is reduced to a tractable size by constructing a canonical metric on the space of chords (which is a variant of the metric used for multi-species chord diagrams in the discussions of $q$-brownian motion) and eliminating the zero norm states.

 Our main results are attained in \textbf{Section 5}, where we calculate the moments using the transfer matrix formalism, and find an analytical expression for the asymptotic density of states. Additionally in this section we present an analysis of the supersymmetric states at both finite $N$ and in the double scaled limit. In particular, the number of ground states constitutes a finite fraction of the total number of states in the double scaled limit. \textbf{Section 6} is dedicated to relating the transfer matrix to the Hamiltonian of the super Liouville theory, showing it agrees with the super-conformal limit of the model. In \textbf{Section 7} we use the transfer matrix formalism to compute 2-point correlation functions of random charged operators in the theory. We provide an exact expression for the additional contribution of the ground states, which does not have the standard falloff behavior usually assumed in the SYK model. Finally, we relate the transfer matrix formalism to the quantum group $sl_q(2|1)$ in \textbf{Section 8}.

\section{Majorana SYK and $\mathcal{N} = 1$ super-symmetric SYK} \label{sec:N=1}

In this section we review the original Majorana SYK model, and its combinatorial solution in the double scaled limit (following \cite{Erdos14,Micha2018,Berkooz_2019}). In parallel we discuss the $\mathcal{N} =1$ super-symmetric version of it, which is similar in nature. Readers who are familiar with the chord diagram method for calculating the spectrum of the SYK model in the double scaled limit can skip to the next section for the $\mathcal{N}=2$ SYK model.

\medskip

{\it Definitions:} The (Majorana) SYK model is a quantum mechanical model of $N$ Majorana fermions that satisfy the canonical anti-commutation relations
\begin{equation}
\{\psi_i,\psi_j\} = \delta_{ij},
\end{equation}
and the Hamiltonian is given by
\begin{equation}
H = i^{p/2} \sum_{1 \le i_1<\cdots <i_p \le N} C_{i_1i_2 \cdots i_p} \psi_{i_1} \cdots \psi_{i_p} \qquad \text{(Majorana SYK)},
\end{equation}
where the $C_{i_1 \cdots i_p} $ are independent random couplings with distribution specified below.
In fact, let us use a shorthand notation where $J = \{i_1,\ldots,i_p\}$ stands for an index set of $p$ distinct sites, and
\begin{equation}
\Psi_J \equiv \psi_{i_1}\psi_{i_2}\ldots \psi_{i_p},\ \ 1\leq i_1<i_2<\ldots<i_p\leq N, \ \ J = \{i_1,\ldots,i_p\}.
\end{equation} In this notation we can write compactly $H=i^{p/2} \sum _J C_J \Psi_J$.

The $\mathcal{N} =1$ super-symmetric SYK model is defined very similarly; the super-symmetric charge takes the form of
\begin{equation}
Q = i^{(p-1)/2} \sum_{J} C_J \Psi_J,
\end{equation}
and the Hamiltonian is simply
\begin{equation}
H = Q^2 = i^{p-1} \sum_{J_1,J_2} C_{J_1}C_{J_2} \Psi_{J_1} \Psi_{J_2} \qquad \text{($\mathcal{N} =1$ SYK)}.
\end{equation}
Evidently the Majorana SYK Hamiltonian and the $\mathcal{N} =1$ supercharge $Q$ take the same form, and therefore we will discuss the two in parallel. The only difference between them is that for Majorana SYK $p$ is even, while for $Q$, $p$ is odd (and the overall phase is chosen appropriately).

The random couplings $C_J$ are taken to be real Gaussian, with zero mean, and variance
\begin{equation}
\langle  C_J C_{J'} \rangle_C =
\begin{cases}
2 \binom{N}{p} ^{-1} \mathcal{J}^2 & \text{for Majorana SYK}, \\
2 \binom{N}{p} ^{-1} \mathcal{J} & \text{for $\mathcal{N} =1$ SYK}.
\end{cases}
\end{equation}

We will work in a double scaled limit in which
\begin{align}
	N\to\infty,\qquad p\to\infty,\qquad\lambda\equiv\frac{2p^{2}}{N}=\text{fixed},
\end{align}
and we will find it useful to define the parameter
\begin{equation}
q\equiv e^{-\lambda}  .
\end{equation}

\begin{figure} [h]
	\centering
	\includegraphics[page=1,width=0.25\textwidth]{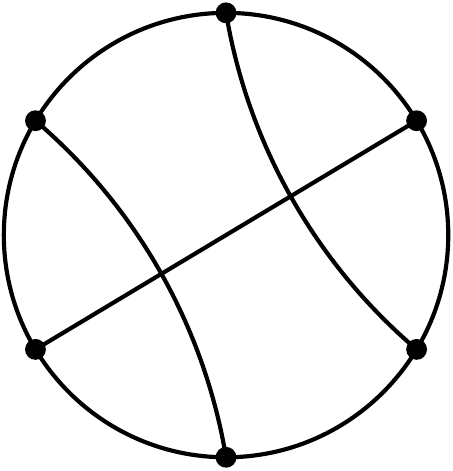}
	\hspace{0.1\columnwidth}
	\includegraphics[page=1,width=0.6\textwidth]{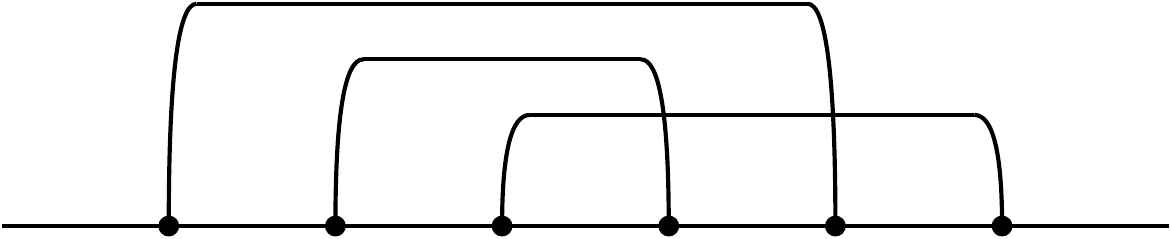}
	\caption{An example of a chord diagram (left: on a circle, right: on a line).}
	\label{fig:real_chord}
\end{figure}

\medskip
{\it Moments and chord diagrams:} Since we are dealing with a random Hamiltonian, we are interested in calculating the expected spectral density function in the double scaled limit. To do so, it is sufficient to calculate the moments $ m_{k}=\left\langle \text{tr \ensuremath{\left(H^{k}\right)}}\right\rangle _{C} $, and from them infer the eigenvalue distribution. The moments $m_k$ are given by
\begin{equation}
m_k = i^{kp/2} \sum _{J_1,J_2,\cdots ,J_k} \langle C_{J_1} \cdots C_{J_k} \rangle _C \tr \left[ \Psi_{J_1} \cdots \Psi_{J_k} \right]  \qquad \text{for Majorana SYK},
\end{equation}
while
\begin{equation}
m_k= i^{2k(p-1)/2} \sum _{J_1,J_2,\cdots ,J_{2k} } \langle C_{J_1} \cdots C_{J_{2k} } \rangle _C \tr \left[ \Psi_{J_1} \cdots \Psi_{J_{2k} }\right]  \qquad \text{for $\mathcal{N} =1$ SYK}.
\end{equation}
By Wick's theorem for the couplings' averaging, we should sum over all possible Wick contractions of the $C_{J_i}$'s. We represent each configuration of Wick contractions by a chord diagram (see left hand side of figure \ref{fig:real_chord}) --- in the Majorana SYK (or $\mathcal{N} =1$ SYK) we mark $k$ ($2k$ respectively) nodes on a circle, each node corresponding to an Hamiltonian (supercharge) insertion, and we connect pairs of nodes by chords, representing the Wick contractions. (The chords here have no orientation, contrary to the case of complex fermions, which we discuss below.)

Each chord diagram is evaluated as follows. We commute $\Psi_J$'s across one another so that contracted pairs appear next to each other. This should therefore be done for each intersection of chords. When commuting $\Psi_J$ with $\Psi_{J'} $, from the fermionic algebra we get a factor of $(-1)^{p^2-|J \cap J'|} = \pm (-1)^{|J \cap J'|} $ where $|J \cap J'|$ is the number of sites in the intersection $J \cap J'$, with the plus sign corresponding to the Majorana SYK and the minus one to $\mathcal{N} =1$ SYK (because of the difference in the parity of $p$). Each such intersection is Poisson distributed (with mean $p^2/N$) as explained in \cite{Erdos14} (or see also \cite{Micha2018,Berkooz_2019}), so that summing over the Poisson weight we get that \emph{each chord intersection is assigned a value of} $ \pm \sum _{|J \cap J'|} P_{\text{Pois}(p^2/N)}(|J \cap J'|)(-1)^{|J \cap J'|} = \pm q$.
Therefore, the $k$'th moment is given by
\begin{equation}
m_k =\langle  \tr H^k\rangle_C = \mathcal{J}^{k} \sum_{\pi}\left(\pm q\right)^{c(\pi)},
\end{equation}
where the sum runs over all chord diagrams (with $k$ vertices for Majorana SYK case, and $2k$ vertices for $\mathcal{N} =1$ SYK), and $c(\pi)$ is the number of intersections in the chord diagram.

\medskip

{\it Transfer matrix:} It will also be useful for us to review the transfer matrix method to evaluate the sum over chord diagrams, following \cite{Micha2018} (see also \cite{Berkooz_2019}). We can draw the same chord diagrams (Wick contractions) on a line rather than a circle (picking an arbitrary starting point), see the right hand side of figure \ref{fig:real_chord} for an example. Then we can provide an effective description of the system in another form. Between each two nodes on the line, the state of the system is determined according to the number of open chords $l$. Thus, we construct an auxiliary Hilbert space spanned by the basis $|l\rangle $ for $l=0,1,2,\cdots $, where the state $|l\rangle $ represents having $l$ chords. As we scan the line, the state $|l\rangle $ can become either $|l-1\rangle $ or $|l+1\rangle $ after passing by a node. Each such transition is assigned a power of $\pm q$ according to the number of chords that intersect in case a chord is closed, that is, going from $|l\rangle $ to $|l-1\rangle $. Thus, each node (Hamiltonian or supercharge insertion) is represented in the auxiliary Hilbert space by a transfer matrix given be
\begin{equation}
T= \begin{pmatrix}
0 & \frac{1-q}{1-q} & 0 & 0 & \cdots \\
1 & 0 & \frac{1-q^2}{1-q} & 0 & \cdots \\
0 & 1 & 0 & \frac{1-q^3}{1-q} & \cdots \\
0 & 0 & 1 & 0 & \cdots \\
\vdots & \vdots & \vdots & \vdots & \ddots
\end{pmatrix} \qquad \text{for Majorana SYK},
\end{equation}
or
\begin{equation}
\mathcal{Q} = \begin{pmatrix}
0 & \frac{1+q}{1+q} & 0 & 0 & \cdots \\
1 & 0 & \frac{1-q^2}{1+q} & 0 & \cdots \\
0 & 1 & 0 & \frac{1+q^3}{1+q} & \cdots \\
0 & 0 & 1 & 0 & \cdots \\
\vdots & \vdots & \vdots & \vdots & \ddots
\end{pmatrix} \qquad \text{for $\mathcal{N} =1$ SYK}.
\end{equation}
(The sign in the numerator of the terms above the diagonal is alternating in $\mathcal{Q} $.) The Hamiltonian in the $\mathcal{N} =1$ case corresponds to the matrix $\mathcal{Q} ^2$.

Since we start and end with no open chords, that is the state $|0\rangle $, the sum over chord diagrams is given by
\begin{equation}
m_k =
\begin{cases}
\mathcal{J}^k \langle 0| T^k |0\rangle & \text{for Majorana SYK},\\
\mathcal{J}^k \langle 0| \mathcal{Q} ^{2k} |0\rangle & \text{for $\mathcal{N} =1$ SYK}.
\end{cases}
\end{equation}

These transfer matrices can be written in terms of $q$-deformed oscillators (see \cite{Berkooz_2019} and \cite{Speicher2019}). For the Majorana SYK we have $T=a_q+a_q^\dagger$, where $a_q,a_q^\dagger$ are $q$-deformed creation-annihilation operators that satisfy $a_q a_q^\dagger - q a_q^\dagger a_q = 1$.
In the $\mathcal{N} = 1$ case $\mathcal{Q}=b_q+b_q^\dagger$ where $b_q,b^\dagger_q$ are $q$-deformed fermionic creation-annihilation operators that satisfy $b_q b_q^\dagger + q b_q^\dagger b_q = 1$.

Thus far we reviewed Majorana SYK and mentioned its analogy in $\mathcal{N} =1$ SYK, and from now on in this section we concentrate on getting the density of states for the $\mathcal{N} =1$ model. The matrices in the auxiliary Hilbert space were diagonalized in \cite{Micha2018,Berkooz_2019} (the analysis there is valid for any sign of $q$), leading to the following result for the $\mathcal{N} =1$ moments
\begin{equation}
m_k = \int _0^{\pi } \frac{d\theta }{2\pi } (-q,e^{\pm 2i\theta } ;-q)_{\infty } \left(\frac{2\sqrt{\mathcal{J}}\cos(\theta )}{\sqrt{1+q}} \right) ^{2k} .
\end{equation}
The energies are therefore
\begin{equation}
E(\theta ) = \frac{4\mathcal{J} \cos(\theta )^2}{1+q} ,
\end{equation}
which are indeed positive definite, and the density of states is
\begin{equation}
  \rho (E)= \frac{1+q}{4\pi \mathcal{J}} (-q,e^{\pm 2i\theta } ;-q)_{\infty } \frac{1}{\sin(2\theta )},
  ~~~~~~~~~~~~~ \theta  = \arccos \left(\sqrt{\frac{E(1+q)}{4\mathcal{J}} } \right).
\end{equation}

In the Majorana double-scaled SYK model, one can take the $q \to 1$ limit (corresponding to the usual SYK model where $p$ is kept fixed) and reproduce the Schwarzian results by taking the energies to be small and scaling them appropriately with $\lambda  \to 0$ \cite{Cotler:2016fpe}. We can do the same for $\mathcal{N} =1$ SYK, by using the triple scaling
\begin{equation}
\lambda  \to 0,\qquad \frac{\sqrt{E/\mathcal{J}}}{\lambda } =\text{fixed} .
\end{equation}
This can be implemented by going to the variable $y$ which is defined by\footnote{Note that the reference point here is $\theta =\pi /2$ rather than $\pi $ which is used in Majorana double-scaled SYK, since the lowest energy here is $E=0$.}
\begin{equation}
\theta  = \frac{\pi }{2} -\lambda y
\end{equation}
so that $y$ is kept fixed as $\lambda  \to 0$. Indeed, its relation to the energy is
\begin{equation}
\sin(\lambda y) = \sqrt{\frac{E(1+q)}{4\mathcal{J}} } \Rightarrow \lambda y \approx \sqrt{\frac{E}{2\mathcal{J}} }.
\end{equation}

Let us evaluate the density of states in this triple scaling. In terms of the $y$ variable
\begin{equation}
\rho  = \frac{1+q}{4\pi \mathcal{J}} \left( -q,-e^{\pm 2i\lambda y} ;-q\right) _{\infty } \frac{1}{\sin(2\lambda y)} ,
\end{equation}
which for small $\lambda $ is approximately\footnote{Recall that in the Schwarzian, describing the low energy of the Majorana SYK model, we find instead that the density of states is a $\sinh$ with an argument proportional to $\sqrt{E}$.}
\begin{equation} \label{eq:low_energy_density_N1}
\rho  \propto \frac{e^{-2\lambda y^2} }{\sin(\lambda y)} \cosh(\pi y) \propto
\frac{1}{\sqrt{E}} \cosh\left( \frac{\pi }{\lambda } \sqrt{\frac{1}{2\mathcal{J}}} \sqrt{E} \right)  .
\end{equation}

As was mentioned, for $p$ being independent of $N$, the low energy of SYK is described by the Schwarzian action. The degrees of freedom of the Schwarzian theory are elements of $\text{Diff}(S^1)/SL(2,R )$, that is, monotonic functions $\phi (\tau )$ such that $\phi (\tau +2\pi )=\phi (\tau )+2\pi $. The $SL(2, \mathbb{R} )$ acts on $f= \tan \frac{\phi }{2} $ by $f \to \frac{af+b}{cf+d} $. This space is a symplectic manifold, and this fact was used in \cite{Stanford_Witten2017} to obtain the exact partition function of the theory, with the result being in agreement with that found using the triple-scaled limit \cite{Cotler:2016fpe}. (The path integral of the theory with the symplectic measure, can be written as the path integral over the original degrees of freedom $\phi (\tau )$ together with additional fermionic fields that behave as $d\phi (\tau )$, with the usual measure; the obtained action has a fermionic symmetry, so that fermionic localization can be used to evaluate it.) The case of the $\mathcal{N} =1$ super Schwarzian theory was evaluated in \cite{Stanford_Witten2017} as well. The density of states that is found there for this case is
\begin{equation}
\rho (E) \propto \frac{\cosh(2\pi  \sqrt{2CE})}{E^{1/2} } , \qquad E \ge 0
\end{equation}
where $C$ is the coupling of the super Schwarzian theory. For energies approaching zero, the density of states grows as $E^{-1/2} $. In the double-scaled limit of Majorana SYK, the spectrum is symmetric around $E=0$, while in the $\mathcal{N} =1$ case it is cut at $E=0$ as we saw, accounting for the decrease in density with increasing energy. We see that the triple-scaled $\mathcal{N} =1$ result \eqref{eq:low_energy_density_N1} indeed agrees with that of $\mathcal{N} =1$ super Schwarzian with the symplectic measure \cite{Stanford_Witten2017}.

We note that any sign of the discrete level spacing in not seen in this analysis, as we only consider single trace quantities that are averaged over the random couplings. Recent progress has been made in analyzing the level spacing in the SYK model,  JT gravity, and their relation to random matrix theory ensembles \cite{Cotler:2016fpe,Kanazawa_2017,saad2018,sun2019periodic,stanford2019jt,Saad:2019lba}. Such analysis requires considering double trace quantities\footnote{This is the same as two replicas of the SYK model, similar to \cite{saad2018}.}, which is beyond the scope of this paper.

\section{Definition and review of the $\mathcal{N}=2$ model}

\subsection{Model definitions}
Consider $ N $ complex fermions $ \psi_{i},i=1,\cdots,N $ which satisfy the canonical anti-commutation relations
\begin{align}
	\left\{ \psi_{i},\overline{\psi}_{j}\right\} =\delta_{ij},\qquad\left\{ \psi_{i},\psi_{j}\right\} =0.
\end{align}
Denote by $ J\equiv\left(j_{1},\cdots,j_{p}\right),\ 1\leq j_{1}<j_{2}\cdots<j_{p}\leq N $, a collection of $ p $ ordered indices, with $ p $ an odd number, and denote the chain $ \Psi_{J}=\psi_{j_{1}}\cdots\psi_{j_{p}},\ \overline{\Psi}_{J}=\overline{\psi}_{j_{p}}\cdots\overline{\psi}_{j_{1}}$. Define the supercharges $ Q,Q^{\dagger} $ to be
\begin{align}
	Q=\sum_{J}C_{J}\Psi_{J},\qquad Q^{\dagger}=\sum_{J}C_{J}^{*}\overline{\Psi}_{J},
\end{align}
where the summation is over all possible ordered index sets $ J $. The coefficients $ C_{J}\in\mathbb{C} $ are independent random Gaussian variables, with zero mean value and normalized variance. We denote the average over the couplings $ C_{J}$ by $\left\langle \cdot\right\rangle _{C} $, with
\begin{align}{\label{CNorm}}
	\left\langle C_{J}\right\rangle _{C}=0,\qquad\left\langle C_{J_{1}}C_{J_{2}}^{*}\right\rangle _{C}=\begin{pmatrix}N\\
	p
\end{pmatrix}^{-1} 2^{p}{\cal J}^{2}\delta_{J_{1}J_{2}}.
\end{align}
Without loss of generality we will set $ {\cal J}=1 $, while noting that ${\cal J}$ can be reintroduced later using dimensional analysis. This specific choice of normalization for $ \left\langle C_{J}^{2}\right\rangle _{C} $ will ensure that $ \left\langle \text{tr}\left(H\right)\right\rangle _{C}=1 $  when we normalize the trace operation by $2^{-N}$ such that $ \text{tr}\left(1\right)=1 $. The $ {\cal N}=2 $ SUSY-SYK model is then defined by the Hamiltonian
\begin{align}
	H=\frac{1}{2}\left\{ Q,Q^{\dagger}\right\} .
\end{align}

This model has a $U(1)$ R symmetry given by $\psi_i \ra e^{i\alpha} \psi_i, \bar{\psi}_i \ra e^{-i\alpha} \bar{\psi}_i$. The symmetry is generated by the operator
\begin{equation}
\gamma = \frac{1}{2p}\sum_{i = 1}^N \left(\bar{\psi}_i\psi_i - \psi_i \bar{\psi}_i\right),
\end{equation}
with the normalization so that the SUSY charge $Q^{\dagger}$, has a fixed $U(1)$ charge of $1$. We will  be interested in coupling this charge to a chemical potential, with a grand canonical Hamiltonian of the form
\begin{equation}
-\beta H_{\text{GC}} = -\frac{\beta}{2}\left\{ Q,Q^{\dagger}\right\} - \mu \gamma.
\end{equation}

In distinction from \cite{SusySYK}, we will work in a double scaled limit in which
\begin{align}
	N\to\infty,\qquad p\to\infty,\qquad\lambda\equiv\frac{2p^{2}}{N}=\text{fixed}.
\end{align}
We will assume that both the chemical potential, $\mu$, and the inverse temperature, $\beta$, are fixed in this limit. We will find it useful to define the parameter
\begin{equation}
q\equiv e^{-\lambda} \ .
\end{equation}

\subsection{A short review of known results}

The $\mathcal{N} =2$ SYK model was introduced in \cite{SusySYK}, which mainly focused on its emergent super-reparametrization symmetry in the IR. The authors showed that in the IR the model can be described by an $\mathcal{N}=2$ super-Schwarzian effective action. They also demonstrated that the model has a large amount of exact ground states that are unbroken at finite $N$, which they computed numerically when $p=3$.

The IR correlation functions of the $\mathcal{N}=2$ SYK model were computed in \cite{Peng_2017}. Using the conformal ansatz, the conformal dimension of a single fermion was found to be $\Delta_f = \frac{1}{2p}$, while its bosonic partner has dimension $\Delta_b = \Delta_f + 1/2$. To compute the four point function, they wrote it as a sum of ladder diagrams, where each rump of the ladder can be either the bosonic or the fermionic field. This results in two kernels that they diagonalized - a diagonal kernel and an antisymmetric kernel\footnote{This is different from the Majorana SYK model where there is a single diagonal kernel.}. We note that the paper did not discuss the large amount of supersymmetric ground states and their effects on correlation functions.

The partition function of the super-Schwarzian action, along with its density of states, were computed independently  in \cite{Stanford_Witten2017} and \cite{Mertens_2017} using different methods. \cite{Stanford_Witten2017} showed that the Schwarzian theory is one loop exact, using fermionic localization arguments, which allowed them to compute the exact partition function for both the original Schwarzian theory, as well as its supersymmetric extensions. In particular they found that the density of states in the $\mathcal{N}=2$ Super-Schwarzian theory is (equation 3.53 in  \cite{Stanford_Witten2017})
\begin{equation}
\rho_n(E) = \frac{\cos(\pi \hat{q} n)}{1-4\hat{q}^2 n^2} \left[ \delta(E) + \sqrt{\frac{a_n}{E}} I_1 \left( 2\sqrt{a_n E}\right)\right] , ~~~~~~~ a_n = 2\pi^2 \left(1-4n^2 q^2\right),
\end{equation}
where $\hat{q}$ is the interaction length in the SYK model which we call $p$, and $n$ is the complex chemical potential. By considering the Fourier transform of the above quantity with respect to $n$, they find that the ground states (which are proportional to $\delta(E)$ ) exist only for charges $|m|<\hat{q}/2$, and that the continuum spectrum has a lowest energy of $E_0 = \frac{1}{2C}\left(\frac{|m|}{2\hat{q}} - \frac{1}{4}\right)^2$. We will replicate these results in the double scaled limit.

  Reference \cite{Mertens_2017} used a different approach, relating quantities in the Schwarzian theory to objects in 2-d Liouville theory. This allowed them to compute the partition function and density of states in the $\mathcal{N} =2$ Schwarzian theory by summing the relevant characters of 2-d super Liouville, taking into account the spectral flow. They matched the density of states with a chemical potential given above, while also finding the density of states at fixed charge sector to be
 \begin{equation}
 \begin{split}
 \rho(E,Q) &= \frac{1}{8N} \frac{\sinh\left(2\pi \sqrt{E-E_0^+} \right)}{E}\Theta(E-E_0^+) ~~+~~ (+ \ra -)\\
 &\qquad\qquad +  \delta(E)\frac{2}{N}\cos\left(\frac{\pi Q}{N}\right)\Theta\left(2|Q|-N\right) ,
 \end{split}
 \end{equation}
where $Q$ is the charge, $N$ is an integer dual to $p$, and $E_0^\pm = \left(\frac{Q}{2N} \pm \frac{1}{4}\right)^2$. We will replicate these results as well in the double scaled limit.

The model was also considered in \cite{Kanazawa_2017}, which analyzed the level spacing statistics of the model using random matrix theory universality classes. \cite{Kanazawa_2017} also computed the number of ground states in each charge sector using a cohomology argument, and verified this numerically. Though \cite{Kanazawa_2017} mostly considered the $p=3$ case, we extend this method to any value of $p$, and in particular show that it agrees with the chord diagram computation in the double scaled limit. We note that the level spacing statistics are not accessible by just considering single trace quantities, and so are tangent to this work.

\section{The chord partition function and the transfer matrix}

Our first interest will be to calculate the expected spectral density function in the double scaled limit, for the ${\cal N}=2$ model. To do so, it is sufficient to calculate the moments $ m_{k}=\left\langle \text{tr \ensuremath{\left(H^{k}\right)}}\right\rangle _{C} $, and from there infer the expected eigenvalue distribution in the large $N$ limit. Calculating the moments will be the objective of the next few sections. In the first step we will carry out the average over the C's, reducing the expression to a sum over chord diagrams, each of them determining a specific trace of fermionic operators. In the second step we will carry out these traces and obtain the appropriate weight on each chord diagram, providing an expression for the moments in terms of specific chord partition function.

The main complication in the computation, relative to the Majorana SYK or the ${\cal N}=1$ models, is that the expression that we need to evaluate is made out of a string of $Q$ and $Q^\dagger$'s. This means that at each stage, in the transfer matrix approach, we can either add one of two types of chords - one type from $Q$ and the other type from a $Q^\dagger$, or close one of two types, i.e., there are two types of basic chords. This means that if we consider states with $n$ chords, there are a-priori $2^n$ different states. This is in contrast to the situation in the Majorana SYK or the ${\cal N}=1$ where there is only one state with a given number of chords. This is actually a situation which arises in a generic multi-dimensional $q$-brownian motion \cite{Speicher1991,Speicher1993,Speicher1997,Speicher2019}, and we borrow from there the notion of a Hermitian metric on the space of multi-species chords. In the first two subsections we rewrite the moments as a chord partition function with two species of chords. In the subsequent subsections we use the chord Hilbert space construction to show that most of the states there are null states, and modding out by them gives a tractable reduced Hilbert space that can be treated using the transfer matrix approach. We will actually encounter another problem that the naive transfer matrix will be non-local (for fixed chemical potential) and we will see how to remedy this.

The outline of this section is the following: We first use Wick contractions to write the moments as a sum of oriented chord diagrams in section \ref{sec:OCD}. The contribution of each oriented chord diagram is evaluated in section \ref{sec:CPF}, resulting in a chord partition function for the moments. We rewrite the moment $m_k$ in terms of sub-moments, which can be assessed locally in section \ref{sec:LocalCPF}. We then define the auxiliary Hilbert space of partial chord diagrams in section \ref{sec:TM}, and construct a transfer matrix that implements the sub-moments discussed before. To make the transfer matrix local, we introduced an auxiliary parameter, $\theta$. We relate this parameter to the charge in section \ref{sec:FCS}, showing that the transfer matrix for every charge sector is local. In section \ref{sec:INP} we define the inner product on the auxiliary Hilbert space using a Hermitian metric on the space of multi-species chords. Finally, in section \ref{sec:PhysHS} we show that under this inner product there are many null states, and by modding them out we obtain a physical Hilbert space that is tractable.

\subsection{Reduction to oriented chord diagrams (OCD) (or X-O diagrams)} \label{sec:OCD}

Plugging the Hamiltonian into the definition of $m_k$ gives

\begin{equation}\label{eq:MomentAsTrace}
\begin{split}
m_{k}&=\langle  \text{tr} H^k \rangle_C = 2^{1-k}\left\langle \text{Tr \ensuremath{\left(\left(QQ^{\dagger}\right)^{k}\right)}}\right\rangle_C \\
&=2^{1-k} \sum_{\begin{matrix}J_{1},\cdots,J_{k}\\
	I_{1},\cdots,I_{k}
	\end{matrix}}\left\langle C_{J_{1}}C_{I_{1}}^{*}\cdots C_{J_{k}}C_{I_{k}}^{*}\right\rangle_C \text{tr}\left(\Psi_{J_{1}}\overline{\Psi}_{I_{1}}\cdots\Psi_{J_{k}}\overline{\Psi}_{I_{k}}\right),
\end{split}
\end{equation}
where we used the nilpotency of $Q,Q^{\dagger}$ to obtain the second equality.

Let us now focus on the term $ \left\langle C_{J_{1}}C_{I_{1}}^{*}\cdots C_{J_{k}}C_{I_{k}}^{*}\right\rangle_C$. If any $ C_{J} $ appears without a corresponding $ C_{J}^{*} $, the entire term will vanish on average. As was shown in \cite{Erdos14}, the only relevant contributions to the moment $ m_{k} $, in the limit $ N\to\infty $, come from summands in which the $ C_{J} $'s are contracted only into pairs - this is just a Wick contraction when the $C$'s are Gaussian, but it also holds under weaker assumptions on the distribution. This means that every index set $ J_{i} $ has a partner $ I_{j} $ such that $ J_{i}=I_{j} $. Higher coincidences, where the ordering is not pair-wise, are suppressed in the large $N$ limit.

\begin{figure} [h]
	\centering
	\includegraphics[page=1,width=0.35\columnwidth]{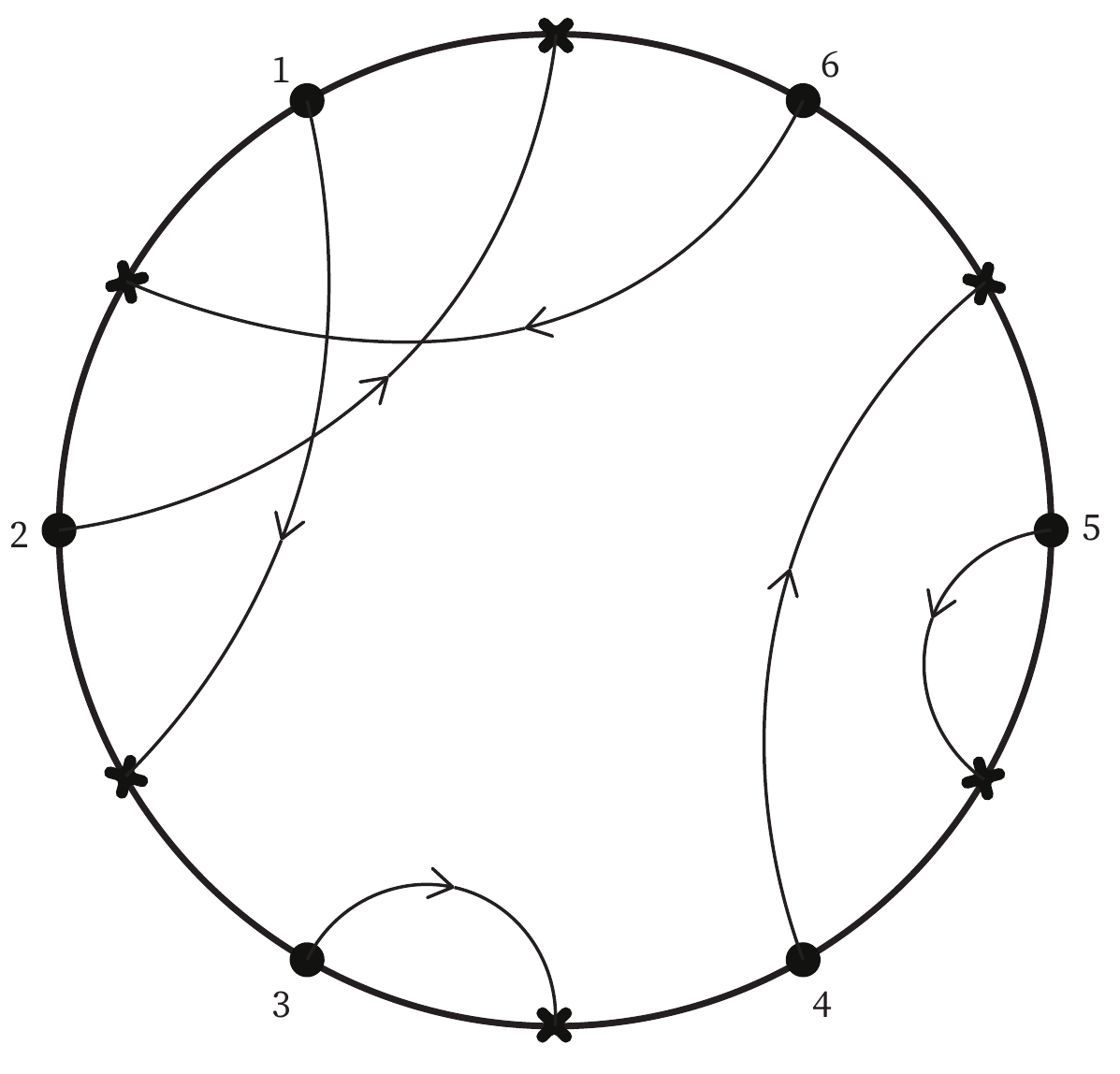}
	\hspace{0.1\columnwidth}
	\includegraphics[page=2,width=0.35\columnwidth]{figures/CDExample.pdf}
	\caption*{(a)\quad\qquad\qquad\qquad\qquad\qquad\qquad\qquad\qquad(b)}
	\includegraphics[page=3,width=0.80\columnwidth]{figures/CDExample.pdf}	
	\caption*{(c)}
	\caption{\\\textbf{(a)} - Chord diagram representing the term $\text{tr}(\Psi_1\overline{\Psi}_6\Psi_2\overline{\Psi}_1\Psi_3\overline{\Psi}_3\Psi_4\overline{\Psi}_5\Psi_5\overline{\Psi}_4\Psi_6)$, contributing to $ m_6 $.
		\\\textbf{(b)} - Disentangled form of the diagram (a). Note that the chord directionality is maintained in the disentangling process.
		\\\textbf{(c)} - The same diagram, represented as an open chord diagram, where the 6'th node is chosen to be the first node.}
	\label{fig:CDexample}
\end{figure}

The averaging over the $ C_{J} $'s depends only on the number of pairs $ k $,  and based on (\refeq{CNorm}) we see that it gives $ 2^{kp}{N \choose p}^{-k} $. Now the moment becomes
\begin{align} \label{MomentAfterAvg}
m_{k}=2^{1-k + kp}\frac{1}{{N \choose p}^{k}}\sum_{\begin{matrix}J_{1},\cdots,J_{k}\\
	\pi \in S_k
	\end{matrix}}\text{tr}\left(\Psi_{J_{1}}\overline{\Psi}_{J_{\pi(1)}}\Psi_{J_{2}}\cdots\overline{\Psi}_{J_{\pi(k)}}\right),
\end{align}
where $S_k $ is the group of permutations of $ \{1,\cdots,k\} $.
\\\\
To understand the terms in this sum better we can present each trace pictorially as an {\it oriented chord diagram (OCD)}, or {\it X-O diagram}, as shown in figure ({\ref{fig:CDexample}}a). Each such diagram represents some ordering of pairs of $ \Psi,\overline{\Psi} $ operators inside a trace, such that
\begin{itemize}
	\item $ O $ nodes correspond to $ \Psi $,
	\item $ X $ nodes correspond to $ \overline{\Psi} $,
	\item a chord drawn between the $ X-O$ nodes means they have the same index set.  We draw arrows in the direction of going from $O$ to $X$ .
\end{itemize}
The cyclical structure of this diagram matches that of the trace.

We can now rewrite (\refeq{MomentAfterAvg}) in a more suggestive form, as
\begin{equation} \label{eq:MomentBeforeTraceEval}
\begin{split}
m_{k}=2^{1-k+kp}\frac{1}{{N \choose p}^{k}}\sum_{\pi\in\text{CD}(k)}\sum_{J_{1},\cdots,J_{k}}\text{tr}\left(\Psi_{J_{1}}\overline{\Psi}_{J_{\pi(1)}}\cdots\Psi_{J_{k}}\overline{\Psi}_{J_{\pi(k)}}\right),
\end{split}
\end{equation}
where CD$(k)$ are chord diagrams with $k$ chords, and $\pi(\cdot)$ is the ordering given by the chord diagram.

\subsection{The chord partition function} \label{sec:CPF}
To assess each such chord diagram, we will need to bring it to a disentangled form, i.e - nodes of the same chord are adjacent, for all chords, as seen in (\ref{fig:CDexample}b). In this form the trace can be easily computed, as will be shown below. This disentangling corresponds to permuting fermion chain operators. The commutation relations between such operators $ \Psi_I,\Psi_J $ are dictated by the number of fermions they share (i.e. $ \left|I\cap J\right| $.)

The number of indices in the intersection of two random index sets of size $p\sim \sqrt{N}$ admits Poisson statistics \cite{Erdos14}. As a result, in the $ N\to\infty $ limit, an index $i$ appears in at most two index sets with finite probability, namely $ J_{a}\cap J_{b}\cap J_{c}=\emptyset $ for almost all index sets $ J_a,J_b,J_c $. Non-zero triple intersections will generate sub-leading corrections of order $1/N$ to the moments. We call this property - \textit{no triple intersections}. This fact alone allows us determine if two specific index sets can share indices or not irrespective of any other set. If two index sets can share indices we will refer to them as \textit{friends}, and otherwise - \textit{enemies}.

To see how such restrictions come about, let $ I,J $ be some index sets. We see that $ \Psi_{I}$ and $\Psi_{J}$ must appear in one of the forms	
\begin{align} \label{eq:FriendEnemyStructure}
\begin{matrix}\begin{matrix}\left(\text{i}\right) & \text{tr}\left(\Psi_{I}\cdots\Psi_{J}\cdots\overline{\Psi}_{I}\cdots\overline{\Psi}_{J}\cdots\right)\\
\left(\text{ii}\right) & \text{tr}\left(\Psi_{I}\cdots\Psi_{J}\cdots\overline{\Psi}_{J}\cdots\overline{\Psi}_{I}\cdots\right)
\end{matrix} & \Big\}\ \text{enemy configuration}\\
\begin{matrix}\left(\text{iii}\right) & \text{tr}\left(\Psi_{I}\cdots\overline{\Psi}_{I}\cdots\Psi_{J}\cdots\overline{\Psi}_{J}\cdots\right)\end{matrix} & \text{friend configuration}
\end{matrix}
\end{align}
with all other forms equivalent due to cyclicality of the trace.
\\\\
Let us now consider the different cases:
\begin{itemize}
	\item (i), (ii) -  Take some fermion $\psi_j$ in the intersection $j\in I\cap J$. Assuming no triple intersections (so $\psi_j$ does not appear in any other fermion chain besides $ \Psi_{I}$ and $\Psi_{J} $), we are free to anti-commute the $\psi_j$'s next to each other, resulting in a trace of zero (as $\psi_j^{2}=0 $). We see that such configurations can contribute to the moment only when $ I\cap J=\emptyset $.
	\item (iii) - In distinction from the above case, there is no problem for $\Psi_{I}$ and $\Psi_{J} $ to share fermions.
\end{itemize}

To make progress, and as in \cite{Micha2018,Berkooz_2019}, it is convenient to use an alternative representation, in which each chord diagram is represented by nodes on a line rather than on a circle, which we will call an open chord diagram. An example of this is presented in figure (\ref{fig:CDexample}c). We note that the cyclicity of the trace is broken in this representation,  but the end of the day result is of course independent of which point is chosen to be the first in line.  Open chord diagrams will allow us to think in terms of nested diagrams. Later on we will also use open chord diagrams to construct a transfer matrix that builds all the possible chord diagrams.

\subsubsection*{Disentangling a chord diagram}

We shall now describe the disentangling process of an oriented chord diagram.

Starting with an open oriented chord diagram, we are assured to have a chord connecting some $(\Psi_J,\overline{\Psi}_J)$ such that it is enemies with all chords opening or closing under it, i.e all  the operators separating $\Psi_J,\overline{\Psi}_J$ are of the form of $\Psi_I$ in (\ref{eq:FriendEnemyStructure}(i),(ii)). We shall refer to corresponding chords of this form as \textit{minimal chords}. For simplicity, let us assume $\Psi_J$ appears to the left of $\overline{\Psi}_J$, meaning this is a right pointing chord\footnote{The process for a left pointing chord is identical to the one described here.}. For each $\Psi_I$ or $\overline{\Psi}_I$ between $\Psi_J$ and $\overline{\Psi}_J$ we have that $\{\Psi_J,\Psi_I\}=\{\Psi_J,\overline{\Psi}_I\}=0$, since they are enemies by assumption. This allows us to (anti-)commute $\overline{\Psi}_J$ to the $\Psi_J$, at the cost of the number of operator crossings, which is the number of chords intersecting the minimal chord. Thus

\begin{align}
\text{tr}(\cdots\Psi_J\cdots\overline{\Psi}_J\cdots)=(-1)^{\overline{\Psi}_J \text{ intersecctions}}\text{tr}(\cdots\Psi_J\overline{\Psi}_J\cdots).
\end{align}

Once the operators $\Psi_J$ and $\overline{\Psi}_J$ are adjacent, we can commute the pair to the far left of the open diagram. Notice that
\begin{equation}
\Psi_I \Psi_J\overline{\Psi}_J  =  \Psi_J\overline{\Psi}_J\Psi_I \delta_{|I\cap J|,0}, \qquad  \qquad
\overline{\Psi}_I \Psi_J\overline{\Psi}_J  =  \Psi_{J/I}\overline{\Psi}_{J/I}\overline{\Psi}_I ,
\end{equation}
where $J/I$ is the set of indices in $J$ that are not in $I$. Thus we may lose some pairs of fermions from $\Psi_J\overline{\Psi}_J$ while commuting them to the left, however as $\psi\bar{\psi}\psi\bar{\psi}=\psi\bar{\psi}$ the value of the trace does not change if a pair of indices appears in more than a single index set.

\begin{figure} [h]
	\centering
	\includegraphics[page=1,width=0.7\columnwidth]{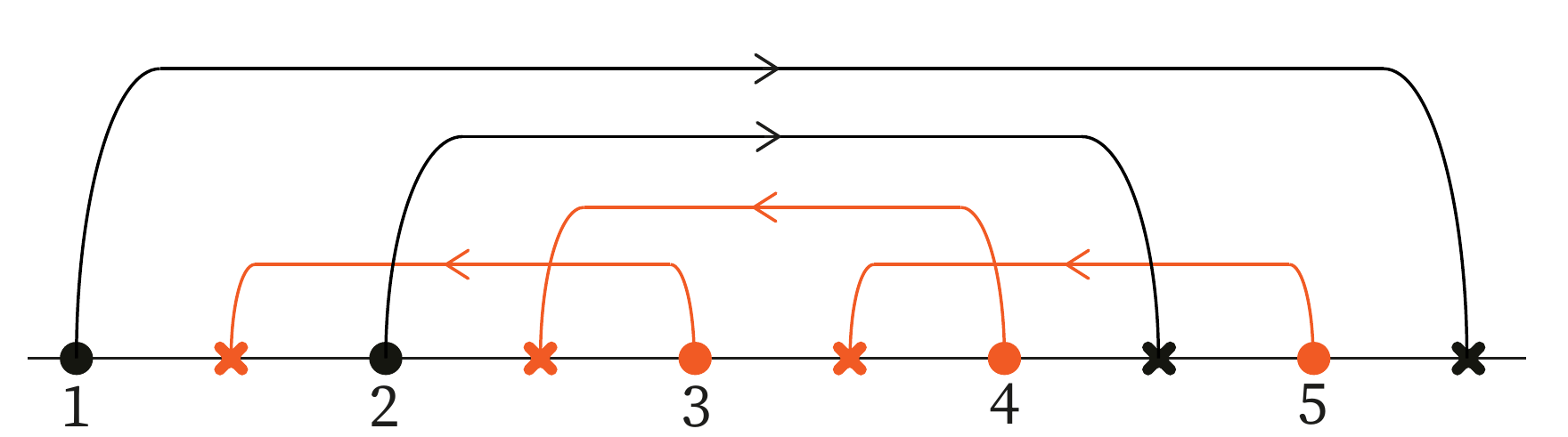}
	\\
	\includegraphics[page=2,width=0.7\columnwidth]{figures/DisentanglingAlgorithm.pdf}
	\\
	\includegraphics[page=3,width=0.7\columnwidth]{figures/DisentanglingAlgorithm.pdf}	
	\caption{Disentanglement of a specific chord diagram, according to the algorithm presented above, going from top to bottom. The minimal chords - ones that are separated only by enemy chords, are colored in orange. In the next step these chords taken to the left edge, and we have new minimal chords.
		For example, see that in the first step chord 2 is not minimal, as it is friends with chord 4, nested in it. In the second step chords 1 and 2 are both minimal, as they are enemies.
		Primed notation means that a chord has the original indices, excluding the ones it shared when passing through friends.
		For example, the indices of $3'$ are the indices of $3$, excluding the ones shared with $1$.}
	\label{fig:DisentanglingAlgo}
\end{figure}

 If the diagram is not completely disentangled, we are now assured to have new minimal chords, and can repeat this process until the diagram is completely disentangled. This process is demonstrated in figure \ref{fig:DisentanglingAlgo}. Once a diagram is disentangled and all the operator pairs are adjacent to each other, we can simply pair the individual fermions up.

 If we denote the total number of intersections in a chord diagram $\pi$ by $\#\text{int}(\pi)$, in the end of the process we get
\begin{align} \label{eq:DisentangledTrace}
\text{tr}\left(\Psi_{J_1}\overline{\Psi}_{J_{\pi(1)}}\cdots\Psi_{J_k}\overline{\Psi}_{J_{\pi(k)}}\right)=
(-1)^{\#\text{int}(\pi)}\text{tr}\left(\prod_{i \in J_1\cup\ldots \cup J_k}\psi_{i}\bar{\psi}_{i}\right)
 = (-1)^{\#\text{int}(\pi)} ~2^{-|J_1\cup\ldots \cup J_k|},
\end{align}
as $\text{tr}(\bar{\psi}_{i}\psi_i) = 1/2$.

Since we assume no triple intersection of index sets, we can express the number of distinct indices $ d = |J_1\cup\ldots \cup J_k|$ as $d=kp-\sum_{1\leq i<j \leq k} m_{ij}$, where $m_{ij}\equiv|J_i\cap J_j|$ is the number of mutual indices in $J_i$ and $J_j$. Combining this with (\ref{eq:DisentangledTrace}) reduces the moments (\ref{eq:MomentBeforeTraceEval}) to
\begin{equation}
\begin{split}
m_{k}=2^{1-k}\frac{1}{{N \choose p}^{k}}\sum_{\pi\in\text{CD}(k)}\sum_{J_{1},\cdots,J_{k}}\left(-1\right)^{\#\text{ int}(\pi)}\prod_{1\leq i<j\leq k}2^{m_{ij}}.
\end{split}
\end{equation}

The final step in evaluating the moment $m_k$ involves the summation over all index sets $ \{J_i\}_{i=1}^{k} $ in a given chord diagram. As mentioned above, in the large $ N $ limit the index overlap $ m_{ij} $ admits Poisson statistics, which allows us to move to a summation over it. That, along with the fact that only friend configurations can have a nontrivial intersection gives us
\begin{equation} \label{eq:mk_almostdone}
\begin{split}
\sum_{J_{1},\cdots J_{k}}\text{tr}\left(\Psi_{J_{1}}\overline{\Psi}_{J_{\pi(1)}}\cdots\Psi_{J_{k}}\overline{\Psi}_{J_{\pi(k)}}\right) &= {N \choose p}^k 2^{-kp} \left(-1\right)^{\text{\# int}(\pi)}\\
&\times \left(\prod_{(i,j) \text{ friends}}\sum_{m_{ij} = 0 }^\infty \frac{\lambda^{m_{ij}}}{m_{ij}!}e^{-\lambda/2}\right)
\left(\prod_{(i,j) \text{ enemies}} e^{-\lambda/2}\right) .
\end{split}
\end{equation}

Summing the above series, we see that each pair of friendly chords gives us a factor of $ q^{-1/2}=e^{\lambda/2} $, while each pair of enemy chords gives a factor of $ q^{1/2}=e^{-\lambda/2} $. The moment $ m_k $ is thus written fully as the \textit{chord partition function}
\begin{equation} \label{eq:chord_partition}
\boxed{m_{k}= 2^{-k}\sum_{\pi\left(k\right)}\left(-1\right)^{\text{\# int}(\pi)}q^{\left(\#_{e}-\#_{f}\right)/2}},
\end{equation}
where $ \pi\left(k\right) $ are chord diagrams with $ k $ chords, and $ \#_{e/f} $ is the number of enemies and friends respectively. We note that we allow chord diagrams to start either with a $Q$ or a $Q^{\dagger}$, hence the additional factor of $1/2$ compared to (\ref{eq:mk_almostdone}).

\begin{figure} [h]
	\centering
	\includegraphics[page=1,width=0.30\columnwidth]{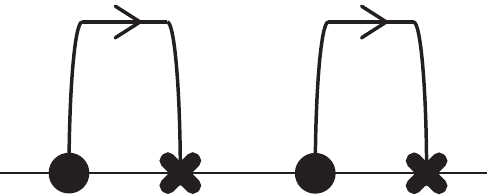}
	\hspace{0.1\columnwidth}
	\includegraphics[page=4,width=0.30\columnwidth]{figures/ChordRelations.pdf}
	
	\caption*{$ \mathrm{I} $ \qquad\qquad\qquad\qquad\qquad\qquad\qquad $ ~~~\mathrm{II} $}
	\includegraphics[page=2,width=0.30\columnwidth]{figures/ChordRelations.pdf}
	\hspace{0.1\columnwidth}
	\includegraphics[page=3,width=0.30\columnwidth]{figures/ChordRelations.pdf}
	\caption*{$ \mathrm{III} $ \qquad\qquad\qquad\qquad\qquad\qquad\qquad $ \mathrm{IV} $}
	\includegraphics[page=6,width=0.30\columnwidth]{figures/ChordRelations.pdf}
	\hspace{0.1\columnwidth}
	\includegraphics[page=5,width=0.30\columnwidth]{figures/ChordRelations.pdf}
	\caption*{$ \mathrm{V} $ \qquad\qquad\qquad\qquad\qquad\qquad\qquad $ \mathrm{VI} $}
	\caption{Six possible chord configurations. According to the chord partition function (\ref{eq:chord_partition}) configurations $ \mathrm{I,II} $ are friends, thus receiving a factor of $q^{-1/2}$. The rest are enemies, and given a factor of $q^{1/2}$. Configurations $ \mathrm{V,VI} $ intersect, so they receive an extra factor of $(-1)$. There are six more diagrams, which can be obtained from the ones above by $ X\leftrightarrow O $.}
	\label{fig:ChordRelations}
\end{figure}

Graphically we can see there are 12 possible configurations for a pair of directional chords, and the chord partition function gives a weight for each such configurations, as shown in the figure \ref{fig:ChordRelations}. Note there are six more relations, not shown in the figure, in which we switch $X\leftrightarrow O$. We denote six configurations in the figure by I$,\cdots,$VI, and the reversed ones by $\bar{\text{I}},\cdots,\bar{\text{VI}}$

\subsection{Localizing the chord partition function} \label{sec:LocalCPF}

We would like to express the chord partition function in terms of a local transfer. However out of the 12 possible configurations $\text{I},\bar{\text{I}},\text{III},\bar{\text{III}}$ are non-local, meaning - when going from left to right we must have information about closed chords in order to account for them properly. It seems that if we want to count the number of these diagrams using a transfer matrix, it must be non-local, meaning - must have information about currently closed chords. Yet, there is more we can do if we use relations between quantities. These will enable us to write the chord partition function in terms of local relations, which in turn can be evaluated using a local transfer matrix.

Notice that the number of pairs of chords is fixed for diagrams contributing to the  $k$'th moment, so
\begin{align}  \label{eq:Conserv1}
\#_e+\#_f={k \choose 2}.
\end{align}
This enables us to compute the moment $m_k$ using only the number of friend chords. This takes care of  diagram III, but we still need to find an alternative way of counting diagrams I and $\bar{\text{I}}$.

If we restrict ourselves to a subspace in which the number of right and left pointing chords, $(n_\rightarrow,n_\leftarrow)$, is fixed,
we can use
\begin{align} \label{eq:Conserv2}
N_{I}+N_{IV}+N_{VI}	={n_{\rightarrow} \choose 2}, &
&&\overline{N}_{I}+\overline{N}_{IV}+\overline{N}_{VI}	={n_{\leftarrow} \choose 2}.
\end{align}

This allows us to express the amount of non-local diagrams using only local ones, as $\#_f=N_I+N_{II}+\bar{N}_{I}+\bar{N}_{II}$. We will find it easier to work with the set of variables $(k,m)$, defined to be
\begin{align}
k = n_{\rightarrow}+n_{\leftarrow},\qquad m=n_{\rightarrow}-n_{\leftarrow}.
\end{align}
Now we can plug the relations (\ref{eq:Conserv1}),(\ref{eq:Conserv2}) into the chord partition function (\ref{eq:chord_partition}) and get
\begin{equation}
\begin{split}
m_{k}	&=\frac{q^{k/4}}{2^{k}}\sum_{m=-k,-k+2,\cdots,k}q^{-m^{2}/4}\sum_{\pi\left(k;m\right)}\left(-1\right)^{\#_{i}}q^{-N_{II}-\overline{N}_{II}+N_{IV}+\overline{N}_{IV}+N_{VI}+\overline{N}_{VI}}\\
&=\frac{q^{k/4}}{2^{k}}\sum_{m=-k,-k+2,\cdots,k}q^{-m^{2}/4}m_{k;m}.
\end{split}
\end{equation}
We see that we have managed to write the non-local chord partition function using a sum over local partition functions in fixed $(k,m)$ subspaces. Now we are in a suitable position to define a Hilbert space and a local transfer matrix that will compute these subspace moments $m_{k,m}$.
We note that having a local transfer matrix is desirable also because we interpret the transfer matrix as the Hamiltonian of the gravitational theory, or at least the generalization of the Hamiltonian of the Schwarzian system, and so we require it to be local in time.

\subsection{Auxiliary Hilbert space and transfer matrix} \label{sec:TM}

\begin{figure} [h]
	\centering
	\includegraphics[width=0.5\columnwidth]{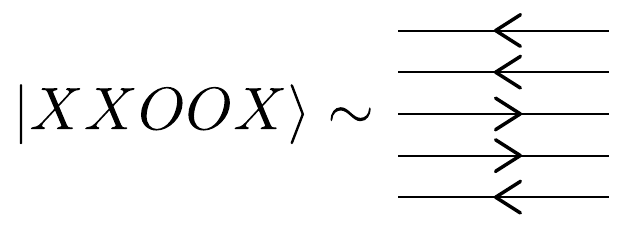}
	\caption{An example of a vector in $ {\cal H}_{\text{aux}} $, and its representation in terms of chords.}
	\label{fig:open_chord}
\end{figure}

Define the auxiliary Hilbert space $ {\cal H}_{\text{aux}}=\bigoplus_{n= 0}^\infty \left\{ \ket{X}, \ket{O}\right\}^{\otimes n} $, where $\ket{O}$ and $\ket{X}$ represent chords emanating from $Q,Q^{\dagger}$ respectively. Denote the empty state to be $\ket{\emptyset}$. The inner product on this vector space will be defined later. An example of a vector in ${\cal H}_{\text{aux}}$ is given in the figure \ref{fig:open_chord}.

Define the transfer matrix $T:{\cal H}_{\text{aux}}\to {\cal H}_{\text{aux}}$. By acting with $T$ on a vector we wish to get all possible results of adding a pair $QQ^{\dagger}$ to a diagram, where any $Q,Q^{\dagger}$ can be either a start or endpoint of a chord.

It remains for us to restrict ourselves to a specific $k,m$ subspace. $k$ is already known, as we act with $T^k$ on $\ket{\emptyset}$, and project onto $\ket{\emptyset}$. We can fix the value of $m$ using an auxiliary real variable $\theta$. Whenever we open a new right pointing chord let us multiply the diagram by a factor of $ e^{i  \theta} $, and whenever we open a new left pointing chord let us multiply by a factor of $ e^{-i  \theta} $. Now we can easily project onto a fixed $m$ subspace using a Fourier transform.

This gives us the defining relation for the transfer matrix $T$
\begin{align} \label{eq:moment_m_km}
m_{k;m}=\sum_{\pi\left(k;m\right)}\left(-1\right)^{\#_{i}}q^{-N_{II}-\overline{N}_{II}+N_{IV}+\overline{N}_{IV}+N_{VI}+\overline{N}_{VI}}=\frac{1}{2\pi}\int_{0}^{2\pi}d\theta ~e^{-im\theta}\left\langle \emptyset\left|T^{k}\left(\theta\right)\right|\emptyset\right\rangle .
\end{align}
\subsubsection*{Rules of the Transfer matrix} \label{sec:TMrules}
Let us now construct $T(\theta)$ explicitly. At each step we have a $Q$ followed by a $Q^\dagger$, so we can split the transfer matrix into two parts.
\begin{enumerate}
	\item When we encounter a $Q$, we can either:
	\begin{enumerate}
		\item Add a "O" to the lowest cell, with a factor of $e^{i\theta}$.
		\item Delete some "X" from the vector, and multiply by the factor
			\begin{equation}
		(-1)^{\#O_{\text{below}}+\#X_{\text{below}}}q^{-\#O_{\text{ above}} + (\#X-1)}  ,
			\end{equation}
where $\#O_{\text{above}}$ is the number of open right-moving chords above the chord we close, and $\#X$ is the total number of open left-moving chords.
	\end{enumerate}
With a slight abuse of notation, we will define the operator $Q$ acting in ${\cal H}_{\text{aux}}$ by these steps.
\item When we encounter a $Q^\dagger$, so we can either:

	\begin{enumerate}
		\item Add a "X" to the lowest cell and multiply by a factor of $e^{-i\theta}$.
		\item Delete some "O" from the vector, and multiply by the factor
		\begin{equation}
		(-1)^{\#O_{\text{below}}+\#X_{\text{below}}}q^{-\#X_{\text{ above}} + (\#O-1)}  ,
		\end{equation}
		where $\#X_{\text{ above}}$ is the number of open left-oriented chords above the chord we close, and $\#O$ is the total number of open right-oriented chords.
	\end{enumerate}
Similarly, we will use these steps to define the operator $Q^{\dagger}$ acting in the auxiliary Hilbert space.
\end{enumerate}

We define the transfer matrix to be $T(\theta)=Q^{\dagger}Q+QQ^{\dagger}.$

%
%
%
%
%
%
%
%
%
%
%
%

\subsection{Chemical potential and fixed charge sectors} \label{sec:FCS}

We shall now add  a chemical potential, and calculate the grand canonical moments
\begin{equation}
m_k(\mu) \equiv \expt{\tr\left[H^k e^{-\mu \gamma } \right] }_C .
\end{equation}
The general derivation of the chord partition function (equation \eqref{eq:chord_partition}) via Wick contractions is still valid for the grand canonical moments, aside from a few extra factors which we derive bellow.

To derive these additional factors, let us focus on some Wick contraction in $m_k(\mu)$:
\begin{equation}
\tr\left[\overline{\Psi}_{J_1} \Psi_{J_{\pi(1)}}\overline{\Psi}_{J_2} \Psi_{J_{\pi(2)}}\ldots
\overline{\Psi}_{J_k} \Psi_{J_{\pi(k)}} \prod_{i=1}^N e^{-\mu \gamma_i }\right],
\end{equation}
where $\pi$ is a permutation of $(1,\ldots ,k)$. Then every fermion index $i=1,\ldots , N$ is in one of the following categories:
\begin{enumerate}
	\item $i \notin J_1 \cup J_2 \cup \ldots \cup J_k$ : In this case $\gamma_i$ commutes with the Wick contraction. We can first evaluate the chord diagram, using the method described above. Then we are left with $\text{tr}(e^{-\mu\gamma_i})$ for each non-participating index. As $\tr(\gamma_i) = 0$ and $\gamma_i^2 = 1/(4p^2)$ we have that
	\begin{equation}
	\tr\left( e^{-\mu \gamma_i}\right) = \sum_{k=0}^\infty \frac{(-\mu/(2p))^k}{k!}\tr\left( (\gamma_i)^k\right)
	= \sum_{k \text{ even}}  \frac{(-\mu/(2p))^k}{k!} = \cosh\left(\frac{\mu}{2p}\right),
	\end{equation}
	and thus the site will contribute a factor of $ \cosh\left(\frac{\mu}{2p}\right)$.
	
	\item  $i \in J_j$ for one $j\in (1,\ldots, k)$: In this case we have two options, if $J_j$ comes in the form $\overline{\Psi}_{J_j}\Psi_{J_j}$ then
	\begin{equation}
	\tr\left( \ldots  \overline{\Psi}_{J_j} \ldots \Psi_{J_j}  \ldots  e^{-\mu\gamma_i}\right)
	= e^{-\mu/(2p)} \tr\left( \ldots  \overline{\Psi}_{J_j} \ldots  \Psi_{J_j}  \ldots \right) .
	\end{equation}
	Similarly, if the pairing $J_j$ comes in the opposite orientation, $\Psi_j \overline{\Psi}_{J_j}$, then we will get a factor of $e^{\mu/2p}$.
	
	\item $i \in J_{j_1},J_{j_2}$ for two index set: In this case the two index sets must be in one of the ``friends'' configurations, and then as $\left(\bar{\psi}_i\psi_i\right)^2 = \bar{\psi}_i\psi_i$ it will contribute one of the same two factors as before.
\end{enumerate}

We can account for factors (1) and (2) by multiplying $m_{k,m}$ by an overall factor of
\begin{equation}
[\cosh\left( \mu/(2p)\right)]^{N-kp}\cdot e^{\mu m/2} \xrightarrow{N \ra\infty}  e^{\frac{ \mu^2}{4 \lambda}+\frac{\mu m}{2}}+O\big(N^{-1}) .
\end{equation}
 The additional correction due to factor (3) is sub-leading in the double scaled limit. Thus the grand canonical moments are given by
\begin{equation}\label{eq:mkmuscaled}
m_k(\mu) = 2^{-k} e^{\frac{ \mu^2}{4 \lambda}} q^{\frac{k}{4}} \sum_{m=-k,-k+2}^k
e^{\mu m /2 }q^{-\frac{m^2}{4} } m_{k;m} .
\end{equation}

We can move to a fixed charge sector by taking a Fourier transform of the moments with respect to $i\mu$, that is
\begin{equation} \label{eq:mk_s_scaled}
m_k(s)\equiv \frac{1}{2\pi} \intinf d\mu~e^{i\mu s} m_k(i\mu)
 = 2^{-k} \sqrt{\frac{\lambda}{\pi}} q^{s^2 + \frac{k}{4}} \sum_{m=-k,-k+2}^k q^{ms} m_{k;m} .
\end{equation}

To continue, we notice that the transfer matrix element $\innn{\emptyset}{T^k(\theta)}{\emptyset}$ has terms proportional to $e^{i n\theta}$ only for $n = -k,-k+2,\ldots,k-2,k$. Thus $m_{k;m} \neq 0$ only for $m = -k,-k+2,\ldots,k-2,k$, so we can extend the sum over $m$ to any additional integers and \eqref{eq:mk_s_scaled} will not change. Then we define $z \equiv e^{i\theta}$ and consider the $\theta$ integral in \eqref{eq:moment_m_km} as a contour integral in the complex plane over the unit circle. This allows us to write
\begin{equation}
m_k(s)  = 2^{-k} \sqrt{\frac{\lambda}{\pi}} q^{s^2 + \frac{k}{4}}\frac{1}{2\pi i} \oint_{|z|=1} \frac{dz}{z} \sum_{m=-N_1}^{N_2} \left(z^{-1}e^{-\lambda s }\right)^m \innn{\emptyset}{T^k(\theta)}{\emptyset} ,
\end{equation}
for arbitrary $N_1,N_2 > k + 1$. For $s > 0$ we can extend $N_2 \ra \infty$ and find that
\begin{equation}
m_k(s)  = 2^{-k} \sqrt{\frac{\lambda}{\pi}} q^{s^2 + \frac{k}{4}}\frac{1}{2\pi i} \oint_{|z|=1} dz \frac{\left(z e^{\lambda s }\right)^{N_1} }{z-e^{-\lambda s}} \innn{\emptyset}{T^k(\theta)}{\emptyset}
=  2^{-k} \sqrt{\frac{\lambda}{\pi}} q^{s^2 + \frac{k}{4}} \innn{\emptyset}{T^k(i\lambda s)}{\emptyset},
\end{equation}
as the only simple pole in the unit circle is at $z = e^{-\lambda s}$. The same result holds for $s<0$ by inverting the contour, and for $s=0$ by noting that $\sum_{m=-\infty}^{\infty} e^{-im\theta} = 2\pi \delta(\theta)$.
This is a surprising result: the transfer matrix in a fixed charge sector is local!\footnote{Whereas this is not the case for a fixed chemical potential.} We therefore define the fixed charge transfer matrix, $T_s \equiv T(i\lambda s)$. It obeys the same rules as the local transfer matrix $T(\theta)$, only whenever we open a new right-pointing chord we multiply the diagram by a factor of $ q^s $, and whenever we open a new left-pointing chord we multiply by a factor of $ q^{-s} $. The fixed charge moments get the compact transfer matrix form
\begin{equation}
m_k(s) = 2^{-k}\sqrt{\frac{\lambda}{\pi}} q^{s^2 + \frac{k}{4}}\innn{\emptyset}{T^k_s}{\emptyset} .
\end{equation}
Furthermore, we can now sum over charge sectors, rather than the auxiliary parameter $\theta$, to calculate the full Hilbert space moments:
\begin{equation}
m_k(\mu) =2^{-k} q^{\frac{k}{4}} \sqrt{\frac{\lambda}{\pi}}\intinf ds~ q^{s^2} e^{-\mu s}\innn{\emptyset}{T^k_s}{\emptyset}.
\end{equation}

Finally we note that the fractional number of states in a given charge subspace is
\begin{align}
\frac{\dim(\Ham_s)}{\dim(\Ham)} =\frac{1}{2^N} {N \choose N/2 + s p }
= \sqrt{\frac{2}{\pi N}} e^{-\lambda s^2} + O(N^{-3/2})
= ds~ \sqrt{\frac{\lambda}{\pi }} q^{ s^2} ,
\end{align}
where the infinitesimal increment $ds$ is just $1/p = \sqrt{2/(N\lambda)}$. Thus we see that the function of $s$ in front of the matrix element is just the measure of the subspace, and we can really think of $T_s$ as the complete transfer matrix in a fixed charge sector.

\subsection{Inner product} \label{sec:INP}

Although the transfer matrix is now strictly local, we still have to deal with the exponential growth of ${\cal H}_{\text{aux}}$ as a function of $n$ - i.e - for $n$ chords there are $2^n$ states. In this section we will show that we can define a semi-positive inner product such that all states apart for 2 for each value of $n>0$ are null states, and that these null states decouple under the action of the transfer matrix. Modding out by the null states we get a physical Hilbert space of a manageable size, which is similar in complexity to the ${\cal N}=0$ and ${\cal N}=1$ cases. In this inner product $Q$ and ${Q}^\dagger$ are Hermitian conjugates of each other.

\begin{figure} [h]
	\centering
	\includegraphics[width=0.8\columnwidth]{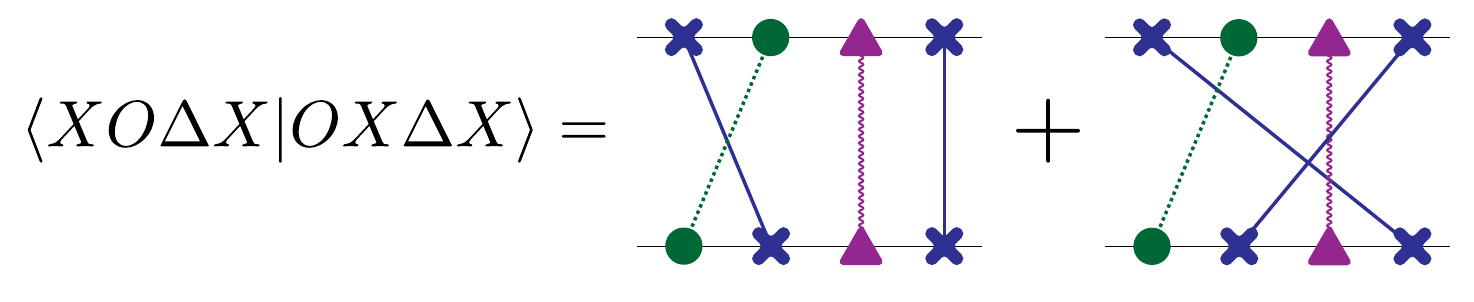}
	\caption{An example for a product of 3 flavors of chords for (\ref{eq:inn_oneq}), denoted by $X,O,\Delta$. The left diagram has a single intersection, and the right one has four, which means that $ \left<XO\Delta X|OX\Delta X\right>=q+q^4 $.}
	\label{fig:GeneralInnerProduct}
\end{figure}

The auxiliary Hilbert space of partial chord diagrams described in section \ref{sec:TM} is similar to the Fock space construction by Pluma and Speicher in \cite{Speicher2019}. There they define an inner product for the auxiliary Hilbert space of multiple copies of the original SYK model. In their paper they consider $r$ identical copies of the regular SYK model, and construct an inner product on the auxiliary Hilbert space of $r$ different flavors of chords, $\mathcal{H}_{\text{aux}} = \bigoplus_{n=0}^\infty \left\{\ket{h_i}_{i=1}^r\right\}^{\otimes n}$, under which $T_i$'s are Hermitian. To compute the inner product of two states, we sum over all possible pairings of chords of the same flavor between the two states, and for each such pairing we assign a weight of $q^{\# \text{ intersections}}$. If the states have a different number of chords of any flavor, then no such pairing exists and the states are orthogonal under this inner product. This inner product has a straightforward pictorial representation, an example of which can be seen in the figure \ref{fig:GeneralInnerProduct}. The explicit formula for the inner product is
\begin{equation} \label{eq:inn_oneq}
\inn{h_{i_1}\otimes \ldots \otimes h_{i_n}}{h_{j_1} \otimes \ldots \otimes h_{j_m}} = \delta_{m,n}
\smashoperator{\sum_{\substack{\text{pairings of $h_{i_k}$'s and $h_{j_{k'}}$'s}\\ \text{ of the same flavor}}} }
q^{\#_{\text{intersections}}} .
\end{equation}

This inner product is derived by constructing the Fock space from $r$ creation and annihilation operators, $a_i^\dagger$ and $a_i$, that satisfy the relations
\begin{equation} \label{eq:oneq_algebra}
a_i a_j^\dagger - q a_j^\dagger a_i = \delta_{ij},
\end{equation}
and demanding that $a_i^\dagger$ is the Hermitian conjugate of $a_i$ (see \cite{Speicher1991,Speicher1997}). Note that at this stage, nothing is assumed about the commutation relations of the $a_i$ among themselves (or the $a^\dagger_i$).

We will follow the procedure in \cite{Speicher1991} to define the inner product on the auxiliary Hilbert space under which $Q^\dagger$ is the Hermitian conjugate of $Q$. We start with vectors $\ket{v} = \ket{e_{1}e_{2}\ldots e_{n}}\in \mathcal{H}_{\text{aux}}$, where $e_i \in \{X,O\}$ represents the two types of chords we have, and $n_v$ is the number of chords in $\ket{v}$. We will denote by $X(v)$ and $O(v)$ the number of $X/O$ chords in $\ket{v}$. We assume that the inner product of $\inn{v}{u}$ is proportional to $ \delta_{O(v), O(u)}\delta_{X(v), X(u)}$, impose that $Q^\dagger$ is the Hermitian conjugate of $Q$, and arrive at the inner product
\begin{equation} \label{eq:inn_real}
\begin{split} \inn{v}{u} = & ~\delta_{O(v), O(u)}~\delta_{X(v), X(u)}
~q^{s(X(v)-O(v))+\frac{(X(v)-O(v))^2-X(v)-O(v)}{2}} \\
&\qquad\times \smashoperator{\sum_{\substack{\text{pairing of $X/O$'s in $\ket{v}$}\\ \text{ with $X/O$'s in $\ket{u}$}}} }
 (-1)^{\#_\text{intersections}} q^{\#_\text{ $X-O$ intersections}} . \end{split}
\end{equation}
This formula can be understood in the same way as \eqref{eq:inn_oneq} up to a normalization factor, we sum over all possible pairings of $X$'s and $O$'s in $\ket{v}$ with $X$'s and $O$'s in $\ket{u}$\footnote{We only connect $X$'s to $X$'s and $O$'s to $O$'s, connecting an $X$ to an $O$ is not allowed.} and to each pairing assign a value which depends on the intersections of chords. Each pairing receives a factor of $(-1)$ for any intersection of chords, and an additional factor of $q$ for each intersection of a chord connecting $X$'s with a chord connecting $O$'s. See appendix \ref{app:inn_general} for the full calculation.

This inner product can be thought of as a generalization of \eqref{eq:inn_oneq}, to a case where we have a more general algebra of creation and annihilation operators. In particular, we can generalize the relations (\ref{eq:oneq_algebra}) to
\begin{equation}\label{eq:gen_algebra}
a_i a_j^\dagger - q_{ij} a_j^\dagger a_i = \delta_{ij},
\end{equation}
with $q_{ij} = q_{ji}$ and $q_{ij} \in [-1,1]$ (see \cite{Speicher1993}). Then the inner product on the Fock space under which $a^\dagger_i$ is the Hermitian conjugate of $a_i$ is
\begin{equation} \label{eq:inn_general}
\inn{h_{i_1}\otimes \ldots \otimes h_{i_n}}{h_{j_1} \otimes \ldots \otimes h_{j_m}} = \delta_{m,n}
 \smashoperator[l]{\sum_{\substack{\text{pairings of $h_{i_k}$'s and $h_{j_{k'}}$'s}\\ \text{ of the same flavor}}} }
~~\smashoperator[r]{  \prod_{1 \leq i\leq j \leq r} }q_{ij}^{\#_{\text{intersections of $i$ and $j$ chords}}} .
\end{equation}

The inner product we found for the auxiliary Hilbert space, (\ref{eq:inn_real}), is of the form (\ref{eq:inn_general}) up to a global normalization of the vectors and with $q_{ij} = (q-1)\delta_{ij}-q$. We will later see in section \ref{sec:algebra} that the algebra of the fermionic creation and annihilation operators indeed satisfies relation (\ref{eq:gen_algebra}) with the given $q_{ij}$.

As a side remark, we note that within the double scaled SYK model (without SUSY) it is possible to create generalized statistics, as given by (\ref{eq:gen_algebra}). Taking multiple SYK operators with different double scaling limits $\lambda_i = \lim_{N \ra \infty} \frac{2p_i^2}{N}$ results in generalized statistics with $q_{ij} = e^{-\sqrt{\lambda_i \lambda_j}}$ (similar to the correlation functions in \cite{Micha2018,Berkooz_2019}). We can also consider the tensor product of $m$ SYK models, and operators that are tensor products of SYK operators, $H_i = H_i^{(1)}\otimes H_i^{(2)} \otimes \ldots \otimes H_i^{(m)}$, each with different double scaling limit $\alpha_i^{(a)} = \lim_{N\ra\infty} \sqrt{2/N} p_i^{(a)}$. In this case the generalized statistics will be $q_{ij} = e^{-\sum_{a=1}^m \alpha_i^{(a)} \alpha_j^{(a)}}$. This is similar to the model considered in \cite{GenSYK}.

\subsection{Reduction to the physical Hilbert space} \label{sec:PhysHS}

Notice that based on the inner product in the auxiliary Hilbert space any vector $v$ with two adjacent X's or O's has the property that $ \inn{v}{w}=0 $, for any vector $w$. This is because for any chord  between $v$ and $w$ there is also the chord diagram where the two adjacent chords are flipped, which has the same weight with an opposite sign. Thus we can define a physical Hilbert space by modding out all these null states, and the inner product will reduce to this physical Hilbert space as well.

\begin{figure} [h]
	\centering
	\includegraphics[page=1,width=0.4\columnwidth]{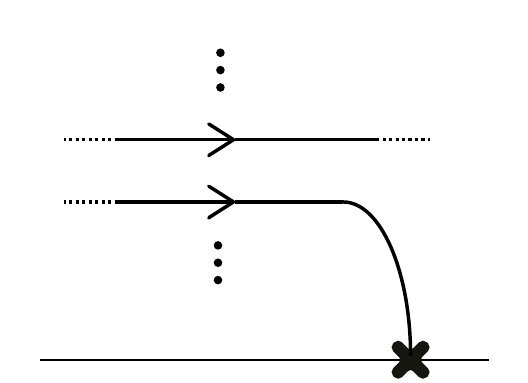}
	\includegraphics[page=2,width=0.4\columnwidth]{figures/NullifyingDiagrams.pdf}
	\caption{An example for two diagrams contributing to the state $ Q^{\dagger} \left |\cdots OO\cdots\right > $. As can be seen in section \ref{sec:TMrules}, the two diagrams have the same contribution, up to a minus sign coming from the intersection in the right diagram. This means that their sum will vanish. We see that we cannot bring diagrams of this type to an empty chord diagram. This means that any diagram with two consecutive $ X $'s or $ O $'s will not contribute to the element $ \left< \emptyset |T^k_s|\emptyset \right> $.}
	\label{fig:NullifyingDiagrams}
\end{figure}

Note that we can ignore vectors with adjacent X's or O's directly from the rules of the transfer matrix. Whenever we have two adjacent open chords of the same type, for every chord diagram there is a corresponding chord diagram in which at the point when one of those chords is closed, we replace it by closing the other chord. This diagram has the same value with an opposite sign because of the additional intersection of chords. This is similar to the inner product argument, but expressed directly in terms of the chord diagrams. This argument is demonstrated in figure \ref{fig:NullifyingDiagrams}.

Thus states with two adjacent $X$'s (or $O$'s) will not contribute to the moments $m_k \sim\innn{\emptyset}{T^k}{\emptyset}$.

We can therefore restrict the calculation of moments to the much smaller physical Hilbert space which only contains states of alternating $O$'s and $X$'s. This space can be characterized by vectors of the form
\begin{equation}
\ket{n,O} \equiv \overbrace{\ket{OXOX\ldots OX}}^{n \text{ pairs}}
\end{equation}
and states $\ket{n,X}$ which start with an $X$ instead of an $O$. We will also include fermionic states $\ket{n+1/2,X/O}$ that start and end with the same chord (of length $2n+1$). All in all we can write the physical Hilbert space as
\begin{equation}
\mathcal{H}_{\text{phys}} =  \left\{ \left|n,X\right>,\left|n,O\right>,\left|n-\frac{1}{2},X\right>,\left|n-\frac{1}{2},O\right>,\left|\emptyset\right>\right\} _{n=1}^{\infty} .
\end{equation}
We note that this is a much smaller space than the original Hilbert space, with only $2L+1$ states up to length $ L $.

We can calculate the inner product formula directly for physical states, however this requires summing over all chords between the vectors, which is complicated. We can instead calculate it directly from the physical Hilbert space, relying on the fact that states with different number of chords are also orthogonal, and that the required inner product is such that $Q$ and $Q^\dagger$ are adjoint of each other. This is done in appendix \ref{app:eigen}, and the result is:
\begin{equation} \label{eq:innerproduct_physical}
\begin{aligned}
&\inn{n,O}{n,O} = q^{-n} \left(q^2;q^2 \right)_{n-1},  && \inn{n,X}{n,X} = q^{-n} \left(q^2;q^2 \right)_{n-1},  \\
&\inn{n+\frac{1}{2},O}{n+\frac{1}{2},O} = q^{-s-n} \left(q^2;q^2 \right)_{n},  &&
 \inn{n+\frac{1}{2},X}{n+\frac{1}{2},X} = q^{s-n} \left(q^2;q^2 \right)_{n},\\
 &\inn{n,O}{n,X} = - \left(q^2;q^2 \right)_{n-1}.
 \end{aligned}
\end{equation}
This inner product is positive definite.

Based on the sub-diagrams we want to count, and the simple structure of this Hilbert space, we can compute how the transfer matrix acts on each of the base states. For the SUSY charge operators, the rules from before imply that when acting with $Q$ on a physical state we get
\begin{equation} \label{eq:Q_action}
\begin{aligned}
	&Q\left|n,X\right> =  -q^{n-1}\left|n-\frac{1}{2},O\right> ,
	&&Q\left|n,O\right> =  q^{-1}\left|n-\frac{1}{2},O\right>+q^s\left|n+\frac{1}{2},O\right>,\\
	&Q\left|n+\frac{1}{2},O\right> = 0,
	&&Q\left|n+\frac{1}{2},X\right> = q^{n}\left|n,O\right>+\left|n,X\right>+q^s\left|n+1,X\right> .
  \end{aligned}
  \end{equation}
And when acting with $Q^{\dagger}$ on a state we get
\begin{equation} \label{eq:Q_dg_action}
\begin{aligned}
	&Q^{\dagger}\left|n,O\right>	=  -q^{n-1}\left|n-\frac{1}{2},X\right>,
	&&Q^{\dagger}\left|n,X\right> =  q^{-1}\left|n-\frac{1}{2},X\right>+q^{-s}\left|n+\frac{1}{2},X\right>,\\
	&Q^{\dagger}\left|n+\frac{1}{2},X\right>	= 0,
	&&Q^{\dagger}\left|n+\frac{1}{2},O\right> = q^{n}\left|n,X\right>+\left|n,O\right>+q^{-s}\left|n+1,O\right>.
\end{aligned}
\end{equation}

The full transfer matrix, $T\equiv QQ^{\dagger}+Q^{\dagger}Q$, can be computed by the same rules. Acting with $T$ on the vacuum gives us
\begin{equation}
T\ket{\emptyset} = \ket{1,X}+\ket{1,O} + \left(q^s+q^{-s} \right)\ket{\emptyset} .
\end{equation}

We can then act on an arbitrary base state $\ket{n,X}$ and $\ket{n,O}$, and see that
\begin{equation}
\begin{split}
T\ket{n,X} &= \ket{n+1,X} + \left(q^sq^{-1}+q^{-s}\right)\ket{n,X}
+\left(q^{-s}q^{n} -q^{-s}q^{n-1}\right)\ket{n,O}\\
&\qquad \qquad +\left(q^{-1}-q^{2(n-1)} \right)\ket{n-1,X}+\left(q^{n-2}-q^{n-1}\right)\ket{n-1,O} ,
\end{split}
\end{equation}
and
\begin{equation}
\begin{split}
T\ket{n,O} &= \ket{n+1,O} + \left(q^{-s}q^{-1}+q^s\right)\ket{n,O}
+\left(q^sq^{n} -q^sq^{n-1}\right)\ket{n,X}\\
&\qquad \qquad + \left(q^{-1}-q^{2(n-1)} \right)\ket{n-1,O}+\left(q^{n-2}-q^{n-1}\right)\ket{n-1,X} .
\end{split}
\end{equation}

\section{Spectrum}

To compute the chord partition function, we will work in the physical sector of the auxiliary Hilbert space of partial chord diagrams. Our main goal will be to compute the matrix elements $\innn{\emptyset}{T_s^k}{\emptyset}$. We define the bosonic sector to be $ \mathcal{B} \equiv\text{Sp}\left\{ \left|\emptyset\right>,\left| n,X\right>,\left| n,O\right>\right\} _{n=1}^{\infty} $, and the fermionic sector to be $ \mathcal{F}\equiv\text{Sp}\left\{ \left|n+\frac{1}{2},X\right>,\left|n+\frac{1}{2},O\right>\right\} _{n=0}^{\infty} $. Note that this has little to do with the definition of bosonic or fermionic in the microscopic theory. Rather it is in the auxiliary space, which we interpret as the Hilbert space of the gravitational excitations.


\subsection{Diagonalization of $ T $}
To find the spectrum of $T$, we can use an asymptotic analysis\footnote{This is true so long as there are no bound states at small values on $n$. Later when we find the eigenvectors of $T$ we will see that this is indeed the case.}. As $0<q<1$, we can look at the asymptotic form of the matrix in the limit $q^n \ra 0$. Notice that the asymptotic form of the matrix decouples the $X$ and the $O$ sectors, giving us
\begin{equation}
\begin{split}
T_{asy}\ket{n,X} &= \ket{n+1,X} + \left(q^{s}q^{-1}+q^{-s}\right)\ket{n,X}
+q^{-1}\ket{n-1,X}, \\
T_{asy}\ket{n,O} &= \ket{n+1,O} + \left(q^{-s}q^{-1}+q^{s}\right)\ket{n,O}
+q^{-1}\ket{n-1,O} .
\end{split}
\end{equation}

This is a tri-diagonal matrix, immediately giving us the eigenvalues
\begin{equation}
\Lambda_{\pm,k} = q^{\pm s-1}+q^{\mp s}-\frac{2}{\sqrt{q}}\cos\left(\frac{\pi k}{L+1}\right)
\ra q^{\pm s-1}+q^{\mp s} -\frac{2}{\sqrt{q}}\cos(\phi),
\end{equation}
for $\phi\in (0,\pi)$ uniformly distributed. The eigenvalue $ \Lambda _+$ is an eigenvalue of the $ X $ sector while $ \Lambda _{-} $ is an eigenvalue of the $ O $ sector. As $ \Lambda_+(s)=\Lambda_-(-s)$, we will call $\Lambda_s(\phi) \equiv \Lambda_+(s)$ with $\Lambda_{-s}(\phi) = \Lambda_-(s)$. An alternative way to write these eigenvalues would be
\begin{align}\label{eq:EVals}
\Lambda_{s}(\phi) = 2q^{-1/2}[\cosh(\lambda s -\lambda/2)-\cos(\phi)].
\end{align}
 Furthermore, notice that $\Lambda_s(\phi) \geq 0$ as expected from a super-symmetric theory.

With the spectrum of $T$ in hand, we move on to diagonalize the transfer matrix. As $T$ is a bosonic operator it is sufficient to diagonalize $T$ on the bosonic sector in order to calculate the desired matrix elements\footnote{The spectrum in the fermionic sector will be identical to the bosonic sector due to SUSY.}. Let us define the subspaces $ B\equiv Q\mathcal{F}$ and $\bar{B}\equiv Q^{\dagger}\mathcal{F}$. As these are $T$ invariant subspaces, $T$ can be diagonalized separately in each subspace. From SUSY considerations, it follows that the only bosonic states not in $B\oplus \bar{B}$ must be ground states. Furthermore, from the asymptotic matrix analysis we see that $ \Lambda(\phi)>0 $ for all states but a set of measure zero (only when $s=\pm1/2$ do we have a single zero energy state at the edge of the spectrum), which shows that $T$ is positive definite. Therefore $T$ has no ground states and $ \mathcal{B}=B\cup \bar{B} $.

Diagonalizing $ T $ over the spaces $ B = \text{Sp}\left\{ Q\left|n+\frac{1}{2},X\right>\right\} _{n=0}^{\infty}$ and $\bar{B} = \text{Sp}\left\{ Q^{\dagger}\left|n+\frac{1}{2},O\right>\right\} _{n=0}^{\infty} $ proves to be relatively simple. Denote
 \begin{align}{\label{b_nStates}}
 \left|b_{n}\right>\equiv Q\left|n+\frac{1}{2},X\right>,\qquad\left|\overline{b}_{n}\right>\equiv Q^{\dagger}\left|n+\frac{1}{2},O\right> .
 \end{align}
 In terms of these states the transfer matrix acts by
 \begin{equation} \label{eq:T_on_bn}
\begin{split}
	T\left|b_{n}\right>	&=q^{-1}\left(1-q^{2n}\right)\left|b_{n-1}\right>+\left(q^{-s}+q^{-1}q^{s}\right)\left|b_{n}\right>+\left|b_{n+1}\right>,\\
T\left|\overline{b}_{n}\right>	&=q^{-1}\left(1-q^{2n}\right)\left|\overline{b}_{n-1}\right>+\left(q^{s}+q^{-1}q^{-s}\right)\left|\overline{b}_{n}\right>+\left|\overline{b}_{n+1}\right>.
\end{split}
\end{equation}

 Using the asymptotic matrices, we see that the eigenvalues of $ T  $ restricted to $ B  $ are $ \Lambda_{s}\left(\phi\right) $, while the eigenvalues of $ T $ restricted to $ \bar{B} $  are $ \Lambda_{-s}\left(\phi\right) $. Denote the eigenvector corresponding to the eigenvalue $ \Lambda_{s}\left(\phi\right) $ by $ \left|v\left(\phi\right)\right> $, which is an element of $B$. Therefore we can write $ \left|v\left(\phi\right)\right>=\sum_{n=0}^{\infty}\alpha_{n}\left|b_{n}\right> $, for some constants $\alpha_n$. The eigenvalue equation $ T\left|v\left(\phi\right)\right>=\Lambda_{s}\left(\phi\right)\left|v\left(\phi\right)\right>  $ gives
 \begin{equation}
\begin{split}
	\sum_{n=0}^{\infty}\Lambda_{s}\left(\phi\right)\alpha_{n}\left|{b}_{n}\right>	&=\sum_{n=0}^\infty \left[\left(q^{-s}+q^{-1}q^{s}\right)\alpha_{n}\left|b_{n}\right>+q^{-1}\left(1-q^{2n}\right)\alpha_{n}\left|b_{n-1}\right>+\alpha_{n}\left|b_{n+1}\right>\right]\\
&=\sum_{n=0}^\infty \left[\left(q^{-s}+q^{-1}q^{s}\right)\alpha_{n}+q^{-1}\left(1-q^{2\left(n+1\right)}\right)\alpha_{n+1}+\alpha_{n-1}\right]\left|b_{n}\right>,
\end{split}
\end{equation}
from which we obtain the recursion relation over the coefficients $ \alpha_{n} $
\begin{align}
	\frac{2}{\sqrt{q}}\cos(\phi)\alpha_{n}=q^{-1}\left(1-q^{2\left(n+1\right)}\right)\alpha_{n+1}+\alpha_{n-1}.
\end{align}
If we momentarily allow $ n=-1 $, and define $ \alpha_{-1}=0 $, we can redefine $ \alpha_{n} $ to be
\begin{align}
	\alpha_{n}=\frac{q^{\frac{n}{2}}}{\left(q^{2};q^{2}\right)_{n}}a_{n}\qquad a_{-1}=0,a_{0}=1,
\end{align}
such that the above relation becomes
\begin{align}
	2\cos(\phi) a_{n}=a_{n+1}+\left(1-q^{2n}\right)a_{n-1},\qquad a_{-1}=0,a_{0}=1.
\end{align}
We see that the $ a $'s hold the recursion relation satisfied by the continuous q-Hermite polynomials $ H_{n}\left(\cos\phi|q^{2}\right) $ \cite{Hypergeometric_Book}, hence the $ \alpha $'s hold
\begin{equation} \label{eq:alphas}
	\alpha_{n}\left(\phi\right)=\frac{q^{n/2}}{\left(q^{2};q^{2}\right)_{n}}H_{n}\left(\cos\phi|q^{2}\right).
\end{equation}
Note that $ T\big|_{B},T\big|_{\bar{B}} $ are symmetric under the assignment $ \left|b_{n}\right> \to \left|\overline{b}_{n}\right>,s\to(-s) $ ; which means that the eigenvector $ \left|u\left(\phi\right)\right>\in\bar{B} $ with the eigenvalue $ \Lambda_{-s}(\phi)$ is given by $\ket{u(\phi)} = \sum \alpha_n \ket{\overline{b}_n}$ , with the same $\alpha_n$'s.

\subsection{Calculating the moments and the density of states}

To calculate the matrix elements $\innn{\emptyset}{T^k_s}{\emptyset}$ we insert a complete set of eigenvectors. However, we first need to normalize the eigenvectors. Trivially $\inn{v(\theta)}{u(\phi)} = 0$ as they live in orthogonal subspaces. We can calculate the inner product of two $\ket{v(\phi)}$ vectors using the orthogonality relations of $q$-Hermite polynomials and the inner product defined in the previous section. See appendix \ref{app:eigen} for the full calculation. The result is
\begin{align}
\inn{v(\phi)}{v(\phi')} &=q^{s}\Lambda_{s}\left(\phi\right)\frac{2\pi\delta\left(\phi-\phi'\right)}{\left(q^2,e^{\pm2i\phi};q^{2}\right)_{\infty}}.
\end{align}
The inner product $\inn{u(\phi)}{u(\phi')}$ is the same only with $s \ra -s$.

Then plugging this in, the matrix element becomes
\begin{align}
\innn{\emptyset}{T^k_s}{\emptyset} =  \int_0^\pi \frac{d\phi}{2\pi}~
\left(q^2,e^{\pm2i\phi};q^{2}\right)_{\infty} \left[q^{-s}\Lambda_{s}^{k-1}(\phi) + q^{s}\Lambda_{-s}^{k-1}(\phi) \right], ~~~\text{for } k>0,
\end{align}
and for $k=0$ the matrix element is just 1.

We can then plug this into the formula for the moments and see that
\begin{equation}{\label{PartitionFnMoment}}
\begin{split}
m_k(\mu ) &= 2^{-k} q^{\frac{k}{4}} \sqrt{\frac{\lambda}{\pi}}\intinf ds~ q^{s^2} e^{-\mu s}
  \int_0^\pi \frac{d\phi}{2\pi}~\left(q^2,e^{\pm2i\phi};q^{2}\right)_{\infty}
 \left[q^{-s}\Lambda_{s}^{k-1}(\phi) + q^{s}\Lambda_{-s}^{k-1}(\phi) \right] \\
 &=  q^{-\frac{k-1}{4}} \sqrt{\frac{\lambda}{\pi}}\intinf ds~ q^{(s+1/2)^2} \cosh(\mu s)
 \int_0^\pi \frac{d\phi}{2\pi}~\left(q^2,e^{\pm2i\phi};q^{2}\right)_{\infty}\\
&~~~~~~~~~\times\left(\cosh[\lambda (s + 1/2)] - \cos(\phi) \right)^{k-1} ,
\\
m_0(\mu)&=1.
\end{split}
\end{equation}

\begin{figure}
	\centering
	\includegraphics[width=0.6\columnwidth]{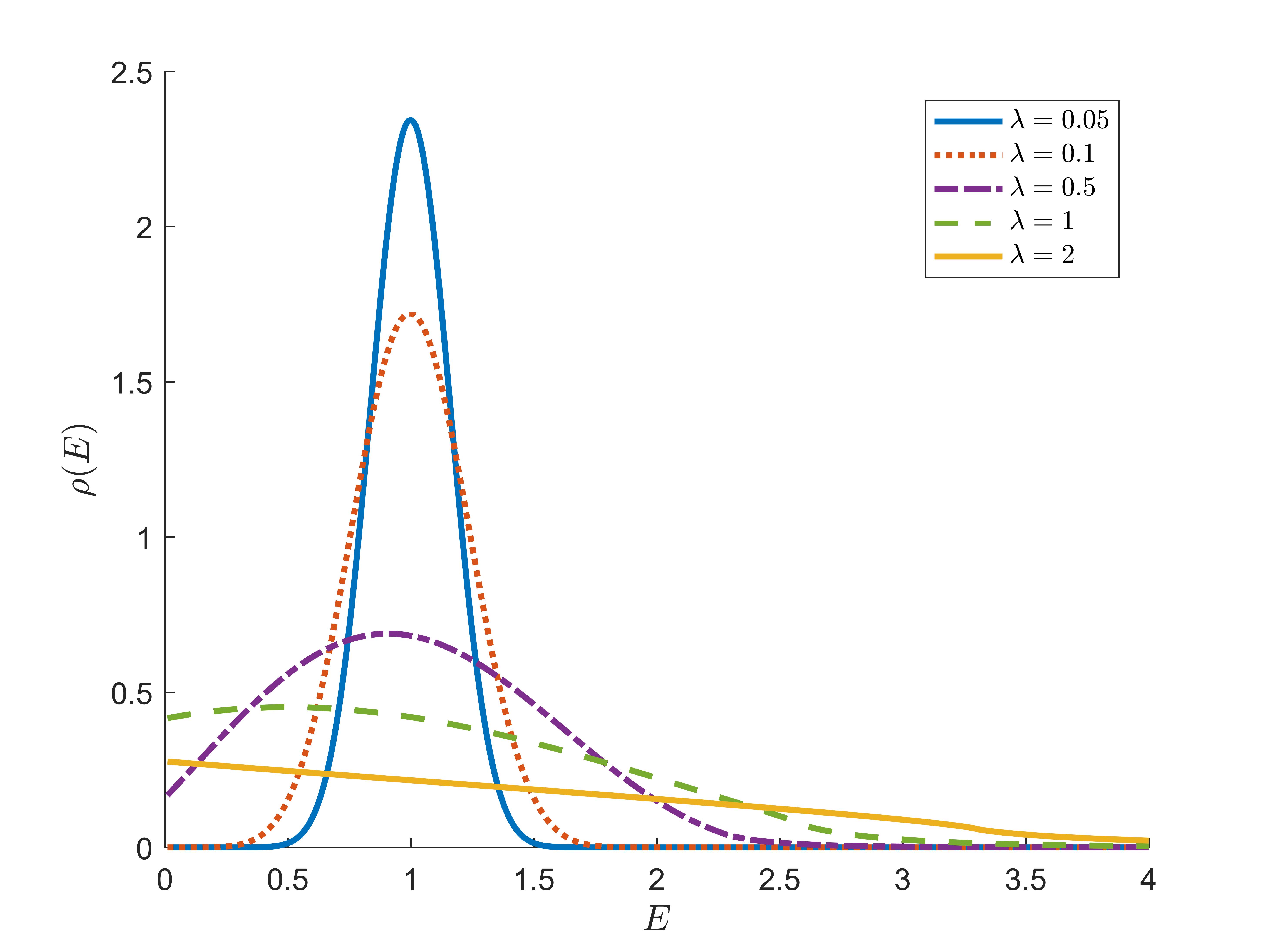}
	\caption{The density of states (continuous part) for various values of $\lambda$.}
	\label{fig:density_of_states}
\end{figure}

The energies depend on the charge $s$ and an angle $\phi$, and are given by
\begin{equation}
E(s,\phi)  = q^{-1/4} \left(\cosh[\lambda (s + 1/2)] - \cos(\phi) \right),
\end{equation}
with the continuous density of states (coupled to a chemical potential) given by
\begin{equation} \label{eq:den_of_states}
\begin{split}
\rho_c(E;\mu) &= \cosh(\mu/2) \frac{q^{1/4}}{\pi^{3/2}\sqrt{\lambda}}
\int_0^\pi d\phi~ \frac{\left(q^2,e^{\pm2i\phi};q^2\right)_{\infty} }{E\sqrt{\left(q^{1/4}E+\cos\phi \right)^2-1}}\\
&~~~~~~~~~~~~~~~\times \exp\left\{-\frac{1}{\lambda}\left[\cosh^{-1}\left(q^{1/4}E+\cos\phi \right) \right]^2\right\} \\
&~~~~~~~~~~~~~~~\times\cosh\left[\frac{\mu\cosh^{-1}\left(q^{1/4}E+\cos\phi \right) }{\lambda}\right] \Theta\left(q^{1/4}E+\cos\phi -1 \right) .
\end{split}
\end{equation}
A plot of the continuous energy distributions (without a chemical potential) for some values of $\lambda$ is given in figure \ref{fig:density_of_states}.

As can be seen by integrating $\rho(E,0)$ over the entire spectrum, the density integral does not amount to $1$. This since the density includes only the continuous part of the spectrum, and misses any $\delta$ functions contributions at zero. Contributions of the form $D\cdot\delta(E)$ appear as a missing density when we integrate the above density of states over $E$, which is simply looking at the zeroth moment of the continuous distribution without a chemical potential. After re-summing this moment (see appendix \ref{app:ground_states} for the complete calculation) we find that
\begin{equation}
1-D(\lambda)=m_0 = \int dE \rho_c(E,0)=2 \sum_{k=0}^\infty(-1)^k \text{erfc}\left(\left(k+\frac{1}{2}\right)\sqrt{\lambda}\right)\ ,
\end{equation}
with the ground state density given by
\begin{equation}
D(\lambda) = \int_{-1/2}^{1/2} ds~\vartheta_2\left(\pi s, e^{-\frac{\pi^2}{\lambda}}\right).
\end{equation}
This is  in agreement with the density of ground states found using a cohomology argument, which is done in the next section.

We can take the Fourier transform in the chemical potential of equation \eqref{eq:den_of_states} to find the continuous spectrum in a fixed charge sector (ignoring the $\delta$ function at zero) to be
\begin{equation} \label{eq:spec_charge_cont}
\begin{split}
\rho_c(E;s) = \frac{\sqrt{\lambda}}{2\pi^{1/2}}&
\int_0^\pi d\phi~ \left(q^2,e^{\pm2i\phi};q^2\right)_{\infty} \\
&\times \left(\frac{q^{(s-1/2)^2}}{E}\delta\left(E - E(s,\phi)\right)
+ \frac{q^{(s+1/2)^2}}{E}\delta\left(E - E(-s,\phi)\right)\right) .
\end{split}
\end{equation}

This continuous spectrum in a fixed charge sector has a minimal energy
\begin{equation}
E_{\min}(s) = q^{-1/4}\left[\cosh\left( \lambda\left(|s|-\frac{1}{2}\right)\right) - 1\right]
 \approx \frac{\lambda^2}{2}\left(|s|-\frac{1}{2}\right)^2,
\end{equation}
in the $\lambda\ra 0$ limit. Translating to the Schwarzian result in \cite{Stanford_Witten2017}, we have that $i\mu = 2\pi n \hat{q}$ and $s \mu = 2 \pi m n$, so $s = \frac{m}{\hat{q}}$, and we obtain the exact same result. This also agrees with \cite{Mertens_2017}. It is interesting that the spectrum starts at zero only for the sector with charge $s= \pm1/2$, which has precisely $p/2$ fermions. This is also the extremal charge sector that contains ground states.

Finally, we look at how the density of states changes when we vary $\lambda$. In the limit $\lambda \ra 0$ we have that $\rho(E) \ra \delta(E-1)$. This concentration of measure around $E=1$ can be seen in figure \ref{fig:density_of_states}, and is also found in the exact computation when setting $q=1$. This is similar behavior to the $\mathcal{N} = 1$ model, when the density is also concentrated at $E=1$. In the limit $\lambda \ra \infty$ almost all states become ground states, and $\rho(E) \ra \delta(E)$. This can be seen in figure \ref{fig:ground_states} which shows the density of ground states approaching $1$ as $\lambda \ra \infty$. We can think of the transition between long interactions and short interaction as a quantum phase transition, similar to \cite{Erdos14}.


\subsection{Supersymmetric ground states}

We shall now present an analysis of the number of ground states in every charge sector. This is an extension of the analysis done in \cite{Kanazawa_2017}. The ground states, or super-symmetric states are the zero energy states $\ket{\psi}$ such that $Q\ket{\psi} = \overline{Q}\ket{\psi} = 0$. These are also the states in the cohomology of $Q$, that is states in $\ker{(Q)}/\text{Im}(Q)$. We can calculate the cohomology directly using combinatorial arguments. We expect this analysis to be valid for almost all realizations of the couplings, except for a small set of coupling of measure zero (in the Gaussian measure of the space of random couplings). We do not have a full proof of this statement, but rather provide a heuristic argument. We will see that this analysis misses an $O(1)$ number  of ground states with charge $s=\pm 1/2$, but otherwise this argument seems exact.

The full Hilbert space of this model, which we shall denote $\Ham$, consists of the tensor product of $N$ complex fermions, and has $2^N$ states. This Hilbert space is spanned by basis states which are represented by a set of spins, up or down, for each site. The $U(1)$ charge of each base state is linearly related to the number of up spins in the state, $m$, with $0\leq m\leq N$, by $s = (m-N/2)/p$. Let us denote the subspace with $m$ up spins as $\Ham_m$, and note that  $\dim\left(\Ham_m\right) = { N \choose m }$.

We will start from a state with $r$ up spins such that $0\leq r <p$, and consider the long exact sequence:
\begin{equation}
0 \xrightarrow{Q} \Ham_r \xrightarrow{Q} \Ham_{r+p} \xrightarrow{Q} \Ham_{r+2p} \xrightarrow{Q}
\ldots \xrightarrow{Q} \Ham_{r+(M-1)p} \xrightarrow{Q} \Ham_{r+Mp} \xrightarrow{Q} 0,
\end{equation}
with $M=\lfloor \frac{N-r}{p}\rfloor$. Let us start by calculating the image of $Q$, and denote
\begin{equation}
l(m;r)\equiv \dim\left(\text{Im} Q\right)\big|_{\Ham_{r+mp}} .
\end{equation}

For $m=0$ we see immediately that $l(0;r)=0$. For $m=1$ notice that as $\dim(\Ham_{r+p})>\dim(\Ham_r)$ and there are no states in $\Ham_r$ that must be sent to zero, so for a generic realization of the couplings $l(1,r) = {N \choose r}$. For $m=2$ the same argument holds, only now we do have a subspace of exact states that must be sent to zero. This tells us that $l(2,r) = {N \choose r+p} - {N \choose r}$. Continuing on with the sequence we see that
\begin{equation}
l(m;r) = \sum_{n=0}^{m-1} (-1)^{m-1-n}{ N \choose r+np} ,
\end{equation}
at least for $r+mp<N/2$, and that all the cohomologies must be zero except for the one closest to $N/2$.

Let us now start calculating the kernel of $Q$ from the other side of the long exact sequence, and denote
\begin{equation}
S(m;r) \equiv \dim\left(\ker(Q)\right)\big|_{\Ham_{r+mp}} .
\end{equation}
This time we get for free that $S(M,r) = {N \choose r+Mp}$. For $m=M-1$ we expect that the image of $\Ham_{r+Mp-p}$ will be all of $\Ham_{r+Mp}$ as it is a larger vector space and $Q$ is random, so we should have that $S(M-1;r)={N \choose r+Mp - p} - {N \choose r+Mp}$. Continuing down the chain with the same argument we get that
\begin{equation}
S(m;r) =  \sum_{n=0}^{M-m} (-1)^{n}{ N \choose r+(m+n)p} ,
\end{equation}
at least for $r+mp>N/2$, and again all the cohomologies must be zero except for the one closest to $N/2$. Thus for  every value of $r$ we get only a single charge sector with a nonzero cohomology and that will be the charge sector with the least charge. We will get a non-zero number of ground states only for charges $|s|<1/2$. The dimension of this cohomology, call it $D(r)$ will be
\begin{equation}
D(r) = S(m_c;r) - l(m_c;r) = (-1)^{m_c-1} \sum_{n=0}^{M} (-1)^{n}{ N \choose r+np} ,
\end{equation}
with $m_c$ the critical $m$ value for which the cohomology is non-zero.

This analysis may break down slightly, missing an $O(1)$ number of states, in the particular case where for a specific $r$ and $N$ we have that $(N\pm p)/2$ is an integer and so both cases are marginal, as there is no closest $m$ value. This is precisely the charge sectors $s=\pm 1/2$. This analysis predicts that the cohomology of this sector will be zero, but in the $p=3$ case it was found numerically to be $0,1,$ or $3$, which in the large $N$ case is negligible. For all other cases this formula replicates exactly the numerical results found for $p=3$ in \cite{SusySYK}, as well as the analytical results for $p=3$ in \cite{Kanazawa_2017}.

We can take the double scale limit of this formula by plugging in the $U(1)$ charge definition, and then normalizing by the size of the Hilbert space. Then we get that
\begin{equation}
\begin{split}
D(s) ds &= 2^{-N} \sum_{n=-\lfloor M/2 \rfloor}^{ \lfloor M/2 \rfloor} (-1)^{n}{ N \choose N/2 + (s+n)p} \\
&=2^{-N}  \sum_{n=-\infty}^{\infty} (-1)^{n} { N \choose N/2 + (s+n)\sqrt{\lambda N/2}} \\
&\xrightarrow{N\ra\infty} \sqrt{\frac{2}{N\pi}} \sum_{n=-\infty}^{\infty} (-1)^{n} e^{-\lambda(n+s)^2} + O(N^{-3/2}) \\
&=  ds \sqrt{\frac{\lambda}{\pi}} \sum_{n=-\infty}^{\infty} (-1)^{n} e^{-\lambda(n+s)^2} ,
\end{split}
\end{equation}
where we substituted $ds  = 1/p = \sqrt{2/(N \lambda)}$. This is just a Jacobi theta function (the conventions we use appear in (\ref{eq:jactheta_4})), and in particular we get
\begin{equation}
D(s) = \sqrt{\frac{\lambda}{\pi}} q^{s^2}\vartheta_4(i\lambda s, q) .
\end{equation}

Using the modular transformation (\ref{eq:theta_modular_trans}) gives us the final form
\begin{equation}
D(s) = \vartheta_2\left(\pi s, e^{-\frac{\pi^2}{\lambda}}\right) ,
\end{equation}
which is the infinitesimal fraction of ground states at charge $s$.

We can now integrate this over $s\in(-1/2,1/2)$ to get the fraction of the Hilbert space that is a ground state:
\begin{equation}
D = \int_{-1/2}^{1/2} ds~\vartheta_2\left(\pi s, e^{-\frac{\pi^2}{\lambda}}\right).
\end{equation}
We expect this to also be the value multiplying $\delta(E)$ in the normalized density of states, which agrees with what we have shown above.

We can also easily take the $\lambda \ra 0$ limit as
\begin{equation}
D(s) \approx 2e^{-\frac{\pi^2}{4\lambda}}\cos(\pi s)\left(1 + O\left(e^{-\frac{2\pi^2}{\lambda}}\right) \right).
\end{equation}
This result agrees with the number of ground states found through the Schwarzian analysis in \cite{Mertens_2017}. We can also integrate this over $s\in [-1/2,1/2]$ to get the full number of ground states in this limit:
\begin{equation}
D(\lambda) = \frac{4}{\pi}e^{-\frac{\pi^2}{4\lambda}} \left(1+O(e^{-4\pi^2/\lambda})\right).
\end{equation}

At finite $N$ and $p$ this would approximate the number of ground states as
\begin{equation}
D(s;p,N) \approx \frac{2}{p}\cos(\pi s) \left(2 e^{-\frac{\pi^2 }{8 p^2}}\right)^N.
\end{equation}
When $p=3$ the actual number of ground states is $2/3*cos(\pi s)~3^{N/2}$. Already we see that $2 e^{-\frac{\pi^2 }{8 p^2}} \approx \sqrt{3}$ to within 1 percent. For $p=5$ the agreement is better, with
\begin{equation}
D(s;p=5) \approx \frac{2}{5}\cos\left( \pi s\right) \left(\frac{5+\sqrt{5}}{2}\right)^{N/2} ,
\end{equation}
at large $N$, up to exponentially small terms. Here the agreement with the infinite $p$ limit is to within less than $0.1\%$.

\begin{figure}
	\centering
	\includegraphics[width=0.6\columnwidth]{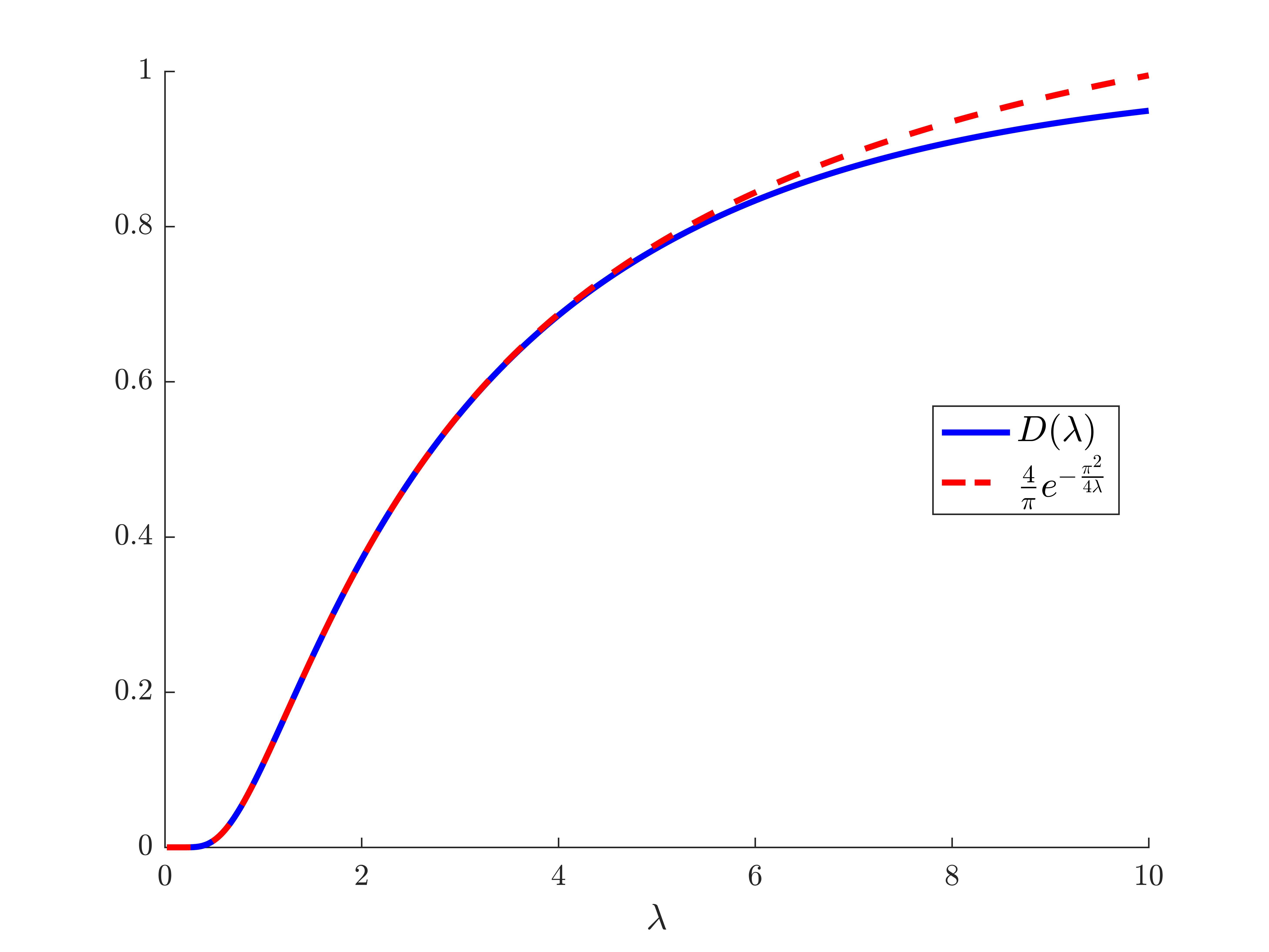}
	\caption{The density of ground states as a function of $\lambda$.}
	\label{fig:ground_states}
\end{figure}

We present a plot of $D(\lambda)$ as a function of $\lambda$ in figure \ref{fig:ground_states}, as well as a comparison to the small $\lambda$ approximation. We see that the small $\lambda$ approximation is a very good approximation up until $\lambda\approx 5$, which is somewhat surprising. We also see that at finite $\lambda$ the number of ground states represents a finite fraction of the total number of states, and that for large $\lambda$ most states are supersymmetric.

\subsection{The Schwarzian limit of the distribution}

We now look at the super conformal limit of the distribution, which is the low energy short interactions limit. We expect our results to reduce to the super--Schwarzian density of states in the so called triple scaling limit (see \cite{Cotler:2016fpe}) $E\ra 0, \lambda \ra 0$.

We start by considering only the density of states $ \rho(E, \mu=0) $ under this double limit. We will take $ E=\epsilon/2 $. The Heaviside forces us to have $ \epsilon/2+\cos\phi \geq 1 $, which limits the integration domain. Since $ \cos\phi $ is decreasing close to the origin we see that the integration limit should be taken up to $ \phi = \sqrt{\epsilon} +O(\epsilon) $, which means that $ E\to 0 \Rightarrow \phi\to 0 $. With this the Heaviside becomes 1.  In this limit we can use two useful approximations \cite{Micha2018}:
\begin{equation}
q^{1/4}\left(q^2,e^{\pm2i\phi};q^2\right)_{\infty}
\approx 8 \sin\phi \sqrt{\frac{\pi}{\lambda}}e^{-\frac{1}{\lambda}\left[\pi^2+\left(\phi-\frac{\pi}{2} \right)^2\right]}
\sinh\left(\frac{\pi\phi}{\lambda}\right)\sinh\left(\frac{\pi(\pi-\phi)}{\lambda}\right),
\end{equation}
and
\begin{equation}
e^{-\frac{1}{\lambda}\left(\cosh^{-1}\left(q^{1/4}E+\cos\phi\right)\right)^{2}}  \approx e^{-\frac{1}{\lambda}\left(\epsilon-\phi^{2}\right)}+O\left(\epsilon^{2}\right),
\end{equation}
under which the density of states, \eqref{eq:den_of_states}, becomes
\begin{align}
\begin{split}
\rho\left(\frac{\epsilon}{2}\right)&=\frac{e^{-\frac{\pi^{2}}{4\lambda}}}{\pi\lambda\epsilon}\int_{0}^{\sqrt{\epsilon}}d\phi\frac{\phi}{\sqrt{\epsilon-\phi^{2}}}\sinh\left(\frac{\pi\phi}{\lambda}\right)
= \frac{e^{-\frac{\pi^{2}}{4\lambda}}}{2\lambda\sqrt{\epsilon}}I_{1}\left(\frac{\pi\sqrt{\epsilon}}{\lambda}\right),
\end{split}
\end{align}
which is the Super-Schwarzian density of states.

Next we will add the chemical potential. To compare to the Schwarzian results in \cite{Stanford_Witten2017} we look at the low energy $\lambda \ra 0$ limit, which as before gives us the density
\begin{equation}
\begin{split}
\rho_0\left(\frac{\epsilon}{2};\mu\right)&\approx
\frac{4 \cosh(\mu/2)}{\sqrt{\lambda}\epsilon} e^{\frac{\mu^2}{4\lambda}-\frac{\pi^2}{4\lambda}}
\intinf dx~e^{-x^2}\frac{1}{2\pi}\int_0^\pi d\phi~ e^{-\frac{\phi^2}{\lambda}}\sin\phi \\
&~~~~~~~~~~\times\sinh\left(\frac{\pi\phi}{\lambda}\right)
\delta\left(\epsilon - \phi^2-\left(x\sqrt{\lambda}+\mu/2 \right)^2 \right) \\
&= \frac{2 \cosh(\mu/2)}{\lambda\sqrt{\epsilon}\pi} e^{-\frac{\pi^2+4\epsilon}{4\lambda}}
\int_0^\pi dt~ \int_0^{\infty} dr~r^2\sin t \cosh\left(\frac{\sqrt{\epsilon}\mu}{\lambda} r \cos t \right)\\
&~~~~~~~~~~\times \sinh\left(\frac{\pi\sqrt{\epsilon}}{\lambda}r \sin t\right)\delta\left(1- r^2 \right) \\
&= \frac{ \cosh(\mu/2)}{\lambda\sqrt{\epsilon}\pi} e^{-\frac{\pi^2+4\epsilon}{4\lambda}}
\int_0^\pi dt~\sin t~ \cosh\left(\frac{\sqrt{\epsilon}\mu}{\lambda} \cos t \right)
\sinh\left(\frac{\pi\sqrt{\epsilon}}{\lambda} \sin t\right) .
\end{split}
\end{equation}
It remains to evaluate the integral
\begin{equation}
I(a,b) \equiv \int_0^\pi dt~\sin t~ \cosh\left(a \cos t \right) \sinh\left(b \sin t\right) .
\end{equation}
We will define $\rho^2 \equiv a^2 + b^2$ and $\tan\varphi \equiv \frac{b}{a}$, so that $a \cos t + b\sin t  = \rho \cos(t+\varphi)$. Then using trigonometric identities we see that
\begin{equation}
\begin{split}
I(a,b) &=  \int_0^\pi dt~\sin t~\sinh\left(\rho \cos( t+\varphi)\right)\\
&=\frac{1}{2} \sin(\varphi) \int_{-\pi}^\pi dt~\exp\left(\rho \cos( t) + it\right)\\
&= \frac{\pi}{\sqrt{1+\frac{a^2}{b^2}}} I_1\left(\sqrt{a^2 +b^2}\right) .
\end{split}
\end{equation}

Together this gives
\begin{equation}
\rho_0\left(\frac{\epsilon}{2};\mu\right)\approx
 \frac{ \cosh(\mu/2)}{\lambda\sqrt{\epsilon\left(1+\frac{\mu^2}{\pi^2}\right)}}
e^{-\frac{\pi^2+4\epsilon}{4\lambda}}
I_1\left(\frac{\pi}{\lambda}\sqrt{\epsilon\left(1+\frac{\mu^2}{\pi^2}\right)}\right) ,
\end{equation}
which is in agreement with the Schwarzian result from \cite{Stanford_Witten2017} when taking $\mu =i2\pi n\hat{q}$.

We can also take the Schwarzian limit of the energy distribution in a fixed charge sector, from equation (\ref{eq:spec_charge_cont}). This involves taking the low energy $\lambda \ra 0$ limit of $\int_0^\pi (q^2,e^{\pm2i\phi};q^2)_\infty \delta(E- E_0(s) +\cos\phi)$ which is the same limit as in the regular SYK model \cite{Cotler:2016fpe}, giving us that the continuous spectrum in a fixed charge sector is
\begin{equation}
\rho_c(E,s) \propto \frac{\sinh(2\pi \sqrt{E-E_0(s)})}{E}\Theta(E-E_0(s)) ~~+ ~~(s \ra -s),
\end{equation}
where $E_0(s) = \frac{\lambda^2}{2}\left(s-\frac{1}{2}\right)^2$. This is in agreement with the Schwarzian results from \cite{Mertens_2017}. We note that the number of ground states in each charge sector is also in agreement with the Schwarzian results from \cite{Mertens_2017}, as was shown in the previous section. Thus the spectrum of the double scaled $\mathcal{N}=2$ SUSY SYK model exactly reduces to the super--Schwarzian density of states in this limit.

\section{Reduction to the Liouville action}

It is well known that the low energy Schwarzian theory of the SYK model reduces to Liouville quantum mechanics (see \cite{Bagrets_2016,Bagrets_2017,Mertens_2017}). As such we expect that the transfer matrix also reduces to the Hamiltonian of Super--Liouville quantum mechanics  in the low energy continuum limit, when we additionally take the limit $q\ra 1$. A similar reduction was seen in the transfer matrix of the regular SYK model \cite{Micha2018}.


We will concentrate on a fixed charge sector $ s $, and work with the basis of exact states $ (\{\ket{b_n}\},\{\ket{\bar{b}_n}\})_{n=0}^{\infty} $ from (\ref{b_nStates}). If we now act on a general vector $ \ket{u}=\sum u_n\ket{\bar{b}_n} $ with the transfer matrix, we see that the $ n $'th component of the new vector is
\begin{align}
	\left(T_{s}\left|u\right>\right)_{n}=\left(q^{-1}-q^{2n+1}\right)u_{n+1}+\left(e^{-\lambda s}+q^{-1}e^{\lambda s}\right)u_{n}+u_{n-1}.
\end{align}

As the asymptotic matrix is a constant tri-diagonal matrix, in order to focus on the low energy states we introduce  the twist (along with a rescaling) $ \tilde{u}_n= (-1)^nq^{-n/2}u_n $,  and get
\begin{align}
	\sqrt{q}\left(T_{s}\left|\tilde u \right>\right)_{n}&=-\left(1-q^{2n+2}\right)\tilde{u}_{n+1}+\left(e^{-\lambda s}q^{1/2}+q^{-1/2}e^{\lambda s}\right)\tilde{u}_{n}-\tilde{u}_{n-1}.
\end{align}
We now take the continuum limit by defining the variable $ q^{2n+2}\equiv e^\phi$, and the continuum function $\tilde{u}_n\equiv u(\phi) $. Using these definitions we have $ \tilde{u}_{n\pm1}=e^{\mp2\lambda\partial_{\phi}}u\left(\phi\right) $. This gives us
\begin{align}
	\sqrt{q}T_{s} u(\phi)=\left[e^{\phi}e^{-2\lambda\partial_{\phi}}+\left(e^{-\lambda\left(s+\frac{1}{2}\right)}+e^{\lambda\left(s+\frac{1}{2}\right)}\right)-\left(e^{-2\lambda\partial_{\phi}}+e^{2\lambda\partial_{\phi}}\right)\right] u(\phi).
\end{align}

Finally, we take the limit $q\ra 1^-$, or $\lambda \ra 0$. We do this by first shifting the variable $\phi$ by defining a variable $\varphi$ through $\phi=\varphi+2\log\lambda$. Then expanding the transfer matrix in the small parameter $\lambda$, and keeping only terms up to order $ \lambda^2 $, we get
\begin{align}
	T_{s} u(\varphi)=\lambda^{2}\left[e^{\varphi}-4\partial_{\varphi}^{2}+\left(s+\frac{1}{2}\right)^{2}\right]u(\varphi) + O(\lambda^3),
\end{align}
which is the Liouville Hamiltonian. If we choose a vector in the other $T$ invariant subspace, namely $ \ket{v}=\sum v_n\ket{{b}_n} $, then the same manipulations would give a Liouville Hamiltonian acting on the continuum function $ v(\phi)$ (which is the continuum analog of the rescaled vectors $ \tilde{v}_n $). This Liouville Hamiltonian would be similar, only with $ s\to(-s) $. This means that in the continuum Schwarzian limit we have two bosonic functions $u$ and $v$, with the transfer matrix acting on each as a  Liouville Hamiltonian:
\begin{equation}
\begin{split}
	T_s v & = \lambda^{2}\left[e^{\varphi}-4\partial_{\varphi}^{2}+\left(s-\frac{1}{2}\right)^{2}\right]v, \\
	T_s u & = \lambda^{2}\left[e^{\varphi}-4\partial_{\varphi}^{2}+\left(s+\frac{1}{2}\right)^{2}\right]u.
\end{split}
\end{equation}

To compare to the Super Liouville, notice that charge $s$ is just the quantum number of a global $U(1)$ symmetry, which can be replaced by a derivative of a different bosonic field, $s\ra \partial_\tau \sigma$. This gives us the Hamiltonian
\begin{equation} \label{eq:bosonT}
T =\lambda^{2}\left[e^{\varphi}+4p_{\varphi}^{2}+\left(p_\sigma+\frac{1}{2}\right)^{2}\right] + O(\lambda^3),
\end{equation}
which is the bosonic part of the Super Liouville Hamiltonian.

We now turn to the fermionic sector of the Hilbert space. We will show that the action of $T$ on a fermionic vector is the same as its action on a bosonic vector. To see this consider a fermionic vector $\ket{n+1/2,X}$.
Recall that $\ket{b_n} = Q\ket{n+1/2,X}$, so we have
\begin{align}
(QT\ket{n+1/2,X})=(T Q\ket{n+1/2,X})= T\ket{b_n}= Q \sum_{m=\pm 1,0} d_{m,n} \ket{m+n+1/2,X},
\end{align}
where $d_{m,n}$ corresponds to the action of $T$ over a bosonic vector, given by \eqref{eq:T_on_bn}. As the space $\text{Span}\{\ket{n+1/2,X}\}_{n\in \ZZ}$ is orthogonal to the kernel of $Q$, it follows that
\begin{align}
T\ket{n+1/2,X}=\sum_{m=\pm 1,0} d_{m,n} \ket{m+n+1/2,X}.
\end{align}
Thus the action of $T$ on fermions is identical to the action of $T$ on bosons, which means that on the fermionic sector we again get (\ref{eq:bosonT}). In other words the fermionic sector is just a fermion zero mode times the bosonic sector. This is in agreement with the Super Liouville quantum mechanics, which has two fermionic zero modes.

%
%
%

\section{2-pt Function}

 Finally we would like to apply the techniques above to the computation of the 2-pt functions. The main upshot of this section is to show how to consistently include the ground states of the theory in the 2-pt function.
 As in  \cite{Micha2018,Berkooz_2019}, we are interested in operators in a similar statistical class as the Hamiltonian. A natural choice would be an operator of the form
\begin{equation} \label{eq:singlechord}
A = \sum_{|I|=p'} \tilde{C}_I \Psi_I.
\end{equation}
 where $\tilde{C}_I$ are independent random Gaussian variables, and the length $p'$ determines the charge of the operator. We will call this class of operators single chord operators. Furthermore, we will be interested in the double scaled limit, namely
 \begin{align}
p'\to\infty, \text{ with } \tilde{\lambda}=\lambda\frac{p'}{p}\text{ fixed}.
 \end{align}
Again we will find it useful to define
\begin{align}
\tilde{q}=e^{-2\frac{p p'}{N}}.
\end{align}
These operators can be generalized into a more generic class of operators of the form
\begin{equation} \label{eq:dubchord}
{\cal O}_{p,p'}=\sum_{|I|=p', |J|=p''} {\cal O}_{I,J} \overline{\Psi}_I \Psi_J,\ \ \ {\cal O_{I,J}}\ i.i.d\ Gaussian .
\end{equation}
The length $p'-p''$ determines the charge of these operators, and we expect that $p''+p'$ - in the IR and for an appropriate class of operators - determines the conformal dimension of the operator. We will call this class of operators double chord operators.

An underlying assumption is that these coefficients are uncorrelated with the ones that appear in the Hamiltonian. We can also compute correlators of descendants of these operators, where the coefficients are clearly correlated with the ones in the supercharges. This is a manageable slight generalization of our computations, which we will not do here.

In either cases we would like to evaluate
\begin{equation}
\expt{\text{tr}\left(e^{-\frac{\beta}{2} H}\overline{\cal O}(\tau) e^{-\frac{\beta}{2} H}{\cal O}(0)\right)}_C = \expt{\text{tr}\left(e^{(\tau-\beta/2) H} \overline{\cal O} e^{-(\tau +\beta/2)H} {\cal O}\right) }_C,
\end{equation}
which amounts to evaluating the moments
\begin{equation}
m_{k_1,k_2;\tilde{q}} = \expt{\text{tr}\left[ \overline{\cal O} H^{k_1} {\cal O} H^{k_2} \right]}_C.
\end{equation}
By the subscript $C$ we also mean averaging over the random coefficient of the operator $A$ or ${\cal O}$. We will focus on calculating the moments without a chemical potential, though our results can be slightly modified by adding a chemical potential term to the integral over fixed charge sectors to include it (say in equation \eqref{eq:2poin_single_1}).

The section is organized as follows. We first compute the moments of simple operators described by \eqref{eq:singlechord}, and then exponentiate these moments to find the full 2-point function in section \ref{sec:single_chord_ops}. Afterwards we analyze the conformal limit of the two point function of such operators. Finally we discuss the 2-point function of more general operators of the form \eqref{eq:dubchord} in section \ref{sec:dub_chord_ops}. The main novelty in the result is that the ground states generate a substantial contribution to the 2-point function, even in the Schwarzian limit. Thus the conformal ansatz for the 2-point function of the form $1/x^{2\Delta}$ is inconsistent, even in this regime, and should be altered to include additional contributions from the ground states.

\subsection{Single chord operators} \label{sec:single_chord_ops}

\begin{figure} [h]
	\centering
	\includegraphics[page=1,width=0.35\columnwidth]{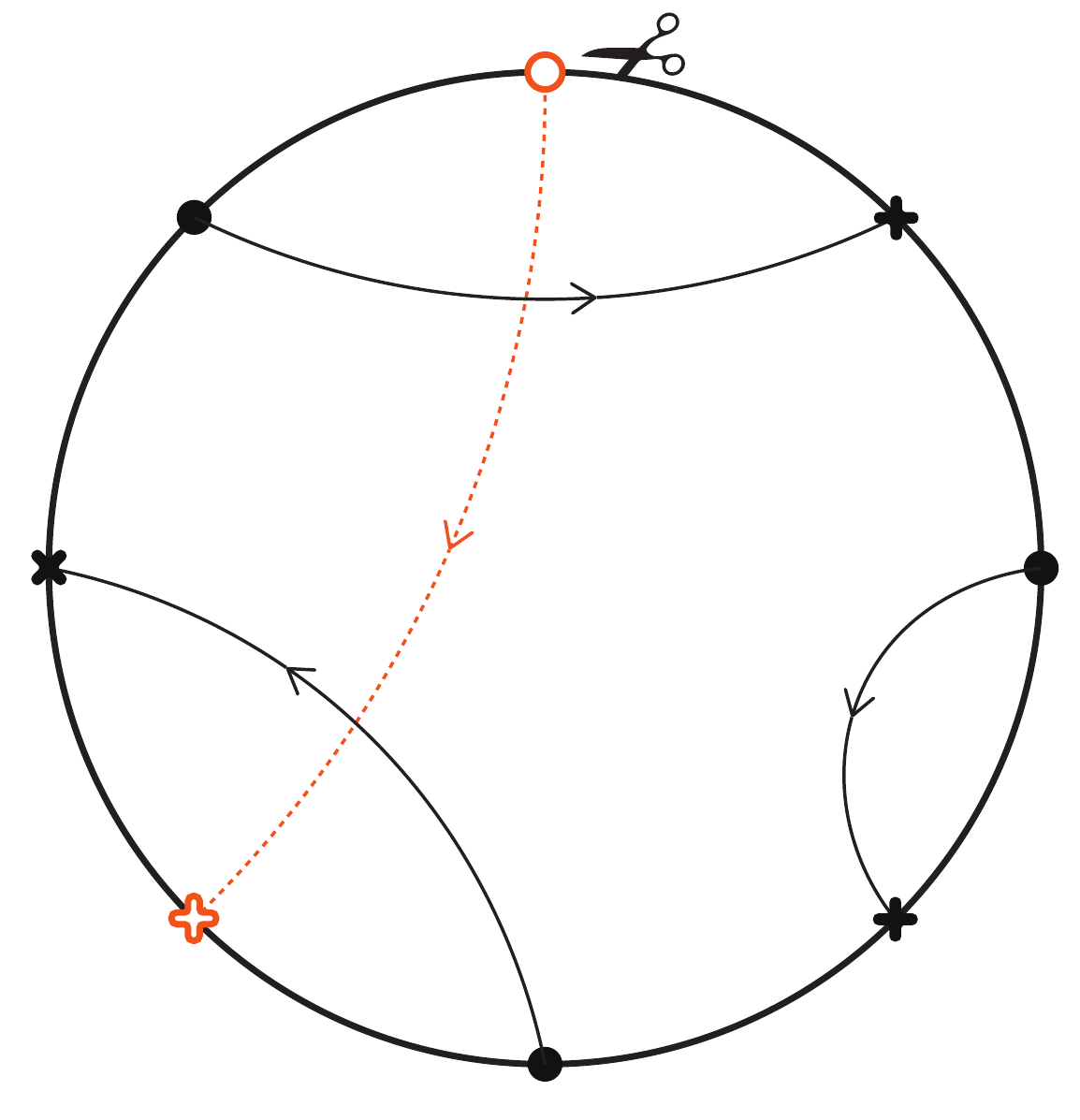}
	\caption*{(a)}
	\includegraphics[page=2,width=0.50\columnwidth]{figures/OperatorInsertionCD.pdf}
	\caption*{(b)}
	\caption{Two different representations of a diagram contributing to the moment $ m_{1,2;\tilde{q}} $. In figure (a) we can see the chord diagram representation, and in (b) we have the same diagram represented as an open chord diagram. The operator region is defined to be the region between the two dashed lines.}
	\label{fig:OperatorInsertionCD}
\end{figure}

We will first compute the two point function of single chord operators. The moments that we consider are of the form
\begin{equation}
m_{k_1,k_2;\tilde{q}} = \expt{\tr\left[\overline{A} H^{k_1} A H^{k_2}\right]}_C.
\end{equation}
Translating to chord diagrams, this amounts to connecting the $A$ and the $\overline{A}$ operators with a new type of chord. It is also convenient to open the chord diagram right before the insertion of the single ``A'' chord. See figure \ref{fig:OperatorInsertionCD} for an example of such a chord diagram.

\begin{figure} [h]
	\centering
	\includegraphics[page=1,width=0.30\columnwidth]{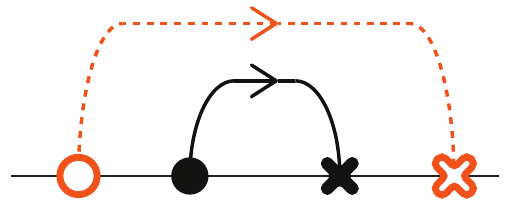}
	\hspace{0.1\columnwidth}
	\includegraphics[page=2,width=0.30\columnwidth]{figures/OperatorInsertionConfigurations.pdf}

\caption*{$ \mathrm{I} $ \qquad\qquad\qquad\qquad\qquad\qquad\qquad $ \mathrm{II} $}
	\includegraphics[page=3,width=0.30\columnwidth]{figures/OperatorInsertionConfigurations.pdf}
	\hspace{0.1\columnwidth}
	\includegraphics[page=4,width=0.30\columnwidth]{figures/OperatorInsertionConfigurations.pdf}
\caption*{$ \mathrm{III} $ \qquad\qquad\qquad\qquad\qquad\qquad\qquad $ \mathrm{IV} $}
	\includegraphics[page=5,width=0.30\columnwidth]{figures/OperatorInsertionConfigurations.pdf}
	\hspace{0.1\columnwidth}
	\includegraphics[page=6,width=0.30\columnwidth]{figures/OperatorInsertionConfigurations.pdf}
\caption*{$ \mathrm{V} $ \qquad\qquad\qquad\qquad\qquad\qquad\qquad $ \mathrm{VI} $}
\caption{Possible chord relations between an operator chord (dashed line) and a regular Hamiltonian chord (solid line). Diagrams $ \mathrm{I,II,IV,VI} $ are enemy configurations, which result in a $ \tilde{q}^{1/2} $ factor. The rest are friend configurations, which result in a $ \tilde{q}^{-1/2} $ factor.}
	\label{fig:OperatorInsertionConfigurations}
\end{figure}

The contribution to each chord diagram will be the contribution of the chord diagram without the additional chord, with an additional factor of $\tilde{q}^{-1/2}$ for every $Q$ chord that is friends with an $A$ chord, and a factor of $\tilde{q}^{1/2}$ for every $Q$ chord that is enemies with an $A$ chord, see figure \ref{fig:OperatorInsertionConfigurations}. These factors change the local transfer matrix.

There is a simple way to change the transfer matrix without a need to re-diagonalize it, at the cost of adding a local operator at the $A$ insertion point. This can be done by changing the factors $ q^{\pm s} $ when we open a chord, which is the same as changing the charge sector. If we open a chord parallel to the operator chord inside the operator region (configurations $ \mathrm{I,II} $ in figure (\ref{fig:OperatorInsertionConfigurations})), it is an enemy of the operator chord. If a chord anti-parallel to the operator chord opens inside the operator region (configurations $ \mathrm{III,IV} $ there), it is friends with the operator chord, if they do not intersect. We can account for the factors due to diagrams $\mathrm{I,II,III}$ by changing the charge inside the operator region (and similarly the factors due to diagrams $\mathrm{V,VI}$ by changing the charge outside the operator region), and correct the factors due to diagram $\mathrm{IV}$ through an operator insertion on the boundary between the two regions.

To implement this idea we should change $ s\to s_{-}\equiv s-\frac{\tilde{\lambda}}{2\lambda}$ inside the operator region, and $ s\to s+\frac{\tilde{\lambda}}{2\lambda}\equiv s_{+} $ outside this region. To count the number of type (IV) intersections we can add the operator $ \tilde{q}^{\hat{N}} $ on the boundary between the two regions, with $\hat{N}$ being the number operator defined by $\hat{N}\ket{n,X/O}=n\ket{n,X/O}$ for $n\in \NN/2$. Note that $\hat{N}$ is defined only on the physical Hilbert space.
This change in the transfer matrix is consistent with the charge formalism, as the $A$ operator has a normalized charge of $ \tilde{\lambda}/\lambda = p'/p$, which is exactly the change in the charge when we insert the $A$ chord.

Now we can compute the moment by
\begin{align} \label{eq:2poin_single_1}
m_{k_{1},k_{2};\tilde{q}}=\frac{q^{(k_1+k_2)/4}}{2^{k_1+k_2}}\sqrt{\frac{\lambda}{\pi}}\int_{-\infty}^{\infty}ds~e^{-\lambda s^{2}}\left\langle \emptyset\left|T_{s_{-}}^{k_{1}}\tilde{q}^{\hat{N}}T_{s_{+}}^{k_{2}}\right|\emptyset\right\rangle,
\end{align}
We can compute the moments in a fixed charge sector by stripping the integral over $s$, or add a chemical potential $e^{i\mu s}$ to the integral.

The number operator is diagonal over the number basis, so we can insert a complete set of number states to get
\begin{align}
m_{k_{1},k_{2};\tilde{q}}&=\frac{q^{(k_1+k_2)/4}}{2^{k_1+k_2}}\sqrt{\frac{\lambda}{\pi}}\int_{-\infty}^{\infty}ds~e^{-\lambda s^{2}}\sum_{n_{\pm}}\tilde{q}^{n}\left\langle \emptyset\left|T_{s_{-}}^{k_{1}}\right|n_{\pm}\right\rangle \left\langle n_{\pm}\left|T_{s_{+}}^{k_{2}}\right|\emptyset\right\rangle ,
\end{align}
Since the number states are not eigenstates of the transfer matrix, it will be easier to find the value of $ \left\langle n_{\pm}\left|T_{s}^{k}\right|\emptyset\right\rangle $ by inserting a complete set of $ T $ eigenstates. This gives us
\begin{equation}
\begin{split}
\left\langle n_{\pm}\left|T_{s}^{k}\right|\emptyset\right\rangle
&=\int\frac{d\phi d\phi'}{\left|\left\langle v_{s}\left(\phi\right)|v_{s}\left(\phi\right)\right\rangle \right|^{2}}\left\langle n_{\pm}|v_{s}\left(\phi\right)\right\rangle \left\langle v_{s}\left(\phi\right)\left|T_{s}^{k}\right|v_{s}\left(\phi'\right)\right\rangle \left\langle v_{s}\left(\phi'\right)|\emptyset\right\rangle +\left(v\to u\right)\\
&=\int d\phi\frac{\left(q^2,e^{\pm2i\phi};q^{2}\right)_{\infty}}{2\pi}q^{-s}\Lambda_{s}^{k-1}\left(\phi\right)\left\langle n_{\pm}|v_{s}\left(\phi\right)\right\rangle +\left(\begin{matrix}v\to u\\
s\to\left(-s\right)
\end{matrix}\right),
\end{split}
\end{equation}
where we use
\begin{equation}
\left\langle v\left(\phi\right)|\emptyset\right\rangle =1,
\qquad \left\langle v_{s}\left(\phi\right)|v_{s}\left(\phi'\right)\right\rangle =q^{s}\Lambda_{s}\left(\phi\right)\frac{2\pi\delta\left(\phi-\phi'\right)}{\left(q^{2},e^{\pm 2i\phi};q^{2}\right)_{\infty}},
\end{equation}
and $ \Lambda_s(\phi) $ are the eigenvalues of the transfer matrix, defined in (\ref{eq:EVals}). To compute the sum $ \sum_{n_{\pm}}\tilde{q}^{n}\left\langle \emptyset\left|T_{s_{-}}^{k_{1}}\right|n_{\pm}\right\rangle \left\langle n_{\pm}\left|T_{s_{-}}^{k_{2}}\right|\emptyset\right\rangle  $ we will use the projection of the $ T $ eigenstates over the number basis (see appendix \ref{app:eigen})
\begin{align}
\left\langle n_{\pm}|u_{s}\left(\phi\right)\right\rangle &=\frac{1}{\sqrt{2\left(q^{2};q^{2}\right)_{n-1}\left(1\mp q^{n}\right)}}\left[H_{n}\left(\cos\phi|q^{2}\right)+\left(1\mp q^{n}\right)q^{-s}\frac{1}{\sqrt{q}}H_{n-1}\left(\cos\phi|q^{2}\right)\right],\\\left\langle n_{\pm}|v_{s}\left(\phi\right)\right\rangle &=\pm\frac{1}{\sqrt{2\left(q^{2};q^{2}\right)_{n-1}\left(1\mp q^{n}\right)}}\left[H_{n}\left(\cos\phi|q^{2}\right)+\left(1\mp q^{n}\right)q^{s}\frac{1}{\sqrt{q}}H_{n-1}\left(\cos\phi|q^{2}\right)\right].
\end{align}
Each summand is composed of four different terms. Using $ q $-Hermite polynomials orthogonality relations we get
\begin{equation}
\begin{split}
\sum_{n_{\pm}}\tilde{q}^{n}\left\langle n_{\pm}|v_{s_{+}}\left(\phi'\right)\right\rangle \left\langle n_{\pm}|v_{s_{-}}\left(\phi\right)\right\rangle &=q^{-s}\tilde{q}^{1/2}{\Lambda}_{s_-}\left(\phi\right)\frac{\left(\tilde{q}^{2},q^{2}\right)_{\infty}}{\left(\tilde{q}e^{i\left(\pm\phi\pm\phi'\right)};q^{2}\right)_{\infty}}\\\sum_{n_{\pm}}\tilde{q}^{n}\left\langle n_{\pm}|u_{s_{+}}\left(\phi'\right)\right\rangle \left\langle n_{\pm}|u_{s_{-}}\left(\phi\right)\right\rangle &=q^{s}\tilde{q}^{1/2}\Lambda_{-s_-}\left(\phi\right)\frac{\left(\tilde{q}^{2},q^{2}\right)_{\infty}}{\left(\tilde{q}e^{i\left(\pm\phi\pm\phi'\right)};q^{2}\right)_{\infty}}\\\sum_{n_{\pm}}\tilde{q}^{n}\left\langle n_{\pm}|v_{s_{-}}\left(\phi\right)\right\rangle \left\langle n_{\pm}|u_{s_{+}}\left(\phi'\right)\right\rangle &=\frac{\left(\tilde{q}^{2};q^{2}\right)_{\infty}}{\left(\tilde{q}qe^{i\left(\pm\phi\pm\phi'\right)};q^{2}\right)_{\infty}}\\\sum_{n_{\pm}}\tilde{q}^{n}\left\langle n_{\pm}|u_{s_{-}}\left(\phi\right)\right\rangle \left\langle n_{\pm}|v_{s_{+}}\left(\phi'\right)\right\rangle &=0,
\end{split}
\end{equation}
This gives us
\begin{equation} {\label{OperatorMoment}}
\begin{split}
m_{k_{1},k_{2};\tilde{q}} &=\frac{q^{(k_1+k_2)/4}}{2^{k_1+k_2}}\sqrt{\frac{\lambda}{\pi}}\left(\tilde{q}^{2},q^{2}\right)_{\infty}
\int ds e^{-\lambda s^2} \frac{d\phi'd\phi}{(2\pi)^2}\left(q^2,q^2,e^{\pm2i\phi},e^{\pm2i\phi'};q^{2}\right)_{\infty}\\&\qquad\times\Big[\frac{q^{s}\tilde{q}^{1/2}}{\left(\tilde{q}e^{i\left(\pm\phi\pm\phi'\right)};q^{2}\right)_{\infty}}\left(\Lambda_{s+\frac{\tilde{\lambda}}{2\lambda}}\left(\phi'\right)+\Lambda_{s-\frac{\tilde{\lambda}}{2\lambda}}\left(\phi\right)\right)\Lambda_{s-\frac{\tilde{\lambda}}{2\lambda}}^{k_{2}-1}\left(\phi\right)\Lambda_{s+\frac{\tilde{\lambda}}{2\lambda}}^{k_{1}-1}\left(\phi'\right)\\&\quad\qquad+\Lambda_{s-\frac{\tilde{\lambda}}{2\lambda}}^{k_{2}-1}\left(\phi\right)\Lambda_{-\left(s+\frac{\tilde{\lambda}}{2\lambda}\right)}^{k_{1}-1}\left(\phi'\right)\frac{\tilde{q}^{-1}}{\left(\tilde{q}qe^{i\left(\pm\phi\pm\phi'\right)};q^{2}\right)_{\infty}}\Big].
\end{split}
\end{equation}
These moments are correct so long as $k_1,k_2 \neq 0$. The full moments read
\begin{equation}
m_{k_1,k_2;\tilde{q}} = \left\{ \begin{array}{cc}
m^c_{k_1,k_2;\tilde{q}}, & k_1,k_2 > 0,\\
m^1_{k_1;\tilde{q}} , & k_1 > 0, k_2 =  0,\\
m^1_{k_2;\tilde{q}} , & k_1 = 0, k_2 > 0,\\
1, & k_1 = k_2 = 0,
\end{array} \right.
\end{equation}
with $m^c_{k_1,k_2;\tilde{q}}$ given by equation \eqref{OperatorMoment}, and
\begin{equation}
m^1_{k;\tilde{q}}=  \frac{q^{1/4}}{2}\sqrt{\frac{\lambda}{\pi}} \int_{-\infty}^{\infty}ds~ q^{s^{2}-s+s_0^2}\int_{0}^{\pi}\frac{d\phi}{2\pi}\left(q^{2},e^{\pm2i\phi};q^{2}\right)_{\infty}\cosh\left(2\lambda s_{0}s\right)\left(\frac{q^{1/4}\Lambda_{s}}{2}\right)^{k-1}.
\end{equation}

To compute the two point function we simply exponentiate the moments, however we will get multiple terms as the moments are not analytically continued from $k_1,k_2 > 0$ to $k_1,k_2 = 0$. This happens as the ground states only contribute to the moments when $k_1=0$ or $k_2 = 0$, resulting in the following expansion:

\begin{equation} \label{eq:2pointfun}
\begin{split}
 &\expt{\tr\left[\overline{A}e^{(\tau-\beta/2)H}Ae^{-H(\tau+\beta/2)}\right] }_C
= \sum_{k_1,k_2 = 0}^\infty \frac{(\tau-\beta/2)^{k_2} (-\tau-\beta/2)^{k_1}}{k_1! k_2!} m_{k_{1},k_{2};\tilde{q}} \\
&\qquad \qquad =1+ I_1(\tau-\beta/2) + I_1(-\tau-\beta/2) +  I_c(\tau-\beta/2,-\tau-\beta/2),
\end{split}
\end{equation}
where we define
\begin{equation} \label{eq:I_c_def}
\begin{split}
I_c(x,y) &\equiv\sum_{k_1,k_2=1}^{\infty}\frac{x^{k_1}y^{k_2}}{k_1!k_2!}m^c_{k_1,k_2;\tilde{q}} \\
 &=\sqrt{\frac{\lambda}{\pi}} \int ds ~e^{-\lambda s^2} \frac{d\phi'd\phi}{(2\pi)^2}\left(\tilde{q}^2,q^2,q^2,e^{\pm2i\phi},e^{\pm2i\phi'};q^{2}\right)_{\infty}\\
 &\quad\times\Bigg[\frac{q^{s}\tilde{q}^{1/2}
 \left(\Lambda^{-1}_{s-\frac{\tilde{\lambda}}{2\lambda}}\left(\phi\right) +
 \Lambda^{-1}_{s+\frac{\tilde{\lambda}}{2\lambda}}\left(\phi'\right)\right)}
 {\left(\tilde{q}e^{i\left(\pm\phi\pm\phi'\right)};q^{2}\right)_{\infty}}
\left( e^{-\frac{yq^{1/4}}{2}\Lambda_{s-\frac{\tilde{\lambda}}{2\lambda}}\left(\phi\right)} - 1\right)
\left(e^{ - \frac{xq^{1/4}}{2}\Lambda_{s+\frac{\tilde{\lambda}}{2\lambda}}\left(\phi'\right)}\right)\\
 &\qquad+\frac{\tilde{q}^{-1}}{\left(\tilde{q}qe^{i\left(\pm\phi\pm\phi'\right)};q^{2}\right)_{\infty}}
\left(e^{-\frac{yq^{1/4}}{2}\Lambda_{s-\frac{\tilde{\lambda}}{2\lambda}}\left(\phi\right)}-1\right)
\left(e^{ - \frac{xq^{1/4}}{2}\Lambda_{-s-\frac{\tilde{\lambda}}{2\lambda}}\left(\phi'\right)}\right)\Bigg],
\end{split}
\end{equation}
and
\begin{equation} \label{eq:I_1_definition}
\begin{split}
I_1(x) & \equiv\sum_{k=1}^{\infty}\frac{x^{k}}{k!}m^1_{k;\tilde{q}} \\
&=  \frac{q^{1/4}}{2}\sqrt{\frac{\lambda}{\pi}}\int_0^x dx' \int_{-\infty}^{\infty}ds~ q^{s^{2}-s+s_0^2}\int_{0}^{\pi}\frac{d\phi}{2\pi}\left(q^{2},e^{\pm2i\phi};q^{2}\right)_{\infty}\\
&\qquad \qquad \times \cosh\left(2\lambda s_{0}s\right)  \exp\left(x' \frac{q^{1/4}\Lambda_{s}(\phi)}{2}\right).
\end{split}
\end{equation}

The different terms in (\ref{eq:2pointfun}) correspond to the different contributions to the 2-point function. The $1$ is simply the zero-zero moment. The $I_1$'s are the sum of the zero-$k$ and $k$-zero moments, and so must involve the ground states. Note that the two time parameters in the two intervals are $\pm\tau-\beta/2$ and an $I_1$ function that depends only on one of them means that we have inserted the ground states as intermediate states in the other interval. $I_c$, on the other hand, involves the rest of the moments and thus contains the contribution from the continuous spectrum. We note that any conformal limit of the two point function must be realized in $I_c$, as the conformal ansatz ignores the large amount of exact ground states.

As a check of \eqref{OperatorMoment}, we can verify that it converges to $ m_{k_1+k_2}  $ given by \eqref{PartitionFnMoment}  under the limit $ \tilde{q}\to1 $ (while keeping $q$ fixed), which corresponds to inserting the identity operator. This trivially reduces to the moments if $k_1=0$ or $k_2 =0$, so we only need to check the continuous part, $m_c$. Since $ \lim_{\tilde{q}\to1}\left(\tilde{q}^{2};q^{2}\right)=0 $, the only non-zero contribution to $ m_{k_1,k_2;\tilde{q}} $ can arise from singular terms. The only such term is $ \left(\tilde{q}e^{i\left(\pm\phi\pm\phi'\right)};q^{2}\right)_{\infty}^{-1} $, and we get
\begin{align}
\lim_{\tilde{q}\to1}\frac{\left(\tilde{q}^{2},q^{2}\right)_{\infty}}{\left(\tilde{q}e^{i\left(\pm\phi -\phi'\right)};q^{2}\right)_{\infty}}=
\frac{2\pi}{\left(q^2,e^{\pm2i\phi};q^{2}\right)}\delta\left(\phi-\phi'\right).
\end{align}
This indeed shows us that $\lim_{\tilde{q}\to1}m_{k_1,k_2;\tilde{q}} =m_{k_1+k_2}$.

\subsubsection{Conformal limit of the 2-pt function}

We now turn to analyze the two point function given by \eqref{eq:2pointfun} in the conformal regime. The conformal limit is attained for low temperatures/long times, together with $ q\to1^-\Leftrightarrow\lambda\to0 $.  In particular we expect to recover the conformal limit when $\lambda \ll  \beta\lambda^2,t\lambda^2\ll1 $. As we are interested in the short interaction length limit, we shall scale the length of the operator with the length of the supercharge, and take $ \tilde{p}=\alpha p $, which gives us $ \tilde{q}=q^\alpha $, with $\alpha$ finite as $\lambda \ra 0$. The calculations mirror those done in \cite{Micha2018} for the Majorana SYK model, and the detailed calculation of this limit is given in appendix \ref{app:2point}. We will simply present the results of this computation.

We shall start by concentrating on the $I_c$ term in \eqref{eq:2pointfun}, as it is connected to the continuum spectrum, and split $ I_c(x,y)$ into two parts, $ I_c(x,y) =I_c^1(x,y)+I_c^2(x,y) $ with $ I_c^1$ and $I_c^2 $ given by the third and forth lines of \eqref{eq:I_c_def} respectively. We focus on each of these separately, and further divide each into a contribution from the continuous spectrum (which depends on the two time separations), a mixed contribution (which depends on a single time separation), and a constant coming solely from the ground states (see \eqref{eq:I_1_as_A_1} and \eqref{eq:I_2_as_A_2}). Note that the only part of $I_c^{1,2}$ that can have a standard conformal form is the contribution from the continuous spectrum.

Focusing on these continuum contributions,  and after a lengthy calculation detailed in appendix \ref{app:2point_con}, we find that the two different terms in $I_c^1$ and $I_c^2$ have a conformal form with conformal dimension $\alpha/2$ and $(\alpha+1)/2$, given by \eqref{eq:A_1_conformal} and \eqref{eq:A_2_conformal} respectively. In general we expect that an operator with $\alpha p$ fermions will have a conformal dimension of $(\alpha p)/(2p) = \alpha/2$, which is indeed the conformal dimension of the first term. At long times it dominates. We can think of the second term as this operator's super-partner, as it has a conformal dimension of $\alpha/2 + 1/2$.

However, these are not the only substantial contributions in the conformal limit. Both $I_c^1$ and $I_c^2$ have contributions relating to the ground states, as is the $I_1$ contribution given by \eqref{eq:I_1_definition}. The conformal limit of these terms is computed in appendix  \ref{app:2point_ground}.

To summarize, we divide the 2-pt function (\ref{eq:2pointfun}) into $I_c^1+I_c^2+I_1$, where in general the $I_c$'s receive contribution both from the continuum states and from the ground states, and $I_1$ only receives contribution from the ground states. To see which contribution dominates the conformal limit we inspect their $\lambda$ scaling. The contribution to each term is as follows:
\begin{align}
\begin{split}
&I_c^1\sim
\begin{cases}
\frac{\lambda^{\alpha}}{\beta^{5/2}}e^{-\frac{\pi^2}{4\lambda}+\frac{\pi^2}{16\tilde{\beta}}} & \text{continuum contribution}\\
\lambda^{1-\alpha}e^{-\frac{\pi^2}{4\lambda}+\frac{\pi^2}{2\tilde{\beta}}} & \text{ground states contiburion}
\end{cases}
\\
&I_c^2\sim
\begin{cases}
\frac{\lambda^{\alpha}}{\beta^{1/2}}e^{-\frac{\pi^2}{4\lambda}+\frac{\pi^2}{16\tilde{\beta}}} & \text{continuum contribution}\\
\lambda^{1-\alpha}e^{-\frac{\pi^2}{4\lambda}+\frac{\pi^2}{2\tilde{\beta}}} & \text{ground states contiburion}
\end{cases}
\\
&I_1\sim\quad e^{-\frac{\pi^2}{4\lambda}+\frac{\pi^2}{2\tilde{\beta}}} \qquad\quad \text{ground states contribution}
\end{split}
\end{align}
By comparison to the $\lambda$ scaling of the continuous part, we see that the ground state contribution cannot be neglected in the conformal limit. Thus it seems that the conformal ansatz for the 2-point function is not consistent in this model, as it fails to account for the large amount of ground states.

\subsection{Double chord operators} \label{sec:dub_chord_ops}

We now turn to analyze the more general double chord operators. We will not fully compute the correlation functions, as we did for single chord operators. Rather we will derive the rules to compute these 2-point functions in terms of the transfer matrix and the auxiliary Hilbert space, and provide a general discussion on the results. We consider general operators of the form \eqref{eq:dubchord}. The charge of such operators is $s_0 = \frac{p'- p''}{p}$, and we define
\begin{align}
\tilde{q} \equiv e^{-\frac{2p(p'+p'')}{N}}.
\end{align}

\begin{figure} [h]
	\centering
	\includegraphics[page=3,width=0.7\columnwidth]{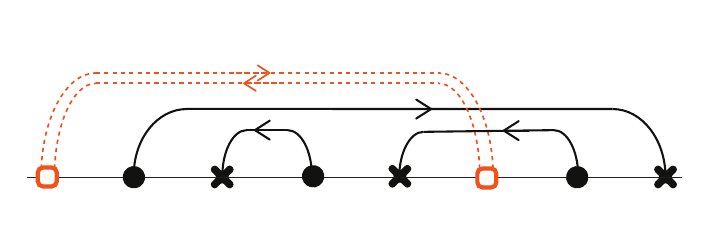}
	\caption{An example for a diagram contributing to the moment $ m_{2,1;\tilde{q}} $ for a double chord diagram.}
	\label{fig:DoubleChordOperator}
\end{figure}

Let us start with the two point function of uncharged double chord operators, as they will end up being simpler than general double chord operators. These operators are of the form \eqref{eq:dubchord} with $p'' = p'$. Again it is sufficient to only consider the moments
\begin{equation}
m_{k_1,k_2;\tilde{q}} = \expt{\tr\left[\overline{\mathcal{O}} H^{k_1} \mathcal{O} H^{k_2}\right]}_C,
\end{equation}
which correspond to opening the chord diagram right before the insertion of the double line. For example, one of the chord diagrams contributing to the $k_1=2, k_2= 1$ moment is given in figure \ref{fig:DoubleChordOperator}.

The contribution to each chord diagram will be the contribution of the chord diagram without the additional double chord, with an additional factor of $\tilde{q}^{-1/4}$ for every $Q$ chord that is friends with a $ \mathcal{O}$ chord, and a factor of $\tilde{q}^{1/4}$ for every $Q$ chord that is enemies with a $ \mathcal{O}$ chord. However every $Q$ chord that doesn't intersect the $ \mathcal{O}$ double--chord is friends with one of the $ \mathcal{O}$ chords and enemies with the other one, and so gives an overall factor of 1. Thus we really only care about the number of intersections, as intersecting chords are enemies with both $\mathcal{O}$ chords. All in all, we simply need to add a factor of $\tilde{q}^{\frac{1}{2}}$ for each intersection between a $Q$ chord and the $ \mathcal{O}$ double chord. We note there is an overall factor of $e^{\frac{{p'}^2}{N}} $ as the two $ \mathcal{O}$ chords are friends which will be omitted as the operators can simply be rescaled.

In the auxiliary Hilbert space language this translates to calculating
\begin{equation}
m_{k_1,k_2;\tilde{q}} =
\frac{q^{\frac{k_1+k_2}{4}}}{2^{k_1+k_2}} \sqrt{\frac{\lambda}{\pi}}\intinf d s~e^{-\lambda s^2}
 \innn{\emptyset}{T^{k_1}_s ~ \tilde{q}^{\hat{N}}~T^{k_2}_s }{\emptyset},
\end{equation}
where $\hat{N}$ is the number operator. Computing this element can be done using insertions of complete sets, the same as in the previous section. The cases $k_1=0$ or $k_2=0$ will again differ from the analytic continuation of the other moments, leading to terms that involve and do not involve the ground states which are similar to those found for single chord operators.

We can now analyze the general double chord operators. Such operators are defined by \eqref{eq:dubchord} with general $p'$ and $p''$. As with single chord operators, the transfer matrix will change in relation to the charge of the operator. Similarly, we will need a factor of $\tilde{q}^{\hat{N}}$ to account for the $Q$ chords that intersect the double $\mathcal{O}$ chord. All in all this will result in the 2-point function
\begin{equation} \label{eq:dub_chord_moments}
m_{k_1,k_2;\tilde{q},s_0} =
\frac{q^{\frac{k_1+k_2}{4}}}{2^{k_1+k_2}} \sqrt{\frac{\lambda}{\pi}}\intinf d s~e^{-\lambda s^2}
 \innn{\emptyset}{T^{k_1}_{s-s_0/2} ~ \tilde{q}^{\hat{N}}~T^{k_2}_{s+s_0/2} }{\emptyset}.
\end{equation}
We note that this result differs from \eqref{eq:2poin_single_1} because in the above $\tilde{q}$ and $s_0$ are independent while in the single chord case $\tilde{q} = q^{s_0}$. The evaluation of \eqref{eq:dub_chord_moments} will be similar to that of \eqref{eq:2poin_single_1}, by inserting a complete set of states and using the orthogonality of $q$-Hermite polynomials.


\section{Deformed algebra and the relation to quantum groups} \label{sec:algebra}

The standard SYK model has a low energy $SL(2,\RR)$ conformal symmetry \cite{MaldecenaStanford}. In the double scaled limit this symmetry undergoes a quantum deformation related to the quantum group $sl_q(2)$, which is seen in the algebra of the transfer matrix, as well as in the structure of the four point function \cite{Berkooz_2019}. Similarly, the low energy theory of the $\mathcal{N} =2 $ SYK model has an $SU(1,1|1)$ super conformal symmetry \cite{SusySYK}. We will show that in the doubled scaled limit this super symmetry also undergoes a quantum deformation, this time related to the quantum super group $sl_q(2|1)$. Specifically we will show that the graded algebra of the transfer matrix is a contraction of the quantum super group $sl_q(2|1)$.

The superalgebra $\mathfrak{sl}(2|1)$ is the algebra of $3\times 3$ matrices with a $(2|1)$ grading and supertrace zero. The bosonic sector forms the Lie algebra $\mathfrak{sl}(2)\oplus \mathfrak{u}(1)$, and the fermionic sector consists of two pairs of generators, one in the fundamental and one in the anti-fundamental representations of $\mathfrak{sl}(2)$. This superalgebra can be described by the Cartan matrix (for the fermionic generators)
\begin{equation} \label{eq:cartanmatrix}
(a_{ij}) = \bpm 0 & 1 \\ -1 & 0 \epm.
\end{equation}
The Cartan matrix contains all the information on the commutation relations of the fermionic generators, $X_{1,2},Y_{1,2}$, and the elements in the Cartan, $H_{1,2}$. We can define the remaining two bosonic raising and lowering operators via the adjoint actions $X_3 = \{X_1,X_2\} $, and $Y_3 = \{Y_1,Y_2\} $. The remaining commutation relations are fixed via the Cartan matrix as well.

One way to construct quantum groups is to start with the Cartan matrix of a simple Lie algebra, and define a $q$ deformed Hopf algebra structure on the generators of the Lie algebra. We will follow this method to construct $sl_q(2|1)$ using the Cartan matrix (\ref{eq:cartanmatrix}). We refer the reader to \cite{quantumgroups} for more information on quantum groups and their constructions from simple Lie algebras.

Using the above Cartan Matrix, (\ref{eq:cartanmatrix}), the quantum super--group  $sl_q(2|1)$ can be described using the same generators $X_{1,2},Y_{1,2},H_{1,2},$ with the (anti--)commutation relations
\begin{equation*}
[H_1,H_2] = \{X_1,Y_2\}= \{X_2,Y_1\} = [H_1,X_1] = [H_2,X_2] = [H_1,Y_1] = [H_2,Y_2] =0,
\end{equation*}
\begin{equation}
\begin{aligned}
&[H_1,X_2 ] = X_2, && [H_1,Y_2] = -Y_2,\\
&[H_2,X_1 ] = -X_1, && [H_2,Y_1] = Y_1,\\
& \{X_1,Y_1\} = \frac{K_1^2 -K_1^{-2}}{q^2 - q^{-2}}, &&
 \{X_2,Y_2\} = \frac{K_2^2 -K_2^{-2}}{q^2 - q^{-2}},
\end{aligned}
\end{equation}
where $X_{1,2},Y_{1,2}$ are still fermionic generators, $H_{1,2}$ are bosonic generators, and $K_{1,2} \equiv q^{H_{1,2}}$. These generators must also obey the generalized Serre relations
\begin{equation}
\begin{aligned}
& X_1^2 X_2 = q^2 X_2 X_1^2,
&&X_2^2 X_1 = q^{-2} X_1 X_2^2,\\
& Y_1^2 Y_2 = q^{-2} Y_2 Y_1^2,
&&Y_2^2 Y_1 = q^{2} Y_1 Y_2^2.
\end{aligned}
\end{equation}

We then define the additional bosonic operators via the $q$-adjoint action:
\begin{equation}
X_3 \equiv X_1X_2 + q X_2 X_1,~~~~~~~
Y_3 \equiv Y_2Y_1 + q^{-1} Y_1 Y_2,~~~~~~~ K_3 \equiv K_1 K_2.
\end{equation}
The remaining relations between $X_3$, $Y_3$, and the rest of the generators are fixed from the above relations. We note that in the limit $q\ra 1$ this algebra reduces to that of $\mathfrak{sl}(2|1)$.We direct the reader to \cite{Kulish1990} and \cite{Floreanini1991} for a full discussion on constructing the $q$ deformed algebra $sl_q(2|1)$, and its quantum oscillator realizations.

Next we will show that the algebra of the transfer matrix is a contraction of the quantum group $sl_q(2|1)$. As we are interested in the graded algebra of $Q$ and $Q^\dagger$, we can break $Q$ (and $Q^\dagger$) into two fermonic creation/annihilation operators:
\begin{equation}
Q = a+ b^\dagger,~~~~~~~~~~~
Q^\dagger = a^\dagger+ b,
\end{equation}
with $a$ and $b$ lowering operators and $a^\dagger$ and $b^\dagger$ raising operators. That is, they are defined by breaking the action of $Q,Q^{^{\dagger} } $ in \eqref{eq:Q_action} and \eqref{eq:Q_dg_action} into operators that lower and raise the number of chords.

%
%
%
%

From this definition, we find the $q$ anti- commutator of $a$ and $a^\dagger$ to be
\begin{equation} \label{eq:relations_a}
aa^\dagger +q a^\dagger a  = q^{-s+ M },
\end{equation}
where $M$ is a ``fermionic'' number operator satisfying
\begin{equation}
M\ket{n,X/O} = 0 ,
~~~~~~ M\ket{n+1/2,X} = \ket{n+1/2,X},
~~~~~~M\ket{n+1/2,O} = -\ket{n+1/2,O}.
\end{equation}
 Similarly we notice that $b$ and $b^\dagger$ satisfy the $q$ commutation relation
\begin{equation} \label{eq:relations_b}
b b^\dagger +q b^\dagger b  = q^{s- M }.
\end{equation}

The operator $M$ and the fermionic operators obey the commutation relations
\begin{equation}
[M,a] = -a,~~~~~~
[M,a^\dagger] = a^\dagger,~~~~~~
[M,b] = b,~~~~~~
[M,b^\dagger] = - b^\dagger .
\end{equation}

As a side note, we can redefine our operators $a$ and $b$ as $a_1 \equiv q^{-M/2+s/2}a$ and $a_2 \equiv q^{M/2-s/2}b $, so that $a_1$ and $a_2$ obey the algebra
\begin{equation}
a_i a_j^\dagger +[q + (1-q)\delta_{ij}] a_j^\dagger a_i = \delta_{ij},
\end{equation}
which is of the form (\ref{eq:gen_algebra}) with $q_{ij} = (q-1)\delta_{ij} - q $. This is in agreement with the inner product formula in the auxiliary Hilbert space, (\ref{eq:inn_real}).

As before, we define the bosonic creation/annihilation operators by the $q$-adjoint action of the fermionic operators,
\begin{equation}
A \equiv ab+q ba,~~~~~~~~~~
A^\dagger \equiv b^\dagger a^\dagger + q a^\dagger b^\dagger.
\end{equation}
They obey the commutation relation
\begin{equation}
A A^\dagger - A^\dagger A = q^{2 N + M} \left(q^{-1} - q\right),
\end{equation}
where $N$ the number operator, $N\ket{n,X/O} = n\ket{n,X/O}$ for $n\in \NN/2$. We can finish the bosonic commutation relations by noting that $M$ commutes with both $N$ and $A$, and that $N$ and $A$ obey the canonical relations $[N,A] = -A$, $[N, A^\dagger] = A^\dagger$. It remains to find the relations between the bosons and the fermions to close the algebra. The number operator acts nicely with the fermions, as they are raising/lowering operators, thus
\begin{equation}
[N,a] = -\frac{1}{2} a,~~~~~~
[N,a^\dagger] = \frac{1}{2} a^\dagger,~~~~~~
[N,b] = -\frac{1}{2} b,~~~~~~
[N,b^\dagger] = \frac{1}{2} b^\dagger.
\end{equation}

We finish closing the algebra by noting the last $q$ relations:
\begin{equation}
\begin{aligned}
&A a - q^{-1} aA = 0, && A^\dagger a - q^{-1} a A^\dagger = 0, \\
&A a^\dagger - q a^\dagger A = 0, && A^\dagger a^\dagger - q a^\dagger A^\dagger = 0, \\
& A b - q b A = 0 , && A^\dagger b - q^{-1} b A^\dagger = q^{s-M}(q - q^{-1}) a^\dagger,\\
& A b^\dagger - q b^\dagger A = q^{s-M}(q^{-1} - q) a , && A^\dagger b^\dagger - q^{-1} b^\dagger A^\dagger = 0.
\end{aligned}
\end{equation}

This graded algebra has four fermionic operators, $\{a,a^\dagger,b,b^\dagger\}$, and four bosonic operators, $\{A,A^\dagger,M,N\}$. This already seems similar to the quantum super group  $sl_q(2|1)$, and indeed this algebra is a contraction of  $sl_q(2|1)$, with the contraction being:
\begin{equation}
\begin{aligned}
&a = \lim_{\epsilon\ra 0}\epsilon q^{-(s+1)/2}\sqrt{q^{-2}-q^2} X_1 q^{N},
&&a^\dagger = \lim_{\epsilon\ra 0}\epsilon q^{-(s+1)/2}\sqrt{q^{-2}-q^2} q^{N} Y_1,\\
&b = \lim_{\epsilon\ra 0}\epsilon q^{(s-1)/2}\sqrt{q^{-2}-q^2} X_2 q^{N},
&& b^\dagger = \lim_{\epsilon\ra 0}\epsilon q^{(s-1)/2}\sqrt{q^{-2}-q^2}  q^{N}Y_2, \\
& H_1 = -N+\frac{M}{2},   && H_2 = N+\frac{M}{2},\\
& q^{-N+M/2} = \lim_{\epsilon\ra 0}\epsilon K_1^{-1},
&& q^{-N-M/2} = \lim_{\epsilon\ra 0}\epsilon K_2 .
\end{aligned}
\end{equation}

We note that the SUSY charges $Q$ and $Q^\dagger$ form the basic operator of the transfer matrix, with
\begin{align}
``T^{1/2}" = Q + Q^{\dagger} =a +b+a^{\dagger} + b^{\dagger},
\end{align}
a kind of super-coordinate operator, in analogy to $T=A+A^\dagger$ in the regular SYK \cite{Berkooz_2019}.

We suspect the connection to quantum groups to be deeper than just the algebra of the transfer matrix, and for the quantum group structure to appear also in higher order correlation functions, like in the regular double scaled SYK \cite{Berkooz_2019}. It remains an open question if this connection can be utilized to reformulate the SYK model in a quantum deformation setting, or if this connection can shed light on the gravity dual of the double scaled SYK.

\section{Summary and Discussion}

The main result of this paper is the derivation of an analytical expression for the asymptotic spectrum of the $\mathcal{N}=2$ SYK model in the double scaled limit, both in fixed charge sectors and in the presence of a chemical potential. Furthermore, we compared our results to an exact calculation of the number of ground states in each charge sector, as well as to the density of states of the super--Schwarzian theory in the relevant limit. We used the same combinatorial methods of chord diagrams to compute exact two point functions at all energy scales. Finally, we connected our results to a quantum deformation related to the quantum group $sl_q(2|1)$.

A future direction of research would be to use the full spectrum and correlation functions in the double scaled limit of this model, to connect it to a full 2-d dual theory of quantum gravity. The added supersymmetry can help constrain the dual 2-d theory, and so may make the process more feasible. In particular, the number of ground states in each charge sector that we computed should be the same as the number of supersymmetric (BPS) states in the dual 2-d theory, which might give a check of such a duality and of the idea that the chord description is the gravitational one. Furthermore, if the techniques above can be pushed to the case of ${\cal N}=4$, say to quiver models, then one can perhaps make contact with the black hole microstate program and perhaps quantify more precisely which states have a gravitational description and which do not.

Additionally, it would be interesting to find simple rules for computing higher order correlation functions, maybe using the rigid structure of the quantum deformation $sl_q(2|1)$. Specifically computing the four point function is of great interest as it allows to find the Lyapunov exponent. It is not clear if the $\mathcal{N}=2$ theory is maximally chaotic in the Schwarzian regime. Computing the chaos exponent using ladder diagrams may be possible, though it is unclear what the correct form of the 2-point function in the ladder kernel should be; as we showed the conformal ansatz of the two point function is inconsistent in the IR due to the large number of ground states.

Finally, the connection between the double scaled limit of the SYK model and quantum deformations is still not well understood. We suspect this connection to also appear in higher order correlation functions, and that this connection may lead to a systematic way to compute them. The quantum deformation may also be related to the gravity dual of the double scaled SYK model, or at least allow us to understand more of its features beyond the gravitational sector.

\acknowledgments

We would like to thank O. Aharony, M. Isachenkov, P. Narayan, H. Raj, M. Rangamani, M. Rozali, R. Spiecher, and H. Verlinde for useful discussions. This work is supported by an ISF center of excellence grant (2289/18). MB is the incumbent of the Charles and David Wolfson Professorial chair of theoretical Physics.

\appendix

\section{Special Functions} \label{app:special_fun}

In this section we will define various special functions we use in the text, and state their relevant properties. We will assume that $|q|<1$ throughout this section.

The $q$-Pochhammer symbol is defined as
\begin{equation}
(a;q)_n \equiv \prod_{k = 1}^n \left(1-a q^{k-1}\right).
\end{equation}
We use the shorthand notation $(a_1,\ldots, a_m;q)_n = (a_1;q)_n \ldots (a_m;q)_n$. The infinite $q$-Pochhammer symbol $(a;q)_{\infty} = \lim_{n\ra \infty} (a;q)_n$ is well defined for $|q|<1$.

The continuous $q$-Hermite polynomials are a set of orthogonal polynomials defined via the $q$-Pochhammer symbol as
\begin{equation}
H_n(\cos \phi|q) \equiv \sum_{k=0}^n \frac{(q;q)_n}{(q;q)_k (q;q)_{n-k}} e^{i(n-k)\phi} .
\end{equation}
$H_n(\cos \phi|q)$ are polynomials in both $\cos \phi$ and $q$. They satisfy the recursion relation
\begin{equation}
2\cos \phi H_n(\cos \phi|q) = H_{n+1}(\cos \phi|q) + (1-q^n) H_{n-1}(\cos \phi|q),~~~~~~H_{-1} = 0, H_1 = 1,
\end{equation}
the $\phi$ orthogonality
\begin{equation}
\int_0^\pi \frac{d\phi}{2\pi} \left(q,e^{\pm 2i\phi};q \right)_{\infty} H_n(\cos \phi|q) H_m(\cos \phi|q)
 = (q;q)_m \delta_{mn},
\end{equation}
and the $n$ orthogonality
\begin{equation}
\sum_{n=0}^\infty \frac{ t^n H_n(\cos \phi|q) H_n(\cos \phi'|q)}{(q;q)_n} = \frac{(t^2;q)_{\infty}}{(te^{i(\pm \phi \pm \phi')};q)_{\infty}} .
\end{equation}

Choosing $t=1$ gives the orthogonality relation
\begin{equation} \label{eq:qhermite_n_orthogonality}
\sum_{n=0}^\infty \frac{ H_n(\cos \phi|q) H_n(\cos \phi'|q)}{(q;q)_n} = \frac{2\pi \left[\delta(\phi - \phi')+\delta(\phi+\phi') \right]}{(q,e^{\pm2i \phi };q)_{\infty}} .
\end{equation}

We will use the following conventions for the Jacobi theta functions (following chapter 21 of \cite{whittaker_watson}):
\begin{align}
\vartheta_1(z,q) &= \sum_{n\in \ZZ + 1/2}  (-1)^{n+1} q^{n^2} e^{2niz} = 2q^{1/4} \sin(z)\left( e^{\pm 2 iz},q^2;q^2 \right)_{\infty} , .\label{eq:jactheta_1}\\
\vartheta_2(z,q) &= \sum_{n\in \ZZ + 1/2}  q^{n^2} e^{2niz} = 2q^{1/4} \cos(z)\left( -e^{\pm 2 iz},q^2;q^2 \right)_{\infty} , \\
\vartheta_3(z,q) &= \sum_{n\in \ZZ}  q^{n^2} e^{2niz} = \left(-q^{-1}e^{\pm 2 iz},q^2;q^2\right)_\infty ,\\
\vartheta_4(z,q) &= \sum_{n\in \ZZ} (-1)^n q^{n^2} e^{2niz} = \left(q^{-1}e^{\pm 2 iz},q^2;q^2\right)_\infty . \label{eq:jactheta_4}
\end{align}
and recall their modular transformation properties
\begin{align} \label{eq:theta_modular_trans}
\vartheta_{1}\left(z,e^{i\pi\tau}\right)&=\frac{-i}{\sqrt{-i\tau}}e^{-\frac{iz^{2}}{\pi\tau}}\vartheta_{1}\left(-\frac{z}{\tau},e^{-\frac{i\pi}{\tau}}\right),
\\\vartheta_2\left(z,e^{ i \pi\tau}\right)&= \frac{1}{ \sqrt{-i \tau}} e^{-\frac{i z^2}{\pi \tau}} \vartheta_4\left(-\frac{z}{\tau},e^{-\frac{i\pi}{\tau}} \right) .
\end{align}

\section{Calculation of the Inner Product in the Auxiliary Hilbert Space} \label{app:inn_general}

Let us allow a non-zero inner product $\langle u|v\rangle$ only between two states $u,v$ with the same number of $X$'s and $O$'s. Let us denote a state $u$ with $n$ $X$'s and $m$ $O$'s by $u_{n,m}$. If we do not assume anything about the inner product other than that, then to check that the representation of $Q^{\dagger}$ in ${\cal H}_{\text{aux}}$ is the conjugate transpose in this inner product of that of $Q$, we need to check four options (since all the others give vanishing matrix elements of $Q$ or $Q^{\dagger}$):
\begin{enumerate}
\item $u_{n,m}$ and $v_{n+1,m}$. In this case there are only two non-zero inner products involving $Q,Q^{\dagger}$, leading to the condition
\begin{equation}
\langle u|Qv\rangle=\langle Q^{\dagger}u|v\rangle.
\end{equation}
In $\langle Q^{\dagger}u|v\rangle$ there is only one (potentially) non-zero product given by adding to $u$ at its end one $X$. In $\langle u|Qv\rangle$ there are several non-zero products obtained by deleting one $X$ from $v$. This gives us the condition
\begin{equation}
\langle uX|v\rangle=q^{s}\sum_{X\in v\text{ deleted}}(-1)^{\#_{\text{below}}}q^{\#_{\text{other X}}-\#_{\text{\text{O above}}}}\langle u|v\text{ (one X deleted)}\rangle\label{eq:condition1}
\end{equation}
where $uX$ means appending an X to $u$ (and all the number of $X$'s or $O$'s below or above the deleted $X$ is with respect to the same vector, which is $v$ in this case).

\item $u_{n,m}$ and $v_{n-1,m}$. In this case the only condition is
\begin{equation}
\langle u|Q^{\dagger}v\rangle=\langle Qu|v\rangle
\end{equation}
leading to
\begin{equation}
\langle u|vX\rangle=q^{s}\sum_{X\in u\text{ deleted}}(-1)^{\#_{\text{below}}}q^{\#_{\text{other X}}-\#_{\text{\text{O above}}}}\langle u\text{ (one X deleted)}|v\rangle.
\end{equation}

\item $u_{n,m}$ and $v_{n,m+1}$. In this case the only condition is
\begin{equation}
\langle Qu|v\rangle=\langle u|Q^{\dagger} v\rangle
\end{equation}
leading to
\begin{equation}
\langle uO|v\rangle=q^{-s}\sum_{O\in v\text{ deleted}}(-1)^{\#_{\text{below}}}q^{\#_{\text{other O}}-\#_{\text{\text{X above}}}}\langle u|v\text{ (one O deleted)}\rangle.
\end{equation}

\item $u_{n,m}$ and $v_{n,m-1}$. In this case the only condition is
\begin{equation}
\langle u|Qv\rangle=\langle Q^{\dagger}u|v\rangle
\end{equation}
leading to
\begin{equation}
\langle u|vO\rangle=q^{-s}\sum_{O\in u\text{ deleted}}(-1)^{\#_{\text{below}}}q^{\#_{\text{other O}}-\#_{\text{\text{X above}}}}\langle u\text{ (one O deleted)}|v\rangle.
\end{equation}
\end{enumerate}

\begin{figure}
\centering
\includegraphics[width=0.4\textwidth]{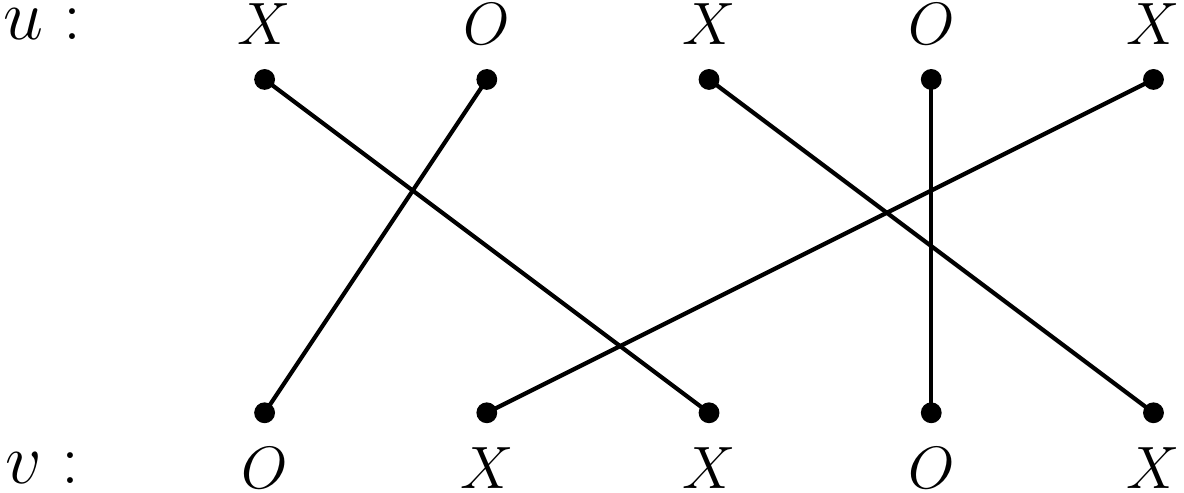}

\caption{An example of a diagram contributing to $\langle u,v\rangle$.}
\label{fig:inner_product_diagram}

\end{figure}

These conditions fix uniquely the inner product up to normalization. We will normalize $\langle\emptyset,\emptyset\rangle=1$. For instance, in conditions 1 and 3, we essentially pair the last $X$ or $O$ with another one (of the same kind) in $v$ (we go over all possibilities in the sum). We can thus represent it in the following way. $u$ is written as a string of $X$'s and $O$'s, and similarly for $v$ placed below it. The pairing above is represented by lines, each one connects an $X$ from $u$ to an $X$ from $v$ and similarly for $O$. One such diagram is shown in figure \ref{fig:inner_product_diagram}.

The solution is then the following. For $u,v$ each having $n$ $X$'s and $m$ $O$'s, $\langle u|v\rangle$ is given by summing all diagrams of this form, each assigned a value
\begin{equation}
q^{s(n-m)}(-1)^{\#\text{intersections}}q^{\frac{n(n-1)}{2}+\frac{m(m-1)}{2}-nm+\#\text{X-O intersections}}\label{eq:diagram_value}
\end{equation}
where ($\#\text{X-O intersections})$ is the number of intersections of a line connecting $X$'s with a line connecting $O$'s.

\begin{figure}
\centering
\subfloat[]{\centering
\includegraphics[height=3cm]{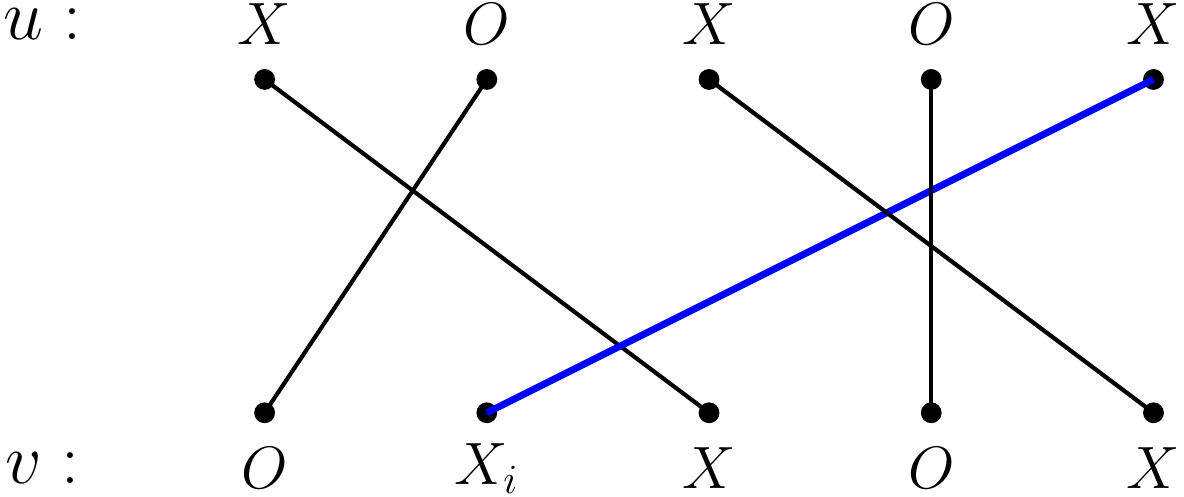}

}\qquad \qquad
\subfloat[]{\centering
\includegraphics[height=3cm]{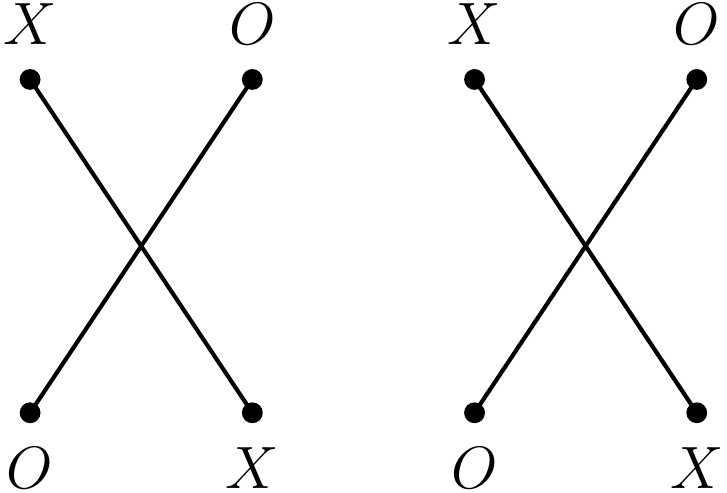}

}\caption{The corresponding diagrams on both sides of (\ref{eq:condition1}).}
\label{fig:condition_1}

\end{figure}

Let us show that this is indeed the solution. Consider the first condition, equation (\ref{eq:condition1}), above. On the LHS, we have a sum over the possible diagrams. Let us pick a particular diagram, such as the one shown in figure \ref{fig:inner_product_diagram}. In the context of condition 1, as mentioned above, the line ending on the
last $X$ of $u$ is singled out; see the left hand side of figure \ref{fig:condition_1}. Denote the $X$ to which it is connected in $v$ by $X_{i}$. Removing this line, we obtain naturally a diagram contributing to the term in the sum on the RHS of (\ref{eq:condition1}) corresponding to removing $X_{i}$ in $v$; see the right hand side diagram of figure \ref{fig:condition_1}. By this, we are matching each diagram on the LHS, to a particular diagram for a term in the
sum on the RHS. We will now show that the value assigned to each such diagram is the same on both sides, and therefore in particular the two sides are equal.

To show this, notice that the diagram on the LHS of (\ref{eq:condition1}) is assigned the value (\ref{eq:diagram_value}). On the RHS, the diagram from which the $X_{i}$ line is removed, comes with a pre-factor
\begin{equation}
q^{s}(-1)^{\#_{\text{below}}}q^{\#_{\text{other X}}-\#_{\text{\text{O above}}}}
\end{equation}
(where all qualifiers are relative to $X_{i}$ in $v$); this is multiplied by the value of the diagram itself which equals
\begin{equation}
q^{s(n-1-m)}(-1)^{\#\text{intersections without }X_{i}}q^{\frac{(n-1)(n-2)}{2}+\frac{m(m-1)}{2}-(n-1)m+\#\text{(X-O intersections without }X_{i}\text{)}}.
\end{equation}
The product of these two expressions indeed equals to (\ref{eq:diagram_value}). The powers of $(-1)$ are the same, because the number of intersections of the line $X_{i}$ is exactly the number of $X$'s and $O$'s below it (note that the state here is represented from left to right, while usually we were drawing it from top to bottom). The ($s$ independent) power of $q$ matches because $\#_{\text{other X}}=n-1$, and
\begin{equation}
\begin{split}
& \#\text{(X-O intersections)}-\#\text{(X-O intersections without }X_{i}\text{)} = \\
 &\qquad  \qquad  =\#\text{(O that }X_{i}\text{ intersects)}=\#_{\text{O below}}=m-\#_{\text{O above}}.
\end{split}
\end{equation}

In addition, since (\ref{eq:diagram_value}) is symmetric between $u,v$, condition 2 follows. Similarly, as (\ref{eq:diagram_value}) is symmetric between $X$ and $O$ (taking $s\to-s$), conditions 3,4 follow as well.

\section{Computations in the Physical Hilbert Space} \label{app:eigen}

\subsection*{The Inner Product in the Physical Hilbert Space}

We can use the inner product formula for the full Hilbert space to find the inner product for physical states, however this requires summing over all chords between the vectors, which is complicated. We can instead calculate it directly from the physical Hilbert space, as we know it is well defined and that states with a different number of chords are orthogonal. This leaves us with five families of undetermined parameters:
\begin{equation}
\begin{aligned}
&\inn{n,O}{n,O} \equiv A_n,  && \inn{n,X}{n,X} \equiv B_n,  &&& \inn{n,O}{n,X} \equiv C_n,\\
&\inn{n+\frac{1}{2},O}{n+\frac{1}{2},O} \equiv a_n,  &&
 \inn{n+\frac{1}{2},X}{n+\frac{1}{2},X} \equiv b_n,.
\end{aligned}
\end{equation}

Notice that the full inner product requires that $ \inn{n+\frac{1}{2},X}{n+\frac{1}{2},O} =0$, as they have a different number of right--moving and left--moving chords. Then we can calculate the matrix elements and require Hermiticity of $Q$ and $Q^\dagger$, resulting in the following relations: \footnote{Note that all the coefficients must be real except $C_n$, and these relations trivially imply $C_n$ is real as well (say from the first equation), so we replaced $C_n^*$ with $C_n$.}
\begin{equation}
\begin{aligned}
&-q^n a_n = q^{-s} C_{n+1}, \qquad \qquad
&& q^s a_n =  q^{n} C_n + A_n,\\
& q^{-1} a_n = q^{-s} A_{n+1},
&& q^n A_n + C_n = 0,\\
& q^n B_n + C_n = 0 ,
&& q^{-s} b_n = q^n C_n + B_n,\\
& -q^n b_n = q^s C_{n+1},
&& q^{-1} b_n = q^s B_{n+1}.
\end{aligned}
\end{equation}

We immediately find the recursion relation for $A_n$:
\begin{equation}
\left(1-q^{2n}\right) A_n = q A_{n+1} ,
\end{equation}
and can relate all the other coefficients to $A_n$ via
\begin{equation}
A_n = B_n = - q^{-n} C_n =\frac{1}{1-q^{2n}}q^{-s} b_n=\frac{1}{1-q^{2n}}q^{s} a_n .
\end{equation}
The solution to the recursion relation is
\begin{equation}
A_n = q^{-n} \left(q^2;q^2 \right)_{n-1} .
\end{equation}
The inner product matrix in the subspace with a given number of chords, $n$, is
\begin{equation}
A_n \bpm 1 & -q^n \\ -q^n & 1 \epm,
\end{equation}
which has the eigenvalues and eigenvectors
\begin{equation}
\lambda_\pm = A_n \left(1 \mp q^n\right),~~~~~~~~
x_\pm = \bpm 1 \\ \pm 1 \epm .
\end{equation}
As these are always positive for any $n$, the inner product is positive definite, and well behaved. It also allows us to define the orthonormal basis
\begin{equation} \label{eq:basis_n_pm}
\ket{n_\pm} \equiv \frac{1}{\sqrt{2A_n(1 \mp q^n)}}\left(\ket{n,O} \pm \ket{n,X}\right) ,
 \qquad \ket{n+\frac{1}{2},\pm} \equiv \frac{q^{\pm s/2 }}{\sqrt{q A_{n+1}}}\ket{n+\frac{1}{2},O/X} ,
\end{equation}
which we use in calculations of the two point function.

\subsection*{Normalization of the Eigenvectors}
We wish to calculate the inner product $\inn{v(\phi')}{v(\phi)}$. Using the expansion of the eigenvectors $\ket{v(\phi)} = \sum_{n=0}^\infty \alpha_n Q\ket{n+1/2,X}$, it follows that
\begin{equation}
\begin{split}
\inn{v(\phi')}{v(\phi)} &=\sum_{n=0}^\infty \alpha_n(\phi')  \innn{n+\frac{1}{2},X}{ Q^\dagger Q \sum_{m=0}^\infty\alpha_m(\phi)}{m+\frac{1}{2},X} \\
&= \sum_{n,m=0}^\infty \alpha_n(\phi') \alpha_m(\phi) \Lambda_s(\phi) \inn{n+\frac{1}{2},X }{m+\frac{1}{2},X} ,
\end{split}
\end{equation}
as this is just the transfer matrix acting on a fermionic eigenvector. Then we use the inner product formula, \eqref{eq:innerproduct_physical}, as well as the definition of $\alpha_m$, \eqref{eq:alphas}, and the orthogonality relation of $q$-Hermite polynomials, \eqref{eq:qhermite_n_orthogonality}, to see that
\begin{equation}
\begin{split}
\inn{v(\phi')}{v(\phi)}&=q^{s}\Lambda_{s}\left(\phi\right)\sum_{n=0}^{\infty}\frac{1}{\left(q^{2};q^{2}\right)_{n}}H_{n}\left(\cos\phi|q^{2}\right)H_{n}\left(\cos\phi'|q^{2}\right)\\
&=q^{s}\Lambda_{s}\left(\phi\right)\frac{2\pi\left(\delta\left(\phi-\phi'\right)+\delta\left(\phi+\phi'\right)\right)}{\left|\left(e^{2i\phi};q^{2}\right)_{\infty}\right|^{2}\left(q^{2};q^{2}\right)_{\infty}}.
\end{split}
\end{equation}
Since the integration domain is $ \phi,\phi'\in[0,\pi] $ we are assured that $ \delta(\phi+\phi')\neq0 $, and we get
\begin{align}
	\left\langle v\left(\phi'\right)|v\left(\phi\right)\right\rangle =q^{s}\Lambda_{s}\left(\phi\right)\frac{2\pi\delta\left(\phi-\phi'\right)}{\left(q^{2},e^{\pm2i\phi};q^{2}\right)_{\infty}}.
\end{align}
The norm $ \left\langle u\left(\phi'\right)|u\left(\phi\right)\right\rangle $ is given by the same expression with $ s\to(-s) $.

\subsection*{The Inner Product of Eigenvectors and Number States:}
Our goal is to compute the overlap $ \inn{n_{\pm}}{v\left(\phi\right)}  $ and $ \inn{n_{\pm}}{u\left(\phi\right)}$. We use the same definition of $\ket{u(\phi)}$, as well as equations \eqref{eq:basis_n_pm}, \eqref{eq:innerproduct_physical}, \eqref{eq:alphas}, to compute the overlap
\begin{equation}
\begin{split}
 \inn{n_{\pm}}{u\left(\phi\right)}&=\frac{\bra{n,O}\pm\bra{n,X}}{\sqrt{2A_{n}\left(1\mp q^{n}\right)}}\left(\sum_{m}\alpha_{m}\left(\phi\right)\left(q^{m}\left|m,X\right>+\left|m,O\right>+e^{\lambda s}\left|m+1,O\right>\right)\right)\\
 &=\frac{\bra{n,O}\pm\bra{n,X}}{\sqrt{2A_{n}\left(1\mp q^{n}\right)}}
 \left[ \alpha_{n}\left(\phi\right)\left(q^{n}\left|n,X\right>+\left|n,O\right>\right)
 +\alpha_{n-1}e^{\lambda s}\left|n,O\right> \right]\\
 &=\frac{q^{-n}\left(q^{2};q^{2}\right)_{n-1}}{\sqrt{2A_{n}\left(1\mp q^{n}\right)}}\left[\left(1-q^{2n}\right)\alpha_{n}\left(\phi\right)+\left(1\mp q^{n}\right)q^{-s}\alpha_{n-1}\left(\phi\right)\right]\\
 &=\frac{H_{n}\left(\cos\phi|q^{2}\right)+\left(1\mp q^{n}\right)q^{-1/2} q^{-s}H_{n-1}\left(\cos\phi|q^{2}\right)}
 {\sqrt{2\left(q^{2};q^{2}\right)_{n-1}\left(1\mp q^{n}\right)}} .
	\end{split}
\end{equation}
Similarly we find that the other overlap is
\begin{equation}
 \inn{n_{\pm}}{v\left(\phi\right)}=\pm\frac{H_{n}\left(\cos\phi|q^{2}\right)+\left(1\mp q^{n}\right)q^{-1/2} q^{s}H_{n-1}\left(\cos\phi|q^{2}\right)}{\sqrt{2\left(q^{2};q^{2}\right)_{n-1}\left(1\mp q^{n}\right)}}.
\end{equation}

\section{The Ground State Density} \label{app:ground_states}

We can integrate the continuous density over $E$ to find the missing density at zero. It is simpler to take the expression for $m_0 = \int \rho(E) dE$, which is
\begin{align}
m_0 = \frac{ q^{1/4}}{\sqrt{\pi}}\intinf dx ~e^{-x^2}~\int_0^\pi \frac{d\phi}{2\pi}~
\left(q^2,e^{\pm2i\phi};q^2\right)_{\infty} \left(\cosh\left(x\sqrt{\lambda}\right)-\cos\phi \right)^{-1} .
\end{align}

Let us take the $\phi$ integral first. In this case we need to compute
\begin{equation}
\begin{split}
I(x;q) & =  \int_0^\pi \frac{d\phi}{2\pi}~\left(q^2,e^{\pm2i\phi};q^2\right)_{\infty}
 \left(\cosh\left(x\sqrt{\lambda}\right)-\cos\phi \right)^{-1} \\
 &=  \sum_{n=0}^\infty \left(\cosh\left(x\sqrt{\lambda}\right)\right)^{-n-1}
\int_0^\pi \frac{d\phi}{2\pi}~ \left(q^2,e^{\pm2i\phi};q^2\right)_{\infty} \cos^n\phi .
\end{split}
\end{equation}

From \cite{Berkooz_2019} appendix B we have that
\begin{equation}
\int_0^\pi \frac{d\phi}{2\pi}~ \left(q^2,e^{\pm2i\phi};q^2\right)_{\infty} \cos^{n}\phi
=\left\{\begin{array}{cc} \frac{1}{2^{n}} c_{n/2,n} ,  & n ~~\text{even}, \\ 0,  & n ~~\text{odd}, \end{array}\right.
\end{equation}
with
\begin{equation}
c_{n,2n} = \sum_{j=0}^{n}(-1)^j q^{2j+2{{j}\choose{2}}}\frac{2j+1}{2n+1}{{2n+1}\choose{n-j}}.
\end{equation}
Thus we get
\begin{equation}
m_0(\lambda) = \frac{ q^{1/4}}{\sqrt{\pi}}\intinf dx ~e^{-x^2}~ \sum_{n=0}^\infty \left(\cosh\left(x\sqrt{\lambda}\right)\right)^{-2n-1} 2^{-2n} c_{n,2n}.
\end{equation}

We can now take the integral over $x$ explicitly by noticing that
\begin{equation}
\begin{split}
\intinf dx~ e^{-x^2}2^{-2n} \cosh^{-2n-1}(ax)
= 2 \sqrt{\pi} & \sum_{m=0}^\infty  {m+2n \choose 2n}  (-1)^m e^{(m+n+1/2)^2a^2}\\
&\times \text{erfc}\left(\frac{(2m+2n+1)a}{2} \right) .
\end{split}
\end{equation}

Thus we can write $m_0$ as the triple sum
\begin{equation}
\begin{split}
m_0(\lambda) &= 2 \sum_{m,n=0}^\infty  \sum_{j=0}^{n} q^{-(m+n)^2 - (m+n)+j(j+1)}
\text{erfc}\left(\left(m+n+\frac{1}{2}\right)\sqrt{\lambda} \right) \\
&~~~~~~~~~~ \times (-1)^{m+j} \frac{2j+1}{2n+1} {m+2n \choose 2n} {{2n+1}\choose{n-j}} \\
&=  2 \sum_{k=0}^\infty(-1)^k \text{erfc}\left(\left(k+\frac{1}{2}\right)\sqrt{\lambda} \right) q^{-k(k+1)} \\
&~~~~~~~~~~ \times \sum_{n=0}^k \sum_{j=0}^{n} q^{j(j+1)}(-1)^{-n+j} \frac{(2j+1)(k+n)!}{(k-n)!(n-j)!(n+j+1)!}\\
&=2 \sum_{k=0}^\infty(-1)^k \text{erfc}\left(\left(k+\frac{1}{2}\right)\sqrt{\lambda} \right) q^{-k(k+1)}
\sum_{j=0}^k q^{j(j+1)}(2j+1) \\
&~~~~~~~~~~ \times  \sum_{l=0}^{k-j}\bigg|_{l=n-j} (-1)^{l} \frac{(k+l+j)!}{(k-l-j)!(l)!(l+2j+1)!} .
\end{split}
\end{equation}

Now notice that the sum
\begin{equation}
 \sum_{l=0}^{k-j} (-1)^{l} \frac{(k+l+j)!}{(k-l-j)!(l)!(l+2j+1)!} = 0, ~~~k\neq j.
\end{equation}
Thus we are left with only the case $j=k$, so the sum simplifies to
\begin{equation}
m_0(\lambda) = 2 \sum_{k=0}^\infty(-1)^k \text{erfc}\left(\left(k+\frac{1}{2}\right)\sqrt{\lambda}\right) .
\end{equation}

We expect this to agree with the exact density of ground states. We can show that this is indeed the case by calculating $1-D$ and showing that it matches $m_0$:
\begin{equation}
\begin{split}
1-D &=1-\sqrt{\frac{\lambda}{\pi}}\int_{-1/2}^{1/2}ds
\sum_{n=-\infty}^\infty (-1)^ne^{-\lambda(n+s)^2}\\
&= \sqrt{\frac{\lambda}{\pi}}\int_{-1/2}^{1/2}ds
\sum_{n=-\infty}^\infty e^{-\lambda(n+s)^2}\left(1-(-1)^n\right)\\
&=4 \sqrt{\frac{\lambda}{\pi}}\int_{-1/2}^{1/2}ds
\sum_{k=0}^\infty e^{-\lambda(2k+1+s)^2}\\
&= 2\sum_{k=0}^{\infty}(-1)^k \text{erfc}\left(\left(k+\frac{1}{2}\right)\sqrt{\lambda}\right) = m_0,
\end{split}
\end{equation}
as expected.

\section{Conformal Limit of 2-point Function: Computations} \label{app:2point}

In this appendix we give the detail calculation of the conformal limit of the 2-point function. Our calculations follow the calculation of the conformal limit in \cite{Micha2018}. The conformal limit is attained for low temperatures together with $ q\to1^-\Leftrightarrow\lambda\to0 $. We shall scale the length of the operator insertion accordingly, and take $ \tilde{p}=\alpha p $, which gives us $ \tilde{q}=q^\alpha $. Following \cite{Micha2018}, we expect to recover the conformal limit when $\lambda \ll  \beta\lambda^2,t\lambda^2\ll1 $. Throughout the computation we will use $ (x,y)\leftrightarrow(\beta +it.\beta-it) $ interchangeably. Furthermore, we shall analyze the terms in \eqref{eq:2pointfun} separately.

We shall start by concentrating on the $I_c$ term in \eqref{eq:2pointfun}, as it is connected to the continuum spectrum whereas the $I_1$ terms are connected to the ground states. We expect the conformal part of the 2-point function to arise from the continuum spectrum, and therefore for $I_c$ to behave like a conformal propagator. After we show this, we analyze the rest of the terms, which are connected to the ground states, in this limit.

\subsection{The conformal part of the 2-point function} \label{app:2point_con}

We shall start by concentrating on the $I_c$ term in \eqref{eq:2pointfun}, and will further split $ I_c(x,y)$ into two parts, $ I_c(x,y) =I_c^1(x,y)+I_c^2(x,y) $ with $ I_c^1$ and $I_c^2 $ given by the third and forth lines of \eqref{eq:I_c_def} respectively.  Furthermore, We can rewrite $I_c^{1}$ as
\begin{equation}  \label{eq:I_1_as_A_1}
I_c^1(x,y)  =A_1(x,y;\alpha) -A_1(x,0;\alpha)-A_1(0,y;\alpha) + A_1(0,0;\alpha),
\end{equation}
where $\alpha = \tilde{\lambda}/\lambda$ is the charge of the operator we inserted, and $A_1$ is given by
\begin{equation} \label{eq:A_1_full}
\begin{split}
A_1(x,y;\alpha) &= 2q^{-1/4} \sqrt{\frac{\lambda}{\pi}} \int ds ~e^{-\lambda s^2} \frac{d\phi'd\phi}{(2\pi)^2}  \left(q^{2\alpha},q^2,q^2,e^{\pm2i\phi},e^{\pm2i\phi'};q^{2}\right)_{\infty}
\frac{ q^{s+\alpha/2}}{\left(q^\alpha e^{i\left(\pm\phi\pm\phi'\right)};q^{2}\right)_{\infty}} \\
&\quad\times \left(\frac{1}{\Lambda_{s-\frac{\alpha}{2}}\left(\phi\right) }+\frac{1}{\Lambda_{s+\frac{\alpha}{2}}\left(\phi'\right) }\right)
  \exp\left\{-\frac{yq^{1/4}}{2}\Lambda_{s-\frac{\alpha}{2}}\left(\phi\right) - \frac{xq^{1/4}}{2}\Lambda_{s+\frac{\alpha}{2}}\left(\phi'\right)\right\}.
 \end{split}
 \end{equation}
Notice that $A_1(x,y;\alpha)$ is the portion that explicitly captures the long time/conformal regime associated with transition between generic states, while the other terms are related to transitions between the continuous spectrum states and the ground states, induced by the inserted operator. We can similarly rewrite $I_c^2$ in the same form:
\begin{equation} \label{eq:I_2_as_A_2}
I_c^2(x,y)  =A_2(x,y;\alpha) -A_2(x,0;\alpha)-A_2(0,y;\alpha) + A_2(0,0;\alpha),
\end{equation}
with
\begin{equation} \label{eq:A_2_full}
\begin{split}
A_2(x,y;\alpha) &= \frac{q^{1/4}}{2}\sqrt{\frac{\lambda}{\pi}} \int ds ~
e^{-\lambda s^2} \frac{d\phi'd\phi}{(2\pi)^2}
\frac{ \left(\tilde{q}^2,q^2,q^2,e^{\pm2i\phi},e^{\pm2i\phi'};q^{2}\right)_{\infty}}
{\left(q^{\alpha+1}e^{i\left(\pm\phi\pm\phi'\right)};q^{2}\right)_{\infty}} \\
 &\quad\times  \frac{q^{-\alpha}}{\Lambda_{s-\frac{\alpha}{2}}\left(\phi\right)
 \Lambda_{-s-\frac{\alpha}{2}}\left(\phi'\right)}
\exp\left\{-\frac{yq^{1/4}}{2}\Lambda_{s-\frac{\alpha}{2}}\left(\phi\right)
- \frac{xq^{1/4}}{2}\Lambda_{-s-\frac{\alpha}{2}}\left(\phi'\right)\right\}.
 \end{split}
 \end{equation}
Our main goal of this subsection  will be to show that $A_{1,2}$ have a conformal form in the late time/Schwarzian regime, with a conformal dimension of $\alpha/2$ and $\alpha/2+1/2$ respectively. In the next subsection we will show that the parts pertaining to the ground states are not negligible in this limit, and must be taken into account.

\subsection*{$A_2(x,y;\alpha)$:}
We will start the analysis from $A_2(x,y;\alpha)$. As both $x,y \gg 1$, the will localize us to small energies, so we can approximate $\Lambda$ by
\begin{align} \label{eq:EValApprox}
\Lambda_{s}\left(\phi\right)\approx\lambda^{2}\left(s-\frac{1}{2}\right)^{2}+\phi^2,
\end{align}
where $\phi$ and $\phi'$ are localized around zero. We then approximate \eqref{eq:A_2_full} in the regime $ \lambda \ll \lambda^2 y,\lambda^2 x \ll 1$ as
\begin{equation} \label{eq:A_2_xy_first}
\begin{split}
A_2(x,y;\alpha) &\approx \frac{2}{\lambda^4} \sqrt{\frac{\lambda}{\pi}}
\left(\tilde{q}^2,q^2,q^2;q^{2}\right)_{\infty}\int ds \frac{d\phi'd\phi}{(2\pi)^2}  \\
 &\quad\times \frac{1}{ \left( (s - \frac{\alpha+1}{2})^2+\varphi^2\right)
 \left( (s + \frac{\alpha+1}{2})^2+{\varphi'}^2\right) }~
  \frac{\left(e^{\pm2i\phi},e^{\pm2i\phi'};q^{2}\right)_{\infty}}
 {\left(q^{\alpha+1}e^{i\left(\pm\phi\pm\phi'\right)};q^{2}\right)_{\infty}} \\
 &\quad \times \exp\left\{-\frac{y}{2}\left(\lambda^2 (s - \frac{\alpha+1}{2})^2+\phi^2\right)
-\frac{x}{2}\left(\lambda^2 (s + \frac{\alpha+1}{2})^2+\phi'^2\right) - \lambda s^2\right\}.
 \end{split}
 \end{equation}
with $\varphi = \phi/\lambda$. We can neglect the exponential terms that couple to $s$, as they are small, and then take the $s$ integral by closing the contour in the complex plane and residues
\begin{equation} \label{eq:s_integral_1}
\intinf ds  \frac{1}{ \left( (s - \frac{\alpha+1}{2})^2+\varphi^2\right) \left( (s + \frac{\alpha+1}{2})^2+{\varphi'}^2\right) }
 = \frac{\pi (\varphi + \varphi')}{\varphi \varphi' \left[ (\alpha+1)^2 + (\varphi + \varphi')^2\right]}.
\end{equation}

Furthermore, in this we can use the approximations
\begin{equation}  \label{eq:conf_approximation_1}
 \frac{\left(e^{\pm2i\phi},e^{\pm2i\phi'};q^{2}\right)_{\infty}}
 {\left(q^{z}e^{i\left(\pm\phi\pm\phi'\right)};q^{2}\right)_{\infty}}
 \approx \left(e^{i\left(\pm\phi\pm\phi'\right)};q^{2}\right)_{z/2}
 \frac{4\phi\phi'}{\phi^{2}-\phi'^{2}}\cdot
 \frac{\sinh\left(\frac{\pi\phi}{\lambda}\right)\sinh\left(\frac{\pi\phi'}{\lambda}\right)}
 {\sinh\left(\frac{\pi\left(\phi-\phi'\right)}{2\lambda}\right)
 \sinh\left(\frac{\pi\left(\phi+\phi'\right)}{2\lambda}\right)}
 e^{-\frac{1}{2\lambda}\left(\phi^{2}+\phi'^{2}\right)},
\end{equation}
as well as
\begin{equation} \label{eq:conf_approximation_2}
\left(e^{i\lambda\left(\pm\varphi\pm\varphi'\right)};q^{2}\right)_{z/2} \approx \lambda^{2z}\left(\varphi^{2}-\varphi'^{2}\right)^{z},
\end{equation}
and
\begin{equation} \label{eq:conf_approximation_3}
\left(q^{2\alpha},q^2,q^2 ;q^2\right)_{\infty} \approx \frac{\pi^{3/2} 2^{\alpha-1}}{\lambda^{\alpha+1/2} \Gamma(\alpha)} e^{-\frac{\pi^2}{4\lambda}} .
\end{equation}

Plugging \eqref{eq:s_integral_1}, \eqref{eq:conf_approximation_1}, \eqref{eq:conf_approximation_2}, and \eqref{eq:conf_approximation_3} into \eqref{eq:A_2_xy_first}, and do the change of variables $\varphi = \phi/\lambda$ and $\varphi' = \phi'/\lambda$ we arrive at
\begin{equation}
\begin{split}
A_2(x,y;\alpha) &\approx\frac{2^\alpha \lambda^{\alpha}}{\Gamma(\alpha)}
e^{-\frac{\pi^2}{4\lambda}} \int_0^\infty d\varphi'd\varphi
\frac{\left(\varphi^2-\varphi'\right)^{\alpha}(\varphi+\varphi')}
{\left[ (\alpha+1)^2 + (\varphi + \varphi')^2\right]}\\
 &\quad\times \frac{\sinh\left(\pi \varphi\right)\sinh\left(\pi \varphi' \right)}
 {\sinh\left(\frac{\pi\left(\varphi-\varphi'\right)}{2}\right)
 \sinh\left(\frac{\pi\left(\varphi+\varphi'\right)}{2}\right)}
 e^{-\frac{\lambda}{2}\left(\varphi^{2}+\varphi'^{2}\right)
 - \frac{\tilde{y} \varphi^2}{2} - \frac{\tilde{x} {\varphi'}^2}{2}},
 \end{split}
 \end{equation}
where $\tilde{x} = \lambda^2 x$ and the same for $\tilde{y}$.

We can neglect the terms proportional to $\lambda$ in the exponent, as $\lambda \ll \tilde{x},\tilde{y} \ll 1$. Then
we can think of $(\varphi^2 - \varphi'^2)^\alpha$ as $\alpha$ derivatives with respect to $\tilde{\tau} = \tilde{x} - \tilde{y}$, so explicitly we get
\begin{equation}
\begin{split}
A_2(x,y;\alpha) &\approx \frac{2^{2\alpha} \lambda^{\alpha}}{\Gamma(\alpha)}
e^{-\frac{\pi^2}{4\lambda}} (-1)^\alpha \partial_{\tilde{\tau}}^\alpha \int_0^\infty d\varphi'd\varphi
\frac{\varphi+\varphi'}{\left[ (\alpha+1)^2 + (\varphi + \varphi')^2\right]} \\
 &\qquad \times \frac{\sinh\left(\pi \varphi\right)\sinh\left(\pi \varphi' \right)}
 {\sinh\left(\frac{\pi\left(\varphi-\varphi'\right)}{2}\right) \sinh\left(\frac{\pi\left(\varphi+\varphi'\right)}{2}\right)}
 e^{ - \frac{\tilde{\beta}}{2}\left( \varphi^2 + \varphi'^2 \right)
 - \frac{\tilde{\tau}}{2}\left( \varphi^2 - \varphi'^2 \right)},
 \end{split}
 \end{equation}
with $\tilde{\beta} = \tilde{x} + \tilde{y}$.

Following \cite{Micha2018}, we move to relative coordinates $\eta =\varphi - \varphi'$, $\rho = \varphi + \varphi'$, and approximating three of the $\sinh$'s as positive exponents, as well as changing the $\eta$ limits to $(-\infty,\infty)$ as it only receives contributions from finite $\eta$, we get
\begin{equation}
A_2(x,y;\alpha) \approx \frac{2^{2\alpha-2} \lambda^{\alpha}}{\Gamma(\alpha)}
e^{-\frac{\pi^2}{4\lambda}}  (-1)^\alpha \partial_{\tilde{\tau}}^\alpha  \int_0^\infty d\rho \intinf d\eta
\frac{\rho}{\left[ (\alpha+1)^2 +\rho^2\right]}
 \frac{1} {\sinh\left(\frac{\pi \eta}{2}\right)}
  e^{ - \tilde{\beta}\rho^2  - \frac{\tilde{\tau}}{2}\eta \rho + \frac{\pi \rho}{2}}.
 \end{equation}

As we are interested in Lorentzian time, we shall Wick rotate $\tau$ taking $i\tilde{t} = \tilde{\tau}$. Then notice that the $\eta$ integral is simply the Fourier transform of $1/\sinh(x)$, which is proportional $i \tanh(\omega)$, and taking one of the $\tilde{t}$ derivatives and get
\begin{equation}
A_2(x,y;\alpha) \approx \frac{ 2^{2\alpha-2} \lambda^{\alpha}}{\Gamma(\alpha)}
e^{-\frac{\pi^2}{4\lambda}}  i^{\alpha-1} \partial_{\tilde{t}}^{\alpha-1}  \int_0^\infty d\rho
\frac{\rho^2   e^{ - \tilde{\beta}\rho^2  + \frac{\pi \rho}{2}}}{\left[ (\alpha+1)^2 +\rho^2\right]}
 \frac{1}{\cosh^2\left(\frac{\tilde{t} \rho}{2}\right)}.
 \end{equation}

We can approximate this last integral in by a saddle point calculation: calling $w = \sqrt{\tilde{\beta}}(\rho-\frac{\pi}{4\tilde{\beta}})$ we see that
\begin{equation} \label{eq:A_2_conformal}
\begin{split}
A_2(x,y;\alpha) &\approx \frac{ 2^{2\alpha-2} \lambda^{\alpha}}{\Gamma(\alpha) \sqrt{\tilde{\beta}}}
e^{-\frac{\pi^2}{4\lambda}+\frac{\pi^2}{16\tilde{\beta}}}
 i^{\alpha-1} \partial_{\tilde{t}}^{\alpha-1}
\int_{-\frac{\pi}{4 \sqrt{\tilde{\beta}}}}^{\infty} dw
\frac{\left(\pi + \sqrt{\tilde{\beta}} w \right)^2   e^{ - w^2}}
{16 \tilde{\beta}^2(\alpha+1)^2 +(\pi+\sqrt{\tilde{\beta}} w)^2 } \\
&\qquad \qquad \qquad \times
 \frac{1}{\cosh^2\left(\frac{\tilde{t}}{8 \tilde{\beta}} \left(\pi+\sqrt{\tilde{\beta}} w \right)\right)} \\
 &\approx \frac{\pi^{1/2} 2^{2\alpha-2} \lambda^{\alpha}}{\Gamma(\alpha)  \sqrt{\tilde{\beta}}}
e^{-\frac{\pi^2}{4\lambda}+\frac{\pi^2}{16\tilde{\beta}}}
 i^{\alpha-1} \partial_{\tilde{t}}^{\alpha-1}
 \left(\frac{1}{\cosh^2\left(\frac{\tilde{t}\pi}{8 \tilde{\beta}} \right)} \right).
 \end{split}
 \end{equation}

From here we can read off the conformal dimension as follows (see \cite{Micha2018}): We shift $t \ra 4i\beta+t$ to turn the $\cosh$ into a $\sinh$, which at late times becomes the power law $\sim 1/t^2$. Then the derivatives in time tell us that we have a power law behavior of $1/t^{1 + \alpha}$, so the conformal dimension of this operator is $\frac{\alpha+1}{2}$.

\subsection*{$A_1(x,y;\alpha)$:}

We now turn to approximate $A_1(x,y;\alpha)$ in the conformal limit. We follow the same procedure conducted above for $A_2$. The energies can still be approximated by \eqref{eq:EValApprox}, reducing \eqref{eq:A_1_full} to
\begin{equation}
\begin{split}
A_1(x,y;\alpha) &\approx 2 \sqrt{\frac{\lambda}{\pi}} \int ds \frac{d\phi'd\phi}{(2\pi)^2}
\frac{\left(\tilde{q}^2,q^2,q^2,e^{\pm2i\phi},e^{\pm2i\phi'};q^{2}\right)_{\infty}}
{\left(q^\alpha e^{i\left(\pm\phi\pm\phi'\right)};q^{2}\right)_{\infty}}  \\
&\quad\times \left(\frac{1}{\lambda^2 \left(s-\frac{\alpha+1}{2}\right)^2 + \phi^2}
+\frac{1}{\lambda^2 \left(s+\frac{\alpha-1}{2}\right)^2 + \phi'^2}\right) \\
&\quad\times \exp\left\{-\frac{y}{2}\left(\lambda^2 \left(s-\frac{\alpha+1}{2}\right)^2 + \phi^2\right)
   - \frac{x}{2}\left(\lambda^2 \left(s+\frac{\alpha-1}{2}\right)^2 + \phi'^2\right)\right\}.
 \end{split}
 \end{equation}

As before, we can take the $s$ integral by neglecting the exponent to leading order, and we get
\begin{equation} \label{eq:A_1_first_approx}
\begin{split}
A_1(x,y;\alpha) &\approx \frac{1 }{2 \lambda^{1/2} \pi^{3/2}}  \int d\phi'd\phi
\frac{\left(\tilde{q}^2,q^2,q^2,e^{\pm2i\phi},e^{\pm2i\phi'};q^{2}\right)_{\infty}}
{\left(q^\alpha e^{i\left(\pm\phi\pm\phi'\right)};q^{2}\right)_{\infty}}
 \frac{(\phi+\phi')}{ \phi \phi'} e^{-\frac{y}{2} \phi^2 - \frac{x}{2}\phi'^2}.
 \end{split}
 \end{equation}

Then the Pochhammer symbols can be approximated using \eqref{eq:conf_approximation_1}, \eqref{eq:conf_approximation_2}, and \eqref{eq:conf_approximation_3}. Plugging these in, equation \eqref{eq:A_1_first_approx} becomes
\begin{equation} \label{eq:A_1_second_approx}
\begin{split}
A_1(x,y;\alpha) &\approx
\frac{2^{\alpha} \lambda^{\alpha}}{\Gamma(\alpha)} e^{-\frac{\pi^2}{4\lambda}}
\int d\varphi'd\varphi   (\varphi+\varphi') \left(\varphi^2-\varphi'^2\right)^{\alpha-1} \\
& \qquad \times\frac{\sinh(\pi \varphi) \sinh(\pi \varphi')}
{\sinh\left(\frac{\pi (\varphi+\varphi')}{2}\right)\sinh\left(\frac{\pi (\varphi-\varphi')}{2}\right)}
e^{-\frac{\lambda}{2}(\varphi^2 + \varphi'^2)-\frac{\tilde{y}}{2} \varphi^2 - \frac{\tilde{x}}{2}\varphi'^2},
 \end{split}
 \end{equation}
where again $\tilde{x} = \lambda^2 x$, $\varphi =  \phi/\lambda$, and the same for $\tilde{y}$ and $\varphi'$.

We now use the same tricks as before: the $\lambda$ term in the exponent is neglected, the term $\left(\varphi^2-\varphi'^2\right)^{\alpha-1}$ becomes derivatives in $\tilde{\tau} = \tilde{x} -\tilde{y} $, and we move to relative coordinate $\rho = \varphi + \varphi'$ and $\eta = \varphi- \varphi'$. Furthermore, we can extend the integration domain of $\eta$ by the same arguments as above and in \cite{Micha2018}. Then \eqref{eq:A_1_second_approx} reduces to
\begin{equation} \label{eq:A_1_third_approx}
\begin{split}
A_1(x,y;\alpha) \approx
\frac{2^{\alpha-2} \lambda^{\alpha}}{\Gamma(\alpha)} e^{-\frac{\pi^2}{4\lambda}}
(-1)^{\alpha-1} \partial_{\tilde{\tau}}^{\alpha-1}
\int_0^\infty d\rho \intinf d\eta  \frac{\rho}{\sinh\left(\frac{\pi\eta}{2}\right)}
e^{-\frac{\tilde{\beta}}{2}\rho^2 -\frac{\tilde{\tau}}{2}\eta \rho + \frac{\pi \rho}{2}}.
 \end{split}
 \end{equation}

As before, we shall move to Lorentzian time $\tilde{\tau} = i\tilde{t}$, allowing the exact evaluation of the $\eta$ integral. Then, after taking one of the derivatives in $t$, \eqref{eq:A_1_third_approx} takes the form
\begin{equation}
\begin{split}
A_1(x,y;\alpha) \approx
\frac{2^{\alpha-2} \lambda^{\alpha}}{\Gamma(\alpha)} e^{-\frac{\pi^2}{4\lambda}}
i^{\alpha-2} \partial_{\tilde{t}}^{\alpha-2}
\int_0^\infty d\rho~\frac{\rho^2}{ \cosh^2\left( \frac{\tilde{t}\rho}{2}\right)}
e^{-\frac{\tilde{\beta}}{2}\rho^2 + \frac{\pi \rho}{2}}.
 \end{split}
 \end{equation}

Following the procedure for $A_2$, we complete the square in the exponent, and approximate this integral by the saddle point $\rho = \pi/(4\tilde{\beta})$, giving us the conformal form of the two point function:
\begin{equation} \label{eq:A_1_conformal}
\begin{split}
A_1(x,y;\alpha) \approx
\frac{2^{\alpha-6} \lambda^{\alpha}}{\Gamma(\alpha)} ~\frac{\pi^{5/2}}{\tilde{\beta}^{5/2}}
e^{-\frac{\pi^2}{4\lambda} + \frac{\pi^2}{16\tilde{\beta}} } i^{\alpha-2} \partial_{\tilde{t}}^{\alpha-2}
\left(\frac{1}{ \cosh^2\left( \frac{\pi\tilde{t}}{8 \tilde{\beta}}\right)} \right).
 \end{split}
 \end{equation}

The conformal dimension of $A_1(x,y)$ can be read from equation \eqref{eq:A_1_conformal} via the same method used for $A_2(x,y;\alpha)$. This results in $A_1$ having a conformal form with dimension $\alpha/2$.

Notice that in the long time regime $A_2(x,y)$ decays much faster than $A_1(x,y)$, and so can be neglected. Thus the leading conformal dimension of this operator is $\alpha/2$. We expect an operator consisting of $\alpha$ fermions to have a leading conformal dimension of $\alpha/2$, and this is indeed what we get.

\subsection{The contributions from the ground states}   \label{app:2point_ground}
Now that we have found the conformal form of the 2-point functions, we turn to analyze the contributions from the ground states in this limit. We will show that these contributions are not negligible in comparison to the conformal part, and thus cannot be excluded from the analysis of correlation functions in this limit. The terms involving the ground states are $A_{1,2}(x,0;\alpha)$, $A_{1,2}(0,y;\alpha)$ (from \eqref{eq:I_1_as_A_1} and \eqref{eq:I_2_as_A_2}), and $I_1(x/y)$ (from \eqref{eq:2pointfun}). We will show that they are all of similar order in this limit, and are also of the same order as the conformal part, and that they do not cancel each other. Thus we must take the into account in the conformal regime.

 Notice that $A_{1,2}(x,0;\alpha)$ and $A_{1,2}(0,y;\alpha)$ have the same exact form in this limit, so we will only focus on later one. Furthermore, we shall ignore the constant contribution of the ground state, as it is independent of time, and in particular of the conformal limit.

\subsection*{$A_{1,2}(0,y;\alpha)$:}
We will start with $A_2(0,y;\alpha)$ in the conformal regime $\lambda \ll \lambda^2 y \ll 0$. Recall that
\begin{equation} \label{eq:A_2_zeros_1}
\begin{split}
A_2(0,y;\alpha) &= 2 q^{-1/4} \sqrt{\frac{\lambda}{\pi}} \int ds ~e^{-\lambda s^2} \frac{d\phi'd\phi}{(2\pi)^2}  \left(\tilde{q}^2,q^2,q^2,e^{\pm2i\phi},e^{\pm2i\phi'};q^{2}\right)_{\infty} \\
 &\quad\times \frac{q^{-\alpha}}{\left(q^{\alpha+1}e^{i\left(\pm\phi\pm\phi'\right)};q^{2}\right)_{\infty}
 ~\Lambda_{s-\frac{\alpha}{2}}\left(\phi\right) ~ \Lambda_{-s-\frac{\alpha}{2}}\left(\phi'\right)}
\exp\left\{-\frac{yq^{1/4}}{2}\Lambda_{s-\frac{\alpha}{2}}\left(\phi\right) \right\}.
 \end{split}
 \end{equation}

In the conformal  regime $\lambda\ll \lambda^2 y \ll 1$ only one of the energies will be localized around zero, while the other will take finite values. Thus the eigenvalues from \eqref{eq:A_2_zeros_1} can be approximate as
\begin{equation} \label{eq:Lambda_zeros}
\Lambda_{s-\frac{\alpha}{2}}\left(\phi'\right)  \approx 2(1-\cos \phi' ),
\quad \Lambda_{s-\frac{\alpha}{2}}\left(\phi\right) \approx
 \lambda^2 \left(s-\frac{\alpha + 1}{2}\right)^2 + \phi^2 .
\end{equation}

We can ignore the exponential factor of $e^{-\lambda s^2}$ as $\lambda \ll \lambda^2 y$, and take the $s$ integral. Thus \eqref{eq:A_2_zeros_1} reduces to
\begin{equation} \label{eq:A_2_zeros_2}
A_2(0,y;\alpha) \approx \frac{1}{4\lambda^{1/2} \pi^{3/2}} \int_0^\pi d\phi'd\phi
 \frac{\left(\tilde{q}^2,q^2,q^2,e^{\pm2i\phi},e^{\pm2i\phi'};q^{2}\right)_{\infty}}{\left(q^{\alpha+1}e^{i\left(\pm\phi\pm\phi'\right)};q^{2}\right)_{\infty}}
 \frac{\text{erfc}\left(\sqrt{y/2}\phi\right) }{ (1-\cos \phi')~\phi} .
 \end{equation}

We can now use the approximation from \cite{Micha2018} for the Pochhammer symbol
\begin{equation} \label{eq:pochammer_approximation_main}
q^{1/4}\left(q^{2},e^{\pm2i\phi};q^{2}\right)_{\infty}\approx 8\sin\phi\sqrt{\frac{\pi}{\lambda}}e^{-\frac{1}{\lambda}\left[\pi^{2}+\left(\phi-\frac{\pi}{2}\right)^{2}\right]}\sinh\left(\frac{\pi\phi}{\lambda}\right)\sinh\left(\frac{\pi\left(\pi-\phi\right)}{\lambda}\right),
\end{equation}
to write \eqref{eq:A_2_zeros_2} as
\begin{equation} \label{eq:A_2_zeros_3}
\begin{split}
A_2(0,y;\alpha) &\approx  \frac{\left(\tilde{q}^2,q^2,q^2;q^2\right)_{\infty}}{4\lambda^{1/2} \pi^{3/2}}
\int_0^\pi  d\phi'd\phi \frac{\text{erfc}\left(\sqrt{y/2}\phi\right)}{\phi (1-\cos\phi')}
\left( e^{i(\pm \phi \pm \phi')};q^2\right)_{\frac{\alpha + 1}{2}} \\
& \qquad \times \frac{\sin \phi ~\sin \phi'}{\sin\left(\frac{\phi+\phi'}{2}\right) \sin\left(\frac{\phi-\phi'}{2}\right)}
\frac{\sinh\left( \frac{\pi \phi}{\lambda}\right) \sinh\left( \frac{\pi \phi'}{\lambda}\right)}
{\sinh\left( \frac{\pi (\phi+\phi')}{2\lambda}\right) \sinh\left( \frac{\pi (\phi-\phi')}{2\lambda}\right)} \\
& \qquad \times \frac{\sinh\left( \frac{\pi (\pi-\phi)}{\lambda}\right) \sinh\left( \frac{\pi (\pi-\phi')}{\lambda}\right)}
{\sinh\left( \frac{\pi (2\pi - \phi -\phi')}{2\lambda}\right) \sinh\left( \frac{\pi (2\pi - \phi+\phi')}{2\lambda}\right)}
e^{-\frac{1}{\lambda}\left( \frac{\phi^2}{2} + \frac{(\phi'-\pi)^2}{2} - \frac{\pi^2}{2}\right)}.
 \end{split}
 \end{equation}

We can see that in the above equation that the main contribution comes from when $\phi$ is localized around zero and $\phi'$ is localized around $\pi$. Thus we define new coordinates $\varphi = \phi/\lambda$, $\varphi' = (\pi - \phi' )/\lambda$, and expand \eqref{eq:A_2_zeros_3} in $\lambda$ by taking the leading order approximations for $\sin$ and $\cos$ and taking the relevant exponent in the $\sinh$'s. Furthermore, we use \eqref{eq:conf_approximation_3}, as well as the Pochhammer symbol approximation
\begin{equation} \label{eq:z_pochammer_pi}
\left(e^{i\left(\pm\lambda\varphi\pm(\pi-\lambda\varphi')\right)};q^{2}\right)_{z/2} \approx 2^{2z},
\end{equation}
under which \eqref{eq:A_2_zeros_3} becomes
\begin{equation} \label{eq:A_2_zeros_4}
A_2(0,y;\alpha) \approx  \frac{ 2^{2\alpha}\lambda^{2-\alpha}}{ \Gamma(\alpha)} e^{-\frac{\pi^2}{4\lambda}}
 \int_0^\infty  d\varphi'd\varphi  ~\varphi' \sinh\left(\pi \varphi \right) \sinh\left( \pi \varphi'\right)
e^{- \frac{\lambda {\varphi'}^2}{2} - \pi \varphi' } \text{erfc}\left(\sqrt{y/2}\lambda\varphi\right).
 \end{equation}

The integrals in \eqref{eq:A_2_zeros_4} can easily be evaluated in the regime $\lambda \ll \lambda^2 y \ll 1$, giving us the final result
\begin{equation} \label{eq:A_2_zero_final}
A_2(0,y;\alpha) \approx  \frac{ 2^{2\alpha-1}\lambda^{1-\alpha}}{\pi \Gamma(\alpha)} e^{-\frac{\pi^2}{4\lambda}}
e^{\frac{\pi^2}{2 \tilde{y}}} ,
 \end{equation}
where, as above, $\tilde{y} = \lambda^2 y$. We see that this expression not negligible in comparison to the conformal parts, \eqref{eq:A_1_conformal} and \eqref{eq:A_2_conformal}, and actually seems parametrically larger.

The analysis for $A_1(0,y;\alpha)$ is identical. Starting with \eqref{eq:A_1_full}, we can use \eqref{eq:Lambda_zeros} to approximate the energies, so
\begin{equation}
\begin{split}
A_1(0,y;\alpha) &\approx 2 \sqrt{\frac{\lambda}{\pi}} \int ds \frac{d\phi'd\phi}{(2\pi)^2}  e^{-\lambda s^2}
\left(\frac{1}{\lambda^2\left(s-\frac{\alpha+1}{2}\right)^2+\phi^2}+\frac{1}{2(1-\cos\phi')}\right) \\
&\quad\times \frac{\left(\tilde{q}^2,q^2,q^2,e^{\pm2i\phi},e^{\pm2i\phi'};q^{2}\right)_{\infty}}
{\left(q^\alpha e^{i\left(\pm\phi\pm\phi'\right)};q^{2}\right)_{\infty}}
  \exp\left\{-\frac{y}{2} \left( \lambda^2\left(s-\frac{\alpha+1}{2}\right)^2+\phi^2 \right) \right\}.
 \end{split}
 \end{equation}

Again we can take the $s$ integral in the approximation $\lambda \ll \lambda^2 y \ll 1$ be neglecting the exponent, leading to
\begin{equation} \label{eq:A_1_zeros_2}
\begin{split}
A_1(0,y;\alpha) &\approx 2 \sqrt{\frac{\lambda}{\pi}} \int\frac{d\phi'd\phi}{(2\pi)^2}  \frac{\left(\tilde{q}^2,q^2,q^2,e^{\pm2i\phi},e^{\pm2i\phi'};q^{2}\right)_{\infty}}
{\left(q^\alpha e^{i\left(\pm\phi\pm\phi'\right)};q^{2}\right)_{\infty}} \\
&\quad \quad\times \left(\frac{\pi}{\lambda\phi} \text{erfc}\left(\sqrt{y/2}\phi\right)
 +e^{-\frac{y}{2} \phi^2} \frac{\sqrt{\pi}}{ \sqrt{2y\lambda^2}(1-\cos\phi')}\right) .
 \end{split}
 \end{equation}

As $\lambda \ll \lambda^2 y \ll 1$, we can neglect the second part of the sum in \eqref{eq:A_1_zeros_2}. Then we use the same approximations from before for the Pochhammer symbols, namely \eqref{eq:z_pochammer_pi}, \eqref{eq:pochammer_approximation_main} and \eqref{eq:conf_approximation_3}. As a result $\phi$ is still localized around zero and $\phi'$ around $\pi$, allowing us to do the change of variable $\varphi = \phi/\lambda$ and $\varphi' = (\pi-\phi')/\lambda$. After all this, and approximating the resulting trigonometric and hyperbolic functions, \eqref{eq:A_1_zeros_2} becomes
\begin{equation}
A_1(0,y;\alpha) \approx \frac{2^{2\alpha}\lambda^{2-\alpha } }{\Gamma(\alpha)}
 e^{-\frac{\pi^2}{4\lambda}}
\int_0^\infty d\varphi'd\varphi \varphi' \sinh(\pi \varphi) \sinh(\pi \varphi' )
  e^{- \frac{\lambda}{2}\left(\varphi^2+\varphi'^2\right) - \pi \varphi' }  \text{erfc}\left(\sqrt{\tilde{y}/2}\phi\right) .
 \end{equation}

This is identical to the result for $A_2(0,y;\alpha)$, \eqref{eq:A_2_zero_final}, so in the conformal regime limit $A_1(0,y;\alpha) \approx A_2(0,y;\alpha)$, and both give significant contributions.

\subsection*{$I_1(x)$}

There is an additional contribution from the ground states from summing the $m_{0,k}$ and $m_{k,0}$  moments, given by \eqref{eq:I_1_definition}. We will analyze it in the same conformal regime $\lambda \ll \lambda^2x,\lambda^2 y \ll 1$. Notice that from  \eqref{eq:I_1_definition} $I_1(x)$ takes the form $I_1(x)  = I_1^{c}(x) +C $ for some constant $C$, and
\begin{equation}
I_1(x)  =  \sqrt{\frac{\lambda}{\pi}} \int_{-\infty}^{\infty}ds~ q^{s^{2}-s+s_0^2}\int_{0}^{\pi}\frac{d\phi}{2\pi}\left(q^{2},e^{\pm2i\phi};q^{2}\right)_{\infty} \frac{\cosh\left(2\lambda s_{0}s\right)}{\Lambda_{s}(\phi)} e^{-x \frac{q^{1/4}\Lambda_{s}(\phi)}{2}}.
\end{equation}
We shall ignore the constant term in the following computations.

As $x$ is large, we can approximate the energies using \eqref{eq:EValApprox}, resulting in
\begin{equation} \label{eq:A_1_conf_1}
I_1(x)  \approx  \sqrt{\frac{\lambda}{\pi}} \int_{-\infty}^{\infty}ds~ q^{s^{2}-s+s_0^2}
\int_{0}^{\pi}\frac{d\phi}{2\pi}\left(q^{2},e^{\pm2i\phi};q^{2}\right)_{\infty}
\frac{\cosh\left(2\lambda s_{0}s\right)}{\lambda^2(s-1/2)^2 + \phi^2}
e^{-x \frac{\lambda^2 (s-1/2)^2 + \phi^2}{2}}.
\end{equation}

As before, in the conformal regime $\phi$ is localized around zero, so we can take the $s$ integral by ignoring the exponents proportional to  $\lambda$, after which \eqref{eq:A_1_conf_1} becomes
\begin{equation} \label{eq:A_1_conf_2}
I_1(x)  \approx \frac{1}{2 \sqrt{\lambda\pi }}
\int_{0}^{\pi} \frac{d\phi}{\phi} \left(q^{2},e^{\pm2i\phi};q^{2}\right)_{\infty}
\text{erfc}\left(\sqrt{x/2}\phi \right).
\end{equation}

The Pochhammer symbol can be approximated using \eqref{eq:pochammer_approximation_main}, and after the change of variable $\varphi = \phi/\lambda$, and approximating one of the $\sinh$'s as a positive exponent \eqref{eq:A_1_conf_2} reduces to
\begin{equation} \label{eq:A_1_conf_3}
I_1(x)  \approx 2 e^{-\frac{\pi^2}{4\lambda}}
\int_{0}^{\infty} d\varphi~\sinh\left(\pi\varphi\right)
e^{ -\lambda \varphi^2} \text{erfc}\left(\sqrt{\tilde{x}/2}\varphi \right),
\end{equation}
with $\tilde{x} = \lambda^2 x$. Neglecting the exponent with $\lambda$, we can integrate \eqref{eq:A_1_conf_3} by parts, resulting in
\begin{equation} \label{eq:A_1_conf_fin}
\begin{split}
I_1(x)  &\approx \frac{2^{3/2}\sqrt{\tilde{x}}}{\pi^{3/2}} e^{-\frac{\pi^2}{4\lambda}}
\int_{0}^{\infty} d\varphi~\cosh\left(\pi\varphi\right) e^{-\frac{\tilde{x}}{2} \varphi^2} \\
&=  \frac{4}{\pi} e^{-\frac{\pi^2}{4\lambda}} e^{\frac{\pi^2}{2\tilde{x}}} .
\end{split}
\end{equation}
The result for $A_1$, given by \eqref{eq:A_1_conf_fin}, is simply the Schwarzian partition function $Z(x)$, which is not surprising as the moments we exponentiated are the regular moments of the distribution. However, we see that this is also of the same order of magnitude as the other terms, and thus is not negligible.

\bibliography{SUSY_SYK}
\bibliographystyle{JHEP}

\end{document}